\documentclass[aps, prd, reprint, twocolumn, nofootinbib, superscriptaddress]{revtex4-2}
\usepackage{amsmath}
\usepackage{amssymb}
\usepackage{amsfonts}
\usepackage{amsthm}
\usepackage{mathrsfs}
\usepackage{booktabs, array}
\usepackage{soul}
\usepackage[T1]{fontenc}
\usepackage[caption=false]{subfig}
\usepackage{graphicx}
\usepackage{float} 
\usepackage{multirow}
\usepackage{makecell}
\usepackage[dvipsnames]{xcolor}
\usepackage{tikz}

\newcommand{\bGR}{{\mbox{\tiny bGR}}}
\newcommand{\geo}{{\mbox{\tiny geo}}}

\newcommand{\nn}{\nonumber\\}
\newcommand{\pa}{\partial}

\definecolor{linkcolor}{rgb}{0.0,0.3,0.5}
\usepackage[
    hypertexnames=false,
    unicode,
    colorlinks=true,
    linkcolor=linkcolor,
    citecolor=linkcolor,
    filecolor=linkcolor,
    urlcolor=linkcolor,
    pdfusetitle
]{hyperref}
\usepackage{url}
\hypersetup{breaklinks=true}

\usepackage{orcidlink}

\newcommand{\UIUC}{Illinois  Center  for  Advanced  Studies  of  the  Universe \&
Department of Physics, University of Illinois at Urbana-Champaign, Urbana, Illinois 61801, USA}
\newcommand{\CAL}{Theoretical Astrophysics 350-17, California Institute of Technology, Pasadena, CA 91125, USA}
\newcommand{\Tsinghua}{Department of Astronomy, Tsinghua University, Beijing 100084, China}

\allowdisplaybreaks
\begin{document}
    \title{Modeling extreme mass-ratio inspirals in astrophysical and dark-matter environments}
    \author{Dongjun Li}
    \affiliation{\UIUC}

    \author{Hao-Jui Kuan}
    \affiliation{\UIUC}

    \author{Colin Weller}
    \affiliation{\CAL}

    \author{Chaoyi Yang}
    \affiliation{\Tsinghua}

     \author{Kelvin Ka-Ho Lam}
    \affiliation{\UIUC}

    \author{Patrick Bourg}
    \affiliation{School of Mathematics \& Statistics, University College Dublin, Belfield, Dublin 4, Ireland, D04 V1W8}
    
    \author{Adrian Ka-Wai Chung}
    \affiliation{St John's College, Cambridge, CB2 1TP, United Kingdom}
    \affiliation{DAMTP, Centre for Mathematical Sciences, University of Cambridge, Wilberforce Road, Cambridge CB3 0WA, United Kingdom}

    \author{Yanbei Chen}
    \affiliation{\CAL}

    \author{Nicol\'as Yunes}
    \affiliation{\UIUC}
    
    \author{Huan Yang}
    \email{hyangdoa@tsinghua.edu.cn} 
    \affiliation{\Tsinghua}

    \date{\today}
    
    \begin{abstract}
        Extreme mass-ratio inspirals offer a powerful probe of massive black holes and their environments, but interpreting their signals requires accurate models of orbital dynamics and radiation. Here we bring together advances in black-hole perturbation theory to establish a unified relativistic framework for adiabatic inspirals in perturbative astrophysical environments. Built on the modified Teukolsky formalism, the framework requires neither post-Newtonian nor slow-rotation approximations. As a first application, we calculate the leading cloud-induced corrections to gravitational-wave fluxes and conservative orbital dynamics for circular, equatorial inspirals around a rotating black hole surrounded by a synchronized ultralight scalar cloud, combining these corrections with the previously calculated scalar response. In our fiducial system, a cloud containing $10^{-4}$ of the black-hole mass produces about $-57.3$ radians of orbital dephasing over two years. The phase contributions from gravitational-wave flux corrections and background-induced orbital shifts each exceed the scalar contribution in magnitude and partially cancel, highlighting the need for coupled matter-gravity modeling. Extensible to other environments and modified gravity theories, our framework provides the relativistic inputs needed to assess detection losses and parameter biases, and to probe the nature of black-hole environments with future space-based gravitational-wave observations.
        \end{abstract}
    \maketitle

\section{Introduction}

\begin{figure*}[t]
    \centering
    \includegraphics[width=1\linewidth]{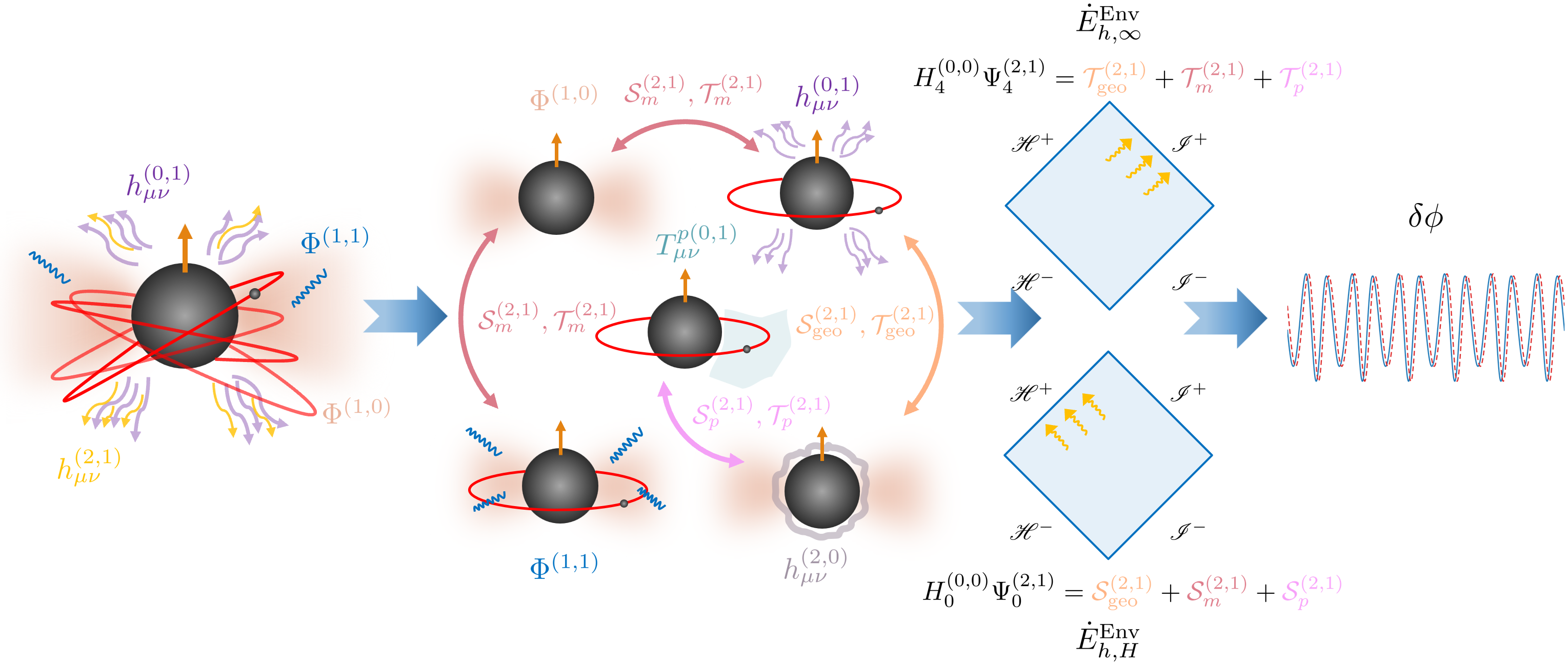}
    \caption{The schematic illustrates the modified Teukolsky formalism applied to an EMRI in an ultralight scalar cloud. 
    {\bf{Left}}: A small companion (grey dot) follows a circular, equatorial orbit around a rotating black hole. Peach shading represents the background cloud $\Phi^{(1,0)}$. Purple, blue and gold wavy arrows represent the vacuum-GR gravitational radiation $h_{\mu\nu}^{(0,1)}$, scalar response $\Phi^{(1,1)}$, and environmental correction to gravitational radiation $h_{\mu\nu}^{(2,1)}$, respectively. 
    {\bf{Middle, left}}: The perturbative ingredients of the same system, including the background metric correction $h_{\mu\nu}^{(2,0)}$ and particle stress-energy $T_{\mu\nu}^{p(0,1)}$. Double-headed arrows indicate couplings contributing to the sources in Eq.~\eqref{eq:MTF}. {\bf{Middle, right}}: The resulting $\Psi_{4}^{(2,1)}$ and $\Psi_{0}^{(2,1)}$, together with lower-order fields, determine the gravitational-wave flux corrections $\dot{E}_{h,\infty}^{\mathrm{Env}}$ and $\dot{E}_{h,H}^{\mathrm{Env}}$. Schematic Penrose diagrams identify future null infinity $\mathscr{I}^{+}$ and the future event horizon $\mathcal{H}^{+}$ as the respective extraction boundaries. 
    {\bf{Right}}: These corrections, the scalar fluxes, and the background-induced orbital-energy and frequency shifts determine the orbital dephasing $\delta\phi$, illustrated schematically by the final curves.}
    \label{fig:MTF}
\end{figure*}

Extreme mass-ratio inspirals (EMRIs), in which stellar-mass compact objects spiral into massive black holes, are key targets for future space-based gravitational-wave detectors \cite{LISA:2017pwj, Hu:2017mde, TianQin:2015yph}. Their hundreds of thousands of orbital cycles make them sensitive probes of strong gravity and of environments \cite{LISA:2022kgy, Barausse:2014pra, Cole:2022yzw}, such as accretion disks \cite{Barausse:2007dy, Yunes:2011ws, Kocsis:2011dr, HegadeKR:2025rpr}, many-body interactions \cite{Yang:2017aht, Bonga:2019ycj, Yang:2019iqa, Santos:2025ass}, dark-matter halos/spikes \cite{Ullio:2001fb, Kavanagh:2020cfn, Cardoso:2022whc, Mitra:2025tag}, and ultralight bosonic clouds \cite{Duque:2023seg, Brito:2023pyl, Dyson:2025dlj, Li:2025ffh, Xu:2026aic, Keijzer:2026vul}. Interpreting these signals requires accurate models of how the environment affects both orbital dynamics and gravitational-wave emission \cite{LISAConsortiumWaveformWorkingGroup:2023arg}.

In vacuum general relativity (GR), black-hole perturbation theory provides a systematic tool to model EMRIs through expansions in the mass ratio \cite{Mino:1996nk, Hughes:1999bq, Barack:2007tm, Hinderer:2008dm, Barack:2018yvs, Pound:2021qin}. Leading-order-in-the-mass-ratio (adiabatic) models evolve a sequence of geodesics using gravitational-wave fluxes obtained from solutions to the Teukolsky equation \cite{Teukolsky:1973ha, Teukolsky:1974yv, Hughes:1999bq, Hinderer:2008dm}, while higher-order self-force calculations extend this description beyond the leading-order approximation \cite{Pound:2019lzj, Wardell:2021fyy, Bourg:2024vre, Mathews:2025nyb}. Extending this description to an environment, however, introduces additional challenges. One must systematically account for the deformation of the background geometry, the response of the surrounding matter to the secondary, and their coupled effects on gravitational radiation.

Much work has gone into the careful modeling of environments in EMRIs. Recent advances include relativistic scalar-cloud EMRI models for nonrotating primaries \cite{Brito:2023pyl, Duque:2023seg, Keijzer:2026vul, Xu:2026aic}, scalar-radiation-only contributions for rotating primaries \cite{Dyson:2025dlj, Li:2025ffh}, and studies of other environmental inspirals \cite{Maselli:2020zgv, Barsanti:2022vvl, Roy:2025kra, HegadeKR:2025rpr, Yang:2017aht, Bonga:2019ycj, Yang:2019iqa, Santos:2025ass, Lui:2026azr, Polcar:2025yto}. Environmental corrections to gravitational-wave fluxes and orbital energy have already been shown to play an important role in spherical configurations \cite{Keijzer:2026vul}. The outstanding task addressed here is to calculate these contributions together for a rotating primary. This requires not only a systematic, reusable relativistic framework analogous to the Teukolsky-based treatment of vacuum GR, but also an efficient numerical implementation for evaluating the complex source terms describing matter--gravity coupling and solving the resulting field and orbital-evolution equations.

This article brings together a sequence of developments in the \emph{modified Teukolsky formalism} \cite{Li:2022pcy, Hussain:2022ins} and its applications to EMRIs \cite{Li:2025ffh, LaHaye:2025ley, Yang:2026erj, Yang:2026csy} to produce a unified framework for adiabatic inspirals in \emph{perturbative environments}, where we can treat non-vacuum effects as a perturbation to the inspiral (Fig.~\ref{fig:MTF}). Our central advance is to calculate how the environmental deformation of the background spacetime and the response of matter to the secondary modify gravitational radiation around rotating black holes. We overcome the computational complexity of these interactions through a reusable numerical strategy, making it possible to incorporate their effects on radiation and orbital motion together. This turns our perturbative framework into a practical tool for connecting black-hole environments to their gravitational-wave imprints. The treatment is fully relativistic: it expands in mass ratio and environmental amplitude, without post-Newtonian or slow-rotation expansions.

As a first application, we calculate the leading-order cloud-induced corrections to gravitational radiation and conservative orbital dynamics for circular, equatorial inspirals around a rotating black hole surrounded by a dipolar ultralight scalar cloud synchronized with its horizon angular velocity. Combining these corrections with the previously calculated scalar response \cite{Li:2025ffh}, we obtain an orbital dephasing of approximately $-57.3\,\mathrm{rad}$ after two years for a cloud containing $10^{-4}$ of the primary black-hole mass, nearly ten times the magnitude of the scalar-radiation-only prediction. The phase contributions from the gravitational-wave flux corrections and background-induced orbital shifts each exceed the scalar contribution in magnitude and partially cancel. This demonstrates that matter-response-only models can substantially misestimate the environmental imprint, even for a dilute environment. Beyond this application, our framework provides a basis for treating other perturbative environments and connects to applications in modified gravity~\cite{Yang:2026erj, Yang:2026csy}. By making the coupled matter--gravity response computationally accessible, it opens a route to using future space-based gravitational-wave observations \cite{LISA:2017pwj,Hu:2017mde,TianQin:2015yph} to reveal the nature of matter or dark matter around massive black holes and disentangle their signatures from departures from GR.

\section{Modified Teukolsky formalism applied to EMRIs in non-vacuum environments}

Consider an EMRI embedded in an environment described by a minimally coupled matter field $\varphi$, with the secondary interacting with the environment only through gravity. The metric and environmental field obey
\begin{subequations} \label{eq:EOM}
\begin{align} 
    & G_{\mu\nu}
    =8\pi G\left(T_{\mu\nu}^{m}+T_{\mu\nu}^{p}\right)\,,
    \quad\nabla^\mu\left(T_{\mu\nu}^{m}+T_{\mu\nu}^{p}\right)=0\,,
    \label{eq:EE} \\
    & \hat{\mathcal{M}}[\varphi]=0\,, \label{eq:EOM_phi}
\end{align}    
\end{subequations}
where $G_{\mu\nu}$ is the Einstein tensor, $T_{\mu\nu}^{m}$ is the stress-energy tensor of the environment field $\varphi$, and $T_{\mu\nu}^{p}$ is the point-particle stress-energy tensor associated with the secondary. The equation of the environment field $\varphi$ in Eq.~\eqref{eq:EOM_phi} depends on the metric implicitly and can be nonlinear in $\varphi$. For the ultralight scalar cloud considered later, Eq.~\eqref{eq:EOM_phi} reduces to the Klein--Gordon equation. Both $T_{\mu\nu}^{m}$ and $T_{\mu\nu}^{p}$ are conserved, and the conservation of the latter determines the secondary's motion. At leading order in the mass ratio, this gives geodesic motion in the environment-deformed background.

To evolve the inspiral adiabatically, we expand Eq.~\eqref{eq:EOM} in the mass ratio $\epsilon=m_p/M\ll1$, where $M$ and $m_p$ are the primary and secondary masses, respectively. We also introduce a dimensionless parameter $\zeta$ tracking the amplitude of the environmental field $\varphi$. Its relation to the characteristic environmental density or mass depends on the matter model and background configuration. We take $\zeta$ sufficiently small that the environmental backreaction remains perturbative. Assuming that $T_{\mu\nu}^{m}$ is quadratic in $\varphi$ to the leading order, the metric and environmental field admit the double expansion
\begin{subequations} \label{eq:expansion}
\begin{align}
    & g_{\mu\nu}
    =g^{(0,0)}_{\mu\nu}+\epsilon h^{(0,1)}_{\mu\nu}
    +\zeta^2 h^{(2,0)}_{\mu\nu}
    +\zeta^2\epsilon h^{(2,1)}_{\mu\nu}+\cdots\,,
    \label{eq:expansion_metric} \\
    & \varphi
    =\zeta\varphi^{(1,0)}
    +\zeta\epsilon\varphi^{(1,1)}+\cdots\,,
\end{align}
\end{subequations}
where the superscripts count the orders in $\zeta$ and $\epsilon$, and $g_{\mu\nu}^{(0,0)}$ is the vacuum background metric, or the Kerr spacetime here.

In vacuum, the secondary sources the gravitational radiation $h_{\mu\nu}^{(0,1)}$. The background environmental field $\varphi^{(1,0)}$ has a stationary stress-energy tensor and induces the background metric deformation $h_{\mu\nu}^{(2,0)}$. This deformation modifies the secondary's geodesic motion, producing $\mathcal{O}(\zeta^2)$ shifts in its orbital frequencies and specific energy. We include these \emph{environmental conservative effects}, which persist in the test-particle limit, but not the conservative self-force corrections arising from the secondary's own gravitational field, which contribute to the orbital phase at post-adiabatic order \cite{Hinderer:2008dm}. As the secondary moves through the environment, it also induces the matter response $\varphi^{(1,1)}$, which could be radiative. The background deformation, the matter response, and the corresponding correction to the particle source, together with the vacuum metric perturbation $h_{\mu\nu}^{(0,1)}$, generate $h_{\mu\nu}^{(2,1)}$, the leading environmental correction to the gravitational radiation.

In vacuum GR, the gravitational perturbation $h_{\mu\nu}^{(0,1)}$ is described by the Weyl scalars $\Psi_0^{(0,1)}$ and $\Psi_4^{(0,1)}$ in the Newman–Penrose formalism \cite{Newman:1961qr}. These scalars satisfy separable, second-order Teukolsky equations \cite{Teukolsky:1973ha} and encode radiation at the primary black-hole horizon and at future null infinity, respectively. For perturbative environments, the corresponding corrections to these NP scalars, $\Psi_{0,4}^{(2,1)}$, satisfy the \emph{modified Teukolsky equations} \cite{Li:2022pcy, Hussain:2022ins, Li:2025ffh, LaHaye:2025ley}:
\begin{subequations} \label{eq:MTF}
\begin{align} 
    & H_{0}^{(0,0)}\Psi_0^{(2,1)}
    =\mathcal{S}_{\geo}^{(2,1)}+\mathcal{S}_{m}^{(2,1)}
    +\mathcal{S}_{p}^{(2,1)}\,, \label{eq:MTF_Psi0}\\
    & H_{4}^{(0,0)}\Psi_4^{(2,1)}
    =\mathcal{T}_{\geo}^{(2,1)}+\mathcal{T}_{m}^{(2,1)}
    +\mathcal{T}_{p}^{(2,1)}\,, \label{eq:MTF_Psi4}
\end{align} 
\end{subequations}
where $H_{0,4}^{(0,0)}$ is the standard Teukolsky operator that acts on $\Psi_{0,4}^{(0,1)}$ in vacuum GR \cite{Teukolsky:1973ha}. The sources $\mathcal{S}_{\geo}^{(2,1)}$ and $\mathcal{T}_{\geo}^{(2,1)}$ drive the corrections to gravitational radiation when the background spacetime deviates from Kerr. The other two sets of sources, $\{\mathcal{S}_{m}^{(2,1)}, \mathcal{T}_{m}^{(2,1)}\}$ and $\{\mathcal{S}_{p}^{(2,1)}, \mathcal{T}_{p}^{(2,1)}\}$, are driven by the matter stress-energy tensor $T_{\mu\nu}^{m}$ and the point-particle stress-energy tensor $T_{\mu\nu}^{p}$. As shown in \cite{Li:2022pcy, LaHaye:2025ley}, all the sources in Eq.~\eqref{eq:MTF} depend only on fields of order lower than $\mathcal{O}(\zeta^2,\epsilon)$, because the coupling constants multiplying them are themselves of $\mathcal{O}(\zeta^2)$ or $\mathcal{O}(\epsilon)$. Thus, Eq.~\eqref{eq:MTF} decouples from other $\mathcal{O}(\zeta^2,\epsilon)$ quantities, largely simplifying the study of gravitational radiation in complicated scenarios. The precise forms of these sources in the Newman-Penrose variables are provided in the Supplemental Material.

\section{Application of the modified Teukolsky formalism to ultralight scalar clouds}

As a first application, we consider a circular, equatorial EMRI embedded in an ultralight scalar cloud described by
\begin{subequations}
\begin{align} 
    & \left(\square-\mu_{\Phi}^2\right)\Phi=0\,, \label{eq:EOM_scalar} \\
    & T_{\mu\nu}^{\Phi}
    =2\partial_{(\mu}\Phi\partial_{\nu)}\bar{\Phi}-g_{\mu\nu}
    \left(\partial^\sigma\Phi\partial_\sigma\bar{\Phi}
    +\mu_{\Phi}^2\Phi\bar{\Phi}\right)\,, \label{eq:stress_cloud}
\end{align}    
\end{subequations}
where $\Phi$ is a complex scalar field without self-interactions. Throughout, we use geometrized units $G=c=1$ and define $\mu_\Phi$ as the inverse reduced Compton wavelength of the scalar field, $\mu_\Phi=m_\Phi c/\hbar$, where $m_\Phi$ is the scalar-particle rest mass and
$\hbar$ is the reduced Planck constant. The combination $\mu_\Phi M$ is therefore dimensionless. Following Refs.~\cite{Dyson:2025dlj, Li:2025ffh}, we define $\zeta=(\mu_{\Phi} M)^3\sqrt{M_c/M}$, which is the square root of the cloud-to-black-hole characteristic density ratio, where $M_c$ is the total cloud mass. We then expand the metric and scalar field according to Eq.~\eqref{eq:expansion}, with $\varphi$ replaced by $\Phi$. Our goal is then to solve Eqs.~\eqref{eq:MTF} and \eqref{eq:EOM_scalar} perturbatively under $\epsilon\ll 1$ and $\zeta\ll 1$.  

From the leading-order part of Eq.~\eqref{eq:EOM_scalar}, which is essentially the massive spin-0 Teukolsky equation on a vacuum Kerr background, we obtain the cloud profile $\Phi^{(1,0)}$ and frequency $\omega_c$ via Leaver's continued-fraction method \cite{Leaver:1985ax, Dolan:2007mj}. In suitable regions of the $(a,\mu_{\Phi})$ parameter space, where $a$ is the primary black-hole spin parameter, Eq.~\eqref{eq:EOM_scalar} admits quasi-bound states that satisfy $|\mathrm{Im}[\omega_c]|\ll\mathrm{Re}[\omega_c]$, decay exponentially at infinity, and can grow into macroscopic clouds in the superradiant regime~\cite{Dolan:2007mj}. For $\mu_{\Phi} M\sim1$, superradiant growth can produce macroscopic clouds~\cite{Dolan:2007mj}, with masses reaching $\sim0.1M$~\cite{Herdeiro:2014goa, Khalvati:2024tzz}. Here, we consider the same fundamental dipolar mode as in Ref.~\cite{Li:2025ffh}, with $\mu_{\Phi}M=0.3$ and $a/M\approx 0.877153$. The spin is chosen to satisfy the synchronization condition $\omega_c=m_c\omega_+$, so $\omega_c$ is effectively real with $\mathrm{Im}[\omega_{c}]\approx 10^{-24}\mathrm{Re}[\omega_{c}]$; this is the sense in which we say we focus on a synchronized scalar cloud.  Since $\omega_c$ is determined by the scalar bound-state problem, a parameter survey of stationary clouds must satisfy the bound-state and synchronization conditions jointly, rather than vary the black-hole and scalar-field parameters independently while retaining the same cloud profile.

With the background scalar field in hand, we can now focus on the full black-hole--cloud spacetime. For our single-frequency cloud, the complex functions $\Phi$ and $\bar{\Phi}$ in Eq.~\eqref{eq:stress_cloud} give a stationary stress-energy tensor when $\omega_c$ is real. We compute the full black-hole--cloud spacetime \cite{Herdeiro:2014goa} using the multi-domain pseudospectral method of Refs.~\cite{Garcia:2023ntf, Kuan:2026pss}. Subtracting the reference Kerr metric and extracting the $\zeta^2$ coefficient gives $h_{\mu\nu}^{(2,0)}$. We assess the perturbative regime by varying the scalar amplitude and checking that the metric deviation scales quadratically. The numerical procedure and these checks are described in the Methods and Supplemental Material.

At $\mathcal{O}(\zeta, \epsilon)$, the vacuum gravitational radiation couples to the cloud, generating the scalar response $\Phi^{(1,1)}$. Expanding Eq.~\eqref{eq:EOM_scalar} to this order shows that its source involves the metric perturbation sourced by the small compact object, $h_{\mu\nu}^{(0,1)}$, explicitly. We obtain this metric perturbation in two steps: we first compute the particle-induced $\Psi_{0,4}^{(0,1)}$ using the \texttt{Teukolsky} package in the BHPToolkit \cite{Teukolsky:1973ha, Hughes:1999bq, BHPToolkit}, and then reconstruct $h_{\mu\nu}^{(0,1)}$ from these NP scalars. For circular, equatorial orbits, we use the Lorenz-gauge \texttt{Kerr-Lorenz-Circ} implementation of Refs.~\cite{Dolan:2021ijg, Dolan:2023enf}, which also underlies the earlier calculations of the scalar response and the associated energy and angular-momentum transfer in Refs.~\cite{Dyson:2025dlj, Li:2025ffh}. Other metric-reconstruction approaches are available \cite{Green:2019nam, Wardell:2024yoi, Nasipak:2025tby, Li:2026rkf}, but adapting them to the present source construction would require greater control over their gauge, completion, and behavior at the location of the secondary.

Our central result is the calculation of the cloud-induced gravitational correction, $\Psi_{0,4}^{(2,1)}$, obtained by combining these ingredients in Eq.~\eqref{eq:MTF}, with Figure~\ref{fig:MTF} summarizing their coupling structure. Obtaining the lower-order fields does not make this step automatic: their couplings produce large source expressions involving higher derivatives and contributions extending to the horizon
and infinity. For example, the fully expanded source expressions contain $\sim10^6$ algebraic terms, while the present
calculation samples $\sim10^4$ harmonic modes. We avoid repeated evaluation of these expressions by organizing them into three reusable modules: angular projection coefficients, auxiliary radial functions, and radial metric and scalar-field factors. Each module is either mode-independent or requires only $\sim10^2$--$10^3$ terms per mode, making
the gravitational-wave flux calculation practical and facilitating higher-resolution calculations. After angular projection, we solve the resulting sourced radial Teukolsky equations using Green's functions. Source construction, boundary regularization, and treatment of the point-particle source are detailed in the Methods and Supplemental Material.

\section{Energy fluxes}
\label{sec:fluxes}

\begin{figure}[t]
    \centering
    \includegraphics[width=\columnwidth]{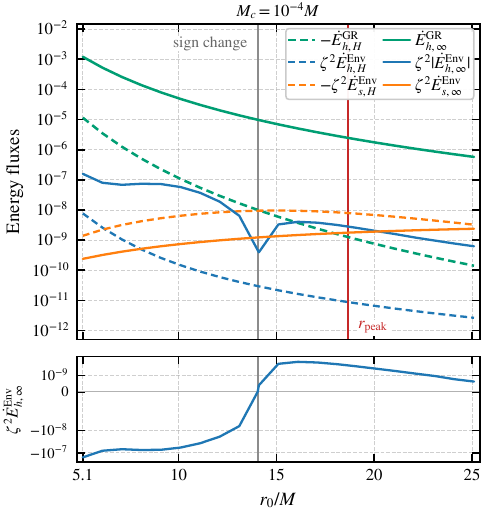}
    \caption{Mode-summed fluxes for circular, equatorial orbits with $5.1\leq r_0/M\leq25.1$, $a/M\approx0.877153$, $\mu_\Phi M=0.3$ and $M_c/M=10^{-4}$ ($\zeta=2.7\times10^{-4}$), with the common factor $\epsilon^2$ scaled out. Teal, blue and orange curves show vacuum-GR gravitational-wave fluxes $\dot{E}_{h,H/\infty}^{\mathrm{GR}}$, their cloud-induced corrections $\dot{E}_{h,H/\infty}^{\mathrm{Env}}$, and scalar fluxes $\dot{E}_{s,H/\infty}^{\mathrm{Env}}$ from Ref.~\cite{Li:2025ffh}, respectively. \textbf{Top:} Flux magnitudes, with dashed and solid curves denoting horizon and infinity channels. Minus signs identify negative quantities, while $\dot{E}_{h,\infty}^{\mathrm{Env}}$ has a sign change near $r_0=14M$, as denoted by the gray solid line. The red solid line marks the cloud-envelope maximum at $r_{\mathrm{peak}}\approx18.6M$. \textbf{Bottom:} The signed correction $\dot{E}_{h,\infty}^{\mathrm{Env}}$. The mode sums include $2\leq\ell\leq5$ for the gravitational-wave fluxes and their cloud-induced corrections, and $0\leq\ell_s\leq6$ for the scalar fluxes.}
    \label{fig:fluxes_total}
\end{figure}

The orbital-energy balance provides the dissipative input for the adiabatic evolution of circular, equatorial orbits, with the corresponding angular-momentum balance fixed by the orbital-energy one. For the particular case of the scalar cloud, we distinguish gravitational-wave and scalar contributions, labeled by $h$ and $s$, respectively, and resolve each into horizon ($H$) and infinity ($\infty$) channels. The energy balance law is then given by \cite{Dyson:2025dlj, Li:2025ffh}:
\begin{equation} \label{eq:energy_balance_channels}
   \dot{E}_{\rm orb}
    =-\epsilon^2\sum_{A=H,\infty}
    \left[\dot{E}_{h,A}^{\mathrm{GR}}
    +\zeta^2\dot{E}_{h,A}^{\mathrm{Env}}
    +\zeta^2\dot{E}_{s,A}^{\mathrm{Env}}\right]\,.
\end{equation}
Here, $\dot{E}_{h,A}^{\mathrm{GR}}$ denotes the vacuum-GR gravitational-wave flux with $\epsilon^2$ scaled out, and $\zeta^2\dot{E}_{h,A}^{\mathrm{Env}}$ is its leading environment-induced correction. For the EMRI--cloud system, the scalar contribution $\zeta^2\dot{E}_{s,A}^{\mathrm{Env}}$, calculated in Ref.~\cite{Li:2025ffh}, combines the scalar energy flux through boundary $A$ with the associated change in cloud energy determined by the particle-number balance. For brevity, we refer to this combined scalar contribution to orbital energy loss as the ``scalar flux''.

We focus here on the gravitational-wave flux corrections. With the common mass-ratio factor $\epsilon^2$ scaled out, the gravitational-wave flux in each mode is $\dot{E}_{h,A}^{\ell m} =\dot{E}_{h,A}^{\mathrm{GR},\ell m}+\zeta^2\dot{E}_{h,A}^{\mathrm{Env},\ell m}$, where $A\in\{H,\infty\}$. For the EMRI--cloud system in Eq.~\eqref{eq:action}, the flux at future null infinity retains the standard relation to $\Psi_4$ derived from the Isaacson stress-energy tensor \cite{Isaacson:1968zza}. At the horizon $r=r_+$, the background deformation and matter coupling modify the relation between $\Psi_0$ and the gravitational-wave flux, i.e.,
\begin{align} \label{eq:flux_schematic}
    \dot{E}^{\ell m}_{h,\infty}
    \propto{}&\lim_{r\rightarrow\infty}
    \frac{1}{\omega^{2}r^{6}}
    \left(\left|{}_{-2}\mathcal{R}_{\ell m}\right|^2
    +2\zeta^2\mathrm{Re}\left[
    {}_{-2}\bar{\mathcal{R}}_{\ell m}\,
    {}_{-2}\mathscr{R}_{\ell m}\right]\right)\,,\nonumber\\
    \dot{E}^{\ell m}_{h,H}
    \propto{}&\lim_{r\rightarrow r_+}
    \frac{\Delta^4\omega}{k(k^2+\varkappa_0^2)}
    \Big[\left|{}_{2}\mathcal{R}_{\ell m}\right|^2\nonumber\\
    & \quad+\zeta^2\Big(2\mathrm{Re}
    \left[{}_{2}\bar{\mathcal{R}}_{\ell m}\,
    {}_{2}\mathscr{R}_{\ell m}\right]
    +\mathcal{F}_{\ell m}\Big)\Big]\,.
\end{align}
Here, ${}_{\pm2}\mathcal{R}_{\ell m}$ and ${}_{\pm2}\mathscr{R}_{\ell m}$ are the radial functions associated with the vacuum Weyl scalars $\Psi_{0,4}^{(0,1)}$ and their environment-induced corrections $\Psi_{0,4}^{(2,1)}$, respectively, with common harmonic frequency $\omega$. We define $k=\omega-m\omega_+$, $\omega_+=a/(2Mr_+)$, and $\Delta=r^2-2Mr+a^2$, where $r_+$, $\omega_+$, and $\varkappa_0$ are the Kerr horizon radius, angular velocity, and surface gravity, respectively. The terms $\propto\left|{}_{\pm2}\mathcal{R}_{\ell m}\right|^2$ give $\dot{E}_{h,\infty/H}^{\mathrm{GR},\ell m}$, while the terms at $\mathcal{O}(\zeta^2)$ are the environmental corrections $\dot{E}_{h,\infty/H}^{\mathrm{Env},\ell m}$, including the interference between ${}_{\pm2}\mathcal{R}_{\ell m}$ and ${}_{\pm2}\mathscr{R}_{\ell m}$ and, at the horizon, the additional correction $\mathcal{F}_{\ell m}$ to the flux relation itself. Summing over $(\ell,m)$ gives $\dot{E}_{h,\infty/H}^{\mathrm{GR}}$ and $\dot{E}_{h,\infty/H}^{\mathrm{Env}}$ entering the orbital-energy balance in Eq.~\eqref{eq:energy_balance_channels}. As both the correction to the gravitational-wave flux and the scalar flux enter at $\mathcal{O}(\zeta^2,\epsilon^2)$, their relative importance cannot be determined from perturbative counting alone. Full expressions of Eq.~\eqref{eq:flux_schematic} and their derivations are given in the Methods and Supplemental Material.

\begin{figure*}[t]
    \centering
    \includegraphics[width=1\linewidth]{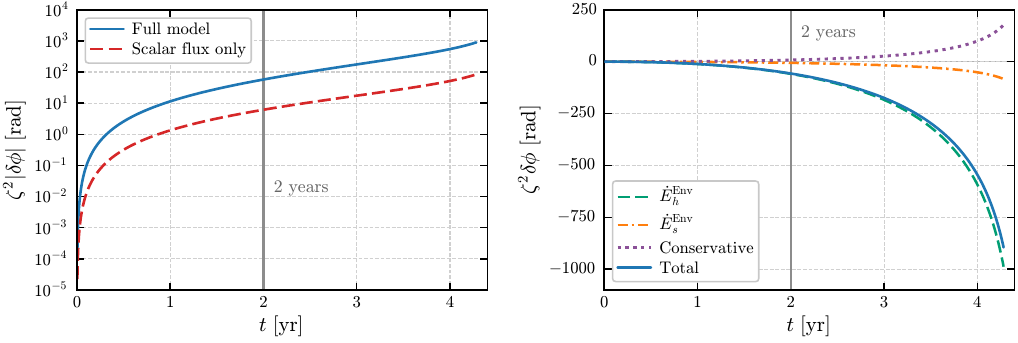}
    \caption{Orbital dephasing $\zeta^2\delta\phi=\phi_{\rm Env}-\phi_{\rm GR}$ for a circular, equatorial, adiabatic inspiral with $M=10^6M_\odot$, $\epsilon=10^{-5}$, and the remaining parameters as in Fig.~\ref{fig:fluxes_total}. \textbf{Left:} The magnitude $\zeta^2|\delta\phi|$, comparing the combined calculation (blue solid) with the scalar-flux-only evolution (red dashed). \textbf{Right:} Signed phase contributions from the environmental corrections to the gravitational-wave fluxes (green dashed), scalar fluxes (orange dash-dotted), and background-induced orbital shifts (purple dotted), together with their sum (blue solid). All evolutions share the initial orbital frequency and phase of the vacuum-GR reference orbit at $r_0\simeq10.6M$ and are compared at equal time until it reaches $r_0=5.1M$. Gray lines mark two years, when the combined and scalar-only orbital dephasing are approximately $-57.3\,\mathrm{rad}$ and $-6.14\,\mathrm{rad}$, respectively.}
    \label{fig:dephasing}
\end{figure*}

Figure~\ref{fig:fluxes_total} compares the vacuum-GR gravitational-wave fluxes $\dot{E}_{h,\infty/H}^{\mathrm{GR}}$, their cloud-induced corrections $\dot{E}_{h,\infty/H}^{\mathrm{Env}}$, and the scalar fluxes $\dot{E}_{s,\infty/H}^{\mathrm{Env}}$ in the horizon and infinity channels for $\mu_{\Phi}M=0.3$ and $M_c/M=10^{-4}$, which corresponds to $\zeta=2.7\times10^{-4}$. The mode sums include $2\leq\ell\leq5$ for the gravitational-wave fluxes and their corrections, and $0\leq\ell_s\leq6$ for the scalar fluxes. The Supplemental Material provides a mode-by-mode analysis and a breakdown of the contributions from the different sources in the modified Teukolsky equations, Eq.~\eqref{eq:MTF}.

Figure~\ref{fig:fluxes_total} shows that the vacuum-GR gravitational-wave fluxes satisfy $\dot{E}_{h,\infty}^{\mathrm{GR}}>0$ and $\dot{E}_{h,H}^{\mathrm{GR}}<0$. The negative horizon flux reflects superradiance, with $\omega/k<0$ for the radiative modes considered. For the cloud-induced corrections to the gravitational-wave fluxes, the horizon correction $\dot{E}_{h,H}^{\mathrm{Env}}$ is positive, whereas the correction at null infinity $\dot{E}_{h,\infty}^{\mathrm{Env}}$ changes from negative to positive with increasing orbital radius, with an interpolated zero near $r_0=14M$ shown in the lower panel. The infinity correction can have either sign because its leading term in Eq.~\eqref{eq:flux_schematic} is an interference term between ${}_{-2}\mathcal{R}_{\ell m}$ and ${}_{-2}\mathscr{R}_{\ell m}$, whose real part need not be positive. The total, unexpanded gravitational-wave flux at infinity remains nonnegative because it is obtained from the squared modulus of the total radiative amplitude. Finally, the scalar fluxes satisfy $\dot{E}_{s,\infty}^{\mathrm{Env}}>0$ and $\dot{E}_{s,H}^{\mathrm{Env}}<0$. Because these fluxes include the associated cloud-energy change, the negative horizon term indicates energy transfer from the black-hole--cloud system to the orbit, not necessarily extraction from the black hole alone \cite{Li:2025ffh}. The magnitude $|\dot{E}_{s,H}^{\mathrm{Env}}|$ peaks near $r_0=15M$~\cite{Dyson:2025dlj,Li:2025ffh}. This non-monotonic behavior may be related to the cloud's radial profile, whose field amplitude peaks near $r=18.6M$, although establishing this connection requires further analysis.

Figure~\ref{fig:fluxes_total} also highlights the importance of the newly calculated environmental corrections to the gravitational-wave energy fluxes, $\dot{E}_{h,H/\infty}^{\mathrm{Env}}$.
At $r_0=15.1M$, $|\dot{E}_{s,H}^{\mathrm{Env}}|$ is approximately $2.9$ times $|\dot{E}_{h,\infty}^{\mathrm{Env}}|$, and the positive $\dot{E}_{h,\infty}^{\mathrm{Env}}$ partially offsets the negative net scalar flux. Moving inward, $|\dot{E}_{h,\infty}^{\mathrm{Env}}|$ overtakes $|\dot{E}_{s,H}^{\mathrm{Env}}|$ near $r_0=13M$ and is about $10^2$ times larger at the innermost sampled radius, $r_0=5.1M$, while $\dot{E}_{h,H}^{\mathrm{Env}}$ remains subdominant over most of the displayed range. In the inner region, the net scalar flux and the net environmental correction to the gravitational-wave flux, each summed over the horizon and infinity channels, are both negative. The gravitational-wave flux correction therefore reinforces the scalar-induced reduction in orbital-energy loss, and a scalar-only treatment misses the dominant environmental contribution in the inner inspiral. Our source decomposition in the Supplemental Material further identifies the terms involving the background metric deformation as the dominant contributions to $\dot{E}_{h,H/\infty}^{\mathrm{Env}}$, highlighting the importance of including the cloud's gravitational backreaction alongside its scalar response.

\section{Orbital Dephasing}

To determine the cloud-induced orbital dephasing, we evolve the inspiral as a sequence of circular, equatorial geodesics in the stationary background $g_{\mu\nu}^{(0,0)}+\zeta^2h_{\mu\nu}^{(2,0)}$. At each orbital radius $r_0$, the radial geodesic equation and four-velocity normalization determine the orbital frequency $\omega_z=d\phi/dt$ and specific orbital energy $\mathcal{E}=E_{\rm orb}/m_p=-u_t$, where $\phi$ is the orbital phase and $u^\mu$ is the secondary's four-velocity. Expanding at fixed $r_0$, we write $\omega_z=\Omega+\zeta^2\delta\omega_z$ and $\mathcal{E}=\mathcal{E}_0+\zeta^2\delta\mathcal{E}$, where $\Omega$ and $\mathcal{E}_0$ are the corresponding Kerr values, and $\delta\omega_z$ and $\delta\mathcal{E}$ are the cloud-induced correction coefficients, with $\zeta^2$ factored out. We then combine these orbital quantities with the energy balance in Eq.~\eqref{eq:energy_balance_channels}, defining $\mathcal{F}$ as its right-hand side with the overall factor $-\epsilon^2$ removed. Introducing the dimensionless variables $x=r_0/M$, $\hat{t}=t/M$, and $\hat{\omega}_z=M\omega_z$, we obtain the orbital evolution equations:
\begin{equation} \label{eq:supp_adiabatic_evolution}
    \frac{dx}{d\hat{t}}
    =-\epsilon\frac{\mathcal{F}}{\mathcal{E}'}\,,
    \qquad
    \frac{d\phi}{d\hat{t}}
    =\hat{\omega}_z\,,
\end{equation}
where a prime denotes $d/dx$. The fluxes determine the energy-loss rate, while the background deformation modifies both its conversion into radial evolution through $\mathcal{E}'$ and the phase accumulation through $\omega_z$. We evolve these equations consistently to first order in $\zeta^2$, as detailed in the Methods.

Figure~\ref{fig:dephasing} shows the total orbital dephasing $\zeta^2\delta\phi=\phi_{\rm Env}-\phi_{\rm GR}$ for a primary of mass $M=10^6M_{\odot}$ and mass ratio $\epsilon=10^{-5}$, with the remaining parameters as in Fig.~\ref{fig:fluxes_total}. The vacuum-GR reference orbit starts at $r_0\simeq10.6M$. All evolutions share the same initial orbital frequency and phase, and are compared at equal elapsed time until the reference orbit reaches $r_0=5.1M$, at $t\simeq4.28\,\mathrm{yr}$. The vacuum-GR gravitational-wave fluxes include modes with $2\leq\ell\leq10$, while the environmental corrections to the gravitational-wave fluxes include $2\leq\ell\leq5$. The left panel compares the combined dephasing, including the scalar fluxes, environmental corrections to the gravitational-wave fluxes, and background-induced orbital shifts, with a scalar-flux-only evolution based on Ref.~\cite{Li:2025ffh}.~\footnote{The time axis of Fig.~12 in \cite{Li:2025ffh} mistakenly included a factor of $2\pi$, and the evolution used a wrong sign for the vacuum-GR horizon flux. After the corrections, the red dashed line here agrees with the one in \cite{Li:2025ffh} if one includes the same number of vacuum-GR modes.} The right panel separates the three contributions. Details of the orbital evolution are given in the Methods.

For $M_c=10^{-4}M$, the total orbital dephasing reaches approximately $\zeta^2\delta\phi=-57.3\,\mathrm{rad}$ after two years, nearly ten times the magnitude of the scalar-only prediction of $-6.14\,\mathrm{rad}$. Furthermore, the phase contributions from the gravitational-wave flux corrections and background-induced orbital shifts each exceed the scalar-flux contribution in magnitude and partially cancel one another, with the gravitational-wave flux corrections dominating the combined dephasing. A scalar-flux-only treatment consequently misses the dominant environmental contribution to the dephasing for this configuration, demonstrating the importance of combined matter--gravity modeling even for a dilute cloud. Similar cancellations have been found for spherical scalar clouds around nonrotating primaries \cite{Keijzer:2026vul}. Here, we demonstrate their importance for a synchronized dipolar cloud around a rotating primary.

The substantial dephasing produced even by a dilute cloud underscores the importance of combined matter--gravity modeling for future space-based gravitational-wave observations \cite{LISA:2017pwj, Hu:2017mde, TianQin:2015yph}. Quantifying potential detection losses and parameter biases from omitted environmental effects requires waveform comparisons that account for detector response, noise, and parameter correlations.

\section{Discussion}

This work turns a sequence of developments in black-hole perturbation theory into a practical relativistic calculation of environmental corrections to EMRI evolution around rotating black holes \cite{Li:2022pcy, Dolan:2023enf, Li:2025ffh, LaHaye:2025ley}. The modified Teukolsky formalism provides the theoretical structure, while the computational strategy developed here makes the resulting matter--gravity interactions tractable. In particular, we reorganize the expanded matter source, containing $\sim10^6$ algebraic terms, into reusable components that make the environmental correction to gravitational radiation computationally accessible. These newly calculated contributions change the leading environmental prediction: for our fiducial system, including the gravitational-wave flux corrections and background-induced orbital shifts yields a two-year orbital dephasing nearly ten times the magnitude of previous, scalar-radiation-only predictions. The phase contributions from these two effects each exceed the scalar contribution in magnitude and partially cancel, extending earlier spherical results \cite{Keijzer:2026vul}. Our work thus demonstrates that combined matter--gravity modeling is both \textit{computationally tractable} without post-Newtonian or slow-rotation approximations, and \textit{necessary} for accurately capturing the leading environmental imprint on inspiral signals.

This capability motivates a more ambitious question: can an EMRI reveal what its environment is made of, rather than merely establish that matter is present? For an ultralight scalar cloud, a common set of field and cloud parameters must account for both the background-induced orbital shifts and the radiative response. Extending our calculation and the scalar-response studies of Ref.~\cite{Xu:2026aic} to eccentric, inclined orbits around a rotating primary would allow us to investigate how resonant features encode these properties and whether they help disentangle cloud parameters from those of the binary. Recovering these correlated signatures from gravitational waveforms could enable measuring the properties of both the scalar field and its macroscopic cloud. Extensions to real scalar fields, other bosonic fields, accretion disks, and dark-matter halos would test whether their dynamical responses can distinguish environments with similar gravitational mass distributions, building on existing environmental inference studies \cite{Cole:2022yzw}. The objective is to infer not only how much matter surrounds the black hole, but also the physics governing its response to the inspiral.

A complementary goal is to distinguish environmental signatures from departures from GR. Combining environmental calculations with modified-gravity applications of the same formalism \cite{Yang:2026erj, Yang:2026csy} would allow competing explanations for deviations from vacuum-Kerr waveforms to be tested within a common relativistic framework. Waveform analyses could establish whether the combined radiative and orbital signatures break degeneracies with binary parameters, or whether cancellations conceal otherwise substantial environmental effects. Incorporating detector response and noise into waveform analyses would connect our framework's predictions to future space-based gravitational-wave observations \cite{LISA:2017pwj, Hu:2017mde, TianQin:2015yph}, opening a route to reveal the nature of matter or dark matter around massive black holes and distinguish their signatures from departures from GR.
\vspace{0.5cm}

\section*{Acknowledgements} 
\label{sec:acknowledgements}
\appendix

We are grateful to Leor Barack, Andrea Maselli, Adam Pound, Takahiro Tanaka, and Barry Wardell for insightful discussions. We thank Sam Dolan for providing the Kerr-Lorenz-Circ {\texttt{Mathematica}} code developed in \cite{Dolan:2021ijg, Dolan:2023enf, Wardell:2024yoi} and clarifying its usage. D.~L., H.-J.~K., K.~L., and N.~Y. acknowledge support from the Simons Foundation through Award No. 896696, NSF Grant No. PHY-2207650, and the National Aeronautics and Space Administration through award 80NSSC22K0806. C.~W. and Y.~C.'s research is supported by the Simons Foundation (Award No. 568762), the Brinson Foundation, and the National Science Foundation (via Grants No. PHY-2011961 and No. PHY-2011968). P.~B. acknowledges support from the Dutch Research Council (NWO) with file number OCENW.M.21.119, and from the ERC grant EMRIWaveforms (DOI: \href{https://doi.org/10.3030/101200625}{10.3030/101200625}). A.~C. acknowledges support from the Herchel Smith Research Fellowship. H.~Y. is supported by the Natural Science
Foundation of China (Grant 12573048). This work makes use of the Kerr-Lorenz-Circ {\texttt{Mathematica}} code developed by Sam Dolan \cite{Dolan:2021ijg, Dolan:2023enf, Wardell:2024yoi} and the Black Hole Perturbation Toolkit \cite{BHPToolkit}. Our calculations use the Illinois Campus Cluster, a computing resource that is operated by the Illinois Campus Cluster Program in conjunction with the National Center for Supercomputing Applications, and is supported by funds from the University of Illinois Urbana-Champaign. Some calculations were also conducted in the Resnick High Performance Computing Center, a facility supported by the Resnick Sustainability Institute at the California Institute of Technology.

\bibliographystyle{apsrev4-2}
\bibliography{reference}

\section*{Methods}

\subsection{Evaluation of the radial sources}

To evaluate the sources of the modified Teukolsky equations in Eq.~\eqref{eq:MTF} for an ultralight scalar cloud, we first make a harmonic decomposition of all the background and perturbed scalar and metric fields. We then replace all the differential operators in the Newman-Penrose formalism with the radial $\{\mathcal{D}_{n},\mathcal{D}^{\dagger}_{n}\}$ and angular $\{\mathcal{L}_{n},\mathcal{L}^{\dagger}_{n}\}$ Chandrasekhar operators \cite{Chandrasekhar_1983, Dolan:2023enf}. This replacement helps us reduce the sources into an almost separable form. After Fourier decomposing the parts having a coupled angular and radial dependence and performing the angular projections, we obtain the radial part of all the sources in Eq.~\eqref{eq:MTF}, namely
\begin{align} \label{eq:radial_source}
    {}_{s}\mathscr{S}_{\geo}^{\ell m}(r)
    =& \sum_{\mathcal{I}_{\rm g}}
    {}_{s}\mathscr{C}_{\mathcal{J}_{\rm g}}^{\ell m}
    {}_{s}\mathcal{F}_{\mathcal{I}_{\rm g}}
    \hat{\mathcal{O}}^{\rm g}_{e}
    \left[{}_{s_b}R_{\ell_b}^{ab}\right]
    \hat{\mathcal{O}}^{\rm g}_{f}
    \left[{}_{s_g}\mathcal{R}_{\ell_g m}^{cd}\right]\,, 
    \nonumber\\
    {}_{s}\mathscr{S}_{A}^{\ell m}(r)
    =& \sum_{\mathcal{I}_A}
    \Big\{{}_{s}\mathscr{C}_{\mathcal{J}_A}^{\ell m}
    {}_{s}\mathcal{F}_{\mathcal{I}_A}
    \hat{\mathcal{O}}^{A}_{d}
    \left[{}_{0}R_{\ell_c m_c}^{\Phi}\right]
    \nonumber\\
    & \hat{\mathcal{O}}^{A}_{e}
    \left[{}_{0}\bar{R}_{\ell_c m_c}^{\Phi}\right]
    \hat{\mathcal{O}}^{A}_{f}
    \left[{}_{s_g}\mathcal{R}_{\ell_g m}^{ab}\right]\Big\}\,, \nonumber\\
    {}_{s}\mathscr{S}_{B}^{\ell m}(r)
    =& \sum_{\mathcal{I}_B}
    \Big\{{}_{s}\mathscr{C}_{\mathcal{J}_B}^{\ell m}
    {}_{s}\mathcal{F}_{\mathcal{I}_{B}}
    \hat{\mathcal{O}}^{B}_{d}
    \left[{}_{0}R_{\ell_c m_c}^{\Phi}\right]
    \hat{\mathcal{O}}^{B}_{e}
    \left[{}_{0}\bar{\mathcal{R}}_{\ell_s m_s}^{\Phi}\right]\nonumber\\
    & +{}_{s}\mathscr{C}_{\mathcal{J}_B}^{\dagger \ell m}
    {}_{s}\mathcal{F}_{\mathcal{I}_{B}}^{\dagger}
    \hat{\mathcal{O}}^{\dagger B}_{d}
    \left[{}_{0}\bar{R}_{\ell_c m_c}^{\Phi}\right]
    \hat{\mathcal{O}}^{\dagger B}_{e}
    \left[{}_{0}\mathcal{R}_{\ell_s m_s}^{\Phi}\right]\Big\}\,,
    \nonumber\\
    {}_{s}\mathscr{S}_{p}^{\ell m}(r)
    =& \sum_{\mathcal{I}_{p}}
    {}_{s}\mathscr{C}_{\mathcal{J}_{p}}^{\ell m}
    {}_{s}\mathcal{F}_{\mathcal{I}_{p}}
    \hat{\mathcal{O}}^{p}_{d}
    \left[{}_{s_b}R_{\ell_b}^{ab}\right]
    \delta^{(n)}(r-r_0)\,. 
\end{align}
For each spin sector, we separated the sources $\mathcal{S}_{m}^{(2,1)}$ and $\mathcal{T}_{m}^{(2,1)}$ into two contributions: the $A$ sector with radial part ${}_{s}\mathscr{S}_{A}^{\ell m}$, which is quadratic in the background scalar field and linear in the vacuum-GR metric perturbation, and the $B$ sector with radial part ${}_{s}\mathscr{S}_{B}^{\ell m}$, which couples the background scalar field to its particle-induced scalar radiation. The functions ${}_{0}R_{\ell_c m_c}^{\Phi}$, ${}_{0}\mathcal{R}_{\ell_s m_s}^{\Phi}$, ${}_{s_b}R_{\ell_b}^{ab}$, and ${}_{s_g}\mathcal{R}_{\ell_g m_g}^{ab}$ are the radial parts of the background scalar field, scalar radiation, background metric correction, and vacuum-GR metric perturbation, respectively. The superscripts of ${}_{s_b}R_{\ell_b}^{ab}$ and ${}_{s_g}\mathcal{R}_{\ell_g m_g}^{ab}$ are the tetrad indices in the unnormalized Kinnersley tetrad $\{l_+,l_-,m_+,m_-\}$ in \cite{Dolan:2023enf}. Since the background metric correction is stationary and axisymmetric, ${}_{s_b}R_{\ell_b}^{ab}$ does not depend on any azimuthal number. The radial operators $\hat{\mathcal{O}}$ and $\hat{\mathcal{O}}^{\dagger}$, with their indices and labels hidden, are sequences of $\{\mathcal{D}_{0},\mathcal{D}^{\dagger}_{0}\}$. In the above, we have also used the selection rule to set $m_g=m$. The procedures for obtaining the radial sources in Eq.~\eqref{eq:radial_source} are provided in the Supplemental Material.

Each radial source in Eq.~\eqref{eq:radial_source} is assembled from three independently evaluated components: angular projection coefficients ${}_{s}\mathscr{C}_{\mathcal{J}}^{\ell m}$ and ${}_{s}\mathscr{C}_{\mathcal{J}}^{\dagger\ell m}$; auxiliary radial functions ${}_{s}\mathcal{F}_{\mathcal{I}}$ independent of the scalar and metric fields; and radial field profiles, such as ${}_{s_b}R_{\ell_b}^{ab}$ and ${}_{s_g}\mathcal{R}_{\ell_g m_g}^{ab}$. The angular projection coefficients and their index sets $\mathcal{J}$ are defined in the Supplemental Material. Each corresponding index set $\mathcal{I}$ extends $\mathcal{J}$ to include the remaining summation indices and mode numbers, excluding the source labels $\mathrm{g},A,B,p$ and the output mode numbers $\ell$ and $m$.

The angular coefficients and radial field profiles are evaluated mode by mode. In contrast, the radial dependence of ${}_{s}\mathcal{F}_{\mathcal{I}}$ is tabulated once on the chosen grid, retaining explicit dependence on the mode numbers so that subsequent mode evaluations require only inexpensive substitutions. The field profiles involve at most $10^2$ independent radial terms per mode, whose spectral representations allow efficient evaluation of their derivatives. Approximately $10^3$ independent angular projection coefficients are required, but these are independent of the radial grid and can therefore be reused when it is refined. This modular organization reduces repeated calculations and facilitates refinement of the source resolution.

To evaluate the sources, we need multiple pieces. The radial parts of the scalar cloud profile, the scalar radiation, and the vacuum-GR gravitational-wave radiation have been computed in \cite{Li:2025ffh}. The only remaining piece is the background metric correction ${}_{s_b}R^{ab}_{\ell_b}$. We obtain this quantity by applying the method in \cite{Garcia:2023ntf}, which solves the full unperturbed Einstein equations along with the scalar field equation in quasi-isotropic coordinates by performing a multi-domain pseudospectral expansion of the solution and utilizing the spectral solver \texttt{Celepha{\"i}s}~\cite{Kuan:2026pss}, which leverages the \texttt{KADATH} and \texttt{FUKA} libraries~\cite{Grandclement:2009ju, Grandclement:2014msa,Papenfort:2021hod}. We then subtract the Kerr spacetime from the full deformed background spacetime to obtain the correction. 

We identify the perturbative regime of our solution by varying the scalar-field amplitude and determining the range over which the background metric correction scales quadratically. Within this range, we choose the smallest amplitude for which the Einstein equations are accurately satisfied, since spectral accuracy deteriorates at very small amplitudes. This procedure also identifies when higher-order nonlinearities invalidate our perturbative treatment. We further verify that the scalar field obtained by spectrally solving the unperturbed Einstein–scalar system agrees with the linearized solution computed using Leaver’s method \cite{Leaver:1985ax, Dolan:2007mj}. Finally, we transform the background metric correction to Boyer–Lindquist coordinates, project it onto the Kinnersley tetrad, and extract the radial functions ${}_{s_b}R^{ab}_{\ell_b}$ using the orthogonality of spin-weighted spherical harmonics. Implementation details and validation checks are provided in the Supplemental Material.

Unlike the scalar radiation equation in \cite{Li:2025ffh}, the operators $\hat{\mathcal{O}}^{\rm g}_{f}$ and $\hat{\mathcal{O}}^{A}_{f}$ in Eq.~\eqref{eq:radial_source} contain radial derivatives up to third order in ${}_{s}\mathcal{R}_{\ell_g m_g}^{ab}$. Since the radial equation of each reconstructed metric component is not explicitly provided in \cite{Dolan:2021ijg, Dolan:2023enf}, we perform multi-domain spectral fits to them so that these higher derivatives can be applied more reliably. We observe that simple interpolation typically results in very noisy data for the derivatives near the horizon or at the outer boundary representing spatial infinity when high-order derivatives are taken. This issue is especially severe for the ingoing components $h^{(0,1)}_{l_-a}$ at small $r_0$, while spectral fitting can use the information close to the asymptotic regions that is not noisy yet to obtain their correct behavior. During the fitting, we also factor the following asymptotic behavior out of these functions:
\begin{align}
    & {}_{s_g}\mathcal{R}^{ab}_{\ell_g m_g}(r\rightarrow r_{+})
    =\Delta^{-\sigma}e^{-ikr^{*}}f^{ab}_H(r)\,,\nonumber\\
    & {}_{s_g}\mathcal{R}^{ab}_{\ell_g m_g}(r\rightarrow\infty)
    =e^{i\omega r^{*}}f^{ab}_{\infty}(r)\,, 
\end{align}
where $\sigma=2$ for the $l_-l_-$ component, $\sigma=1$ for the ${l_-m_+}$ and ${l_-m_-}$ components, and $\sigma=0$ for the remaining components. Similarly, powers of $r^{-1}$ can be extracted from ${}_{s_g}\mathcal{R}_{\ell_g m_g}^{ab}$ at infinity, but it does not significantly affect our fitting in that region, so we do not perform this factorization. 

To control near-horizon divergences, we further introduce the rescaled radial functions ${}_{s_g}\tilde{\mathcal{R}}^{ab}_{\ell_g m}=\Delta^{\sigma}\,{}_{s_g}\mathcal{R}^{ab}_{\ell_g m}$ and rewrite Eq.~\eqref{eq:radial_source} in terms of them. Then, we replace each $\mathcal{D}_{0}$ with $\mathcal{D}_{0}^{\dagger}$ in the operators $\hat{\mathcal{O}}^{\rm g}_{f}$ and $\hat{\mathcal{O}}^{A}_{f}$, using
\begin{align} \label{eq:D_dagger_near_horizon}
    & \mathcal{D}_{0}^{\dagger}
    \left[e^{-ikr^{*}-i\omega t+im\phi}f(r,\theta)\right]
    \nonumber\\
    & \quad =
    e^{-ikr^{*}-i\omega t+im\phi}
    \left(\partial_r+
    im\omega_{+}\frac{r+r_{+}}{r-r_{-}}\right)
    f(r,\theta)\,,
\end{align}
where $r_-$ is the inner horizon of the Kerr metric, and $f(r,\theta)$ is a regular function near the horizon. Since the right-hand side of Eq.~\eqref{eq:D_dagger_near_horizon} is regular near the horizon, this ensures that, after these replacements, all the near-horizon divergences within the $\hat{\mathcal{O}}^{\rm g}_{f}$ and $\hat{\mathcal{O}}^{A}_{f}$ terms are factored out explicitly, facilitating the evaluation of the sources near the horizon. In contrast, $\mathcal{D}_{0}$ acting on the same term in Eq.~\eqref{eq:D_dagger_near_horizon} introduces an additional factor of $\Delta^{-1}$ near the horizon. Similar procedures are also applied to the terms involving the background metric correction ${}_{s}R_{\ell_b}^{ab}$.

The radial derivatives entering the sources must also be interpreted distributionally at the orbital radius. The metric perturbation reconstructed via the method in \cite{Dolan:2021ijg, Dolan:2023enf} is continuous at $r=r_0$, whereas its first radial derivative generally has finite jumps \cite{Dolan:2023enf}. Differentiating the spectral fits separately on $r<r_0$ and $r>r_0$ supplies only the regular, piecewise derivatives, while higher derivatives additionally generate distributions supported at $r_0$. We compute these interface contributions separately for the sources $\mathcal{S}_{\geo}^{(2,1)}$, $\mathcal{T}_{\geo}^{(2,1)}$, $\mathcal{S}_{m}^{(2,1)}$, and $\mathcal{T}_{m}^{(2,1)}$, with details summarized in the Supplemental Material.

After these simplifications, we evaluate the regular, piecewise parts of the sources on a uniform grid of $5001$ points in $r_*$ from $r=r_{+}+10^{-3}M$ to $r=389M$. We have included all the modes in Eqs.~\eqref{eq:harmonic_scalar_metric} and \eqref{eq:harmonic_Weyl_scalar} up to $\ell_b=14$, $\ell_g=8$, $\ell_s=6$, and $\ell=5$ for $5.1M\leq r_0\leq 25.1M$ with a separation of $dr_0=1M$. In the Supplemental Material, we show the convergence of the energy flux calculations when varying these mode cutoffs.

\subsection{Solving the modified Teukolsky equations}

After obtaining the radial sources following the prescription above, we are then able to reduce the modified Teukolsky equations to standard sourced Teukolsky equations of the form
\begin{align} \label{eq:MTF_radial}
    & \left[\Delta^{-s}\frac{d}{dr}
    \left(\Delta^{s+1}\frac{d}{dr}\right)+{}_{s}V_{\ell m}\right]
    {}_{s}\mathscr{R}_{\ell m}(r)
    ={}_{s}\mathscr{S}^{\ell m}(r)\,, \nonumber\\
    & {}_{s}V_{\ell m}
    =\frac{K^2-2is(r-M)K}{\Delta}+4is\omega r
    -{}_{s}\lambda_{\ell m}\,,
\end{align}
where $K=(r^2+a^2)\omega-am$, ${}_{s}\lambda_{\ell m}$ is the radial separation constant in the Teukolsky equation \cite{Teukolsky:1973ha}, and ${}_{\pm2}\mathscr{R}_{\ell m}$ are the radial parts of $\Psi_{0,4}^{(2,1)}$, respectively. The radial function ${}_{s}\mathscr{S}^{\ell m}$ corresponds to the sources in Eq.~\eqref{eq:radial_source}, the solution to each of which can be obtained independently, as described above. 

The standard way to solve Eq.~\eqref{eq:MTF_radial} uses Green's function. We construct the Green's function from two independent homogeneous solutions ${}_{s}\mathcal R_{\ell m}^{\rm in}$ and ${}_{s}\mathcal R_{\ell m}^{\rm up}$, which are respectively ingoing at the future horizon and outgoing at spatial infinity \cite{Teukolsky:1973ha, Srivastava:2021uku}:
\begin{align} \label{eq:hom_asymp}
    & {}_{s}\mathcal{R}^{\rm in}_{\ell m}= 
    \begin{cases}
    {}_{s}B_H\Delta^{-s}e^{-ikr_*}\,, 
    & r \rightarrow r_{+} \\
    {}_{s}B_{\rm out}r^{-2s-1}e^{i\omega r_*}
    +{}_{s}B_{\rm in}r^{-1}e^{-i\omega r_*}\,, 
    & r\rightarrow\infty
    \end{cases}\,, \nonumber\\
    & {}_{s}\mathcal{R}^{\rm up}_{\ell m}=
    \begin{cases}
    {}_{s}D_{\rm in}\Delta^{-s}e^{-ikr_*}
    +{}_{s}D_{\rm out}e^{ikr_*}\,, 
    & r \rightarrow r_{+} \\
    {}_{s}D_{\infty}r^{-2s-1}e^{i\omega r_*}\,, 
    & r\rightarrow\infty
    \end{cases}\,,
\end{align}
respectively, where $\{B_H,B_{\rm in},B_{\rm out},D_{\infty},D_{\rm in},D_{\rm out}\}$ are constants, such that the solution to Eq.~\eqref{eq:MTF_radial} is
\begin{align} \label{eq:Green_function}
    & \begin{aligned}
    {}_{s}\mathscr{R}_{\ell m}(r)
    =& \;\frac{1}{{}_{s}W_{\ell m}}
    \Big[{}_{s}\mathcal{R}^{\rm up}_{\ell m}(r)
    \int_{r_+}^{r}\Delta^s\,
    {}_{s}\mathcal{R}^{\rm in}_{\ell m}\,
    {}_{s}\mathscr{S}^{\ell m}\,dr' \\
    & \;+{}_{s}\mathcal{R}^{\rm in}_{\ell m}(r)
    \int_r^\infty\Delta^s\,
    {}_{s}\mathcal{R}^{\rm up}_{\ell m}\,
    {}_{s}\mathscr{S}^{\ell m}\,dr'\Big]\,,   
    \end{aligned} \nonumber\\
    & {}_{s}W_{\ell m}
    =\Delta^{s+1}
    \left({}_{s}\mathcal{R}^{\rm in}_{\ell m}
    \partial_r\,{}_{s}\mathcal{R}^{\rm up}_{\ell m}
    -{}_{s}\mathcal{R}^{\rm up}_{\ell m}
    \partial_r\,{}_{s}\mathcal{R}^{\rm in}_{\ell m}\right)\,,
\end{align}
where we hide the dependence on $r'$ in the integrands above for simplicity. Both homogeneous solutions ${}_{s}\mathcal{R}^{\rm in}_{\ell m}$ and ${}_{s}\mathcal{R}^{\rm up}_{\ell m}$ can be obtained directly from the BHPToolkit \cite{BHPToolkit}.

Evaluating Eq.~\eqref{eq:Green_function} requires a separate treatment of particle-localized distributions and contributions near the horizon and infinity. The distributions, arising from the particle source and derivative jumps of the lower-order fields, are integrated analytically. Their contributions are then combined with those obtained by numerically integrating the piecewise-smooth source terms on either side of the particle's orbit. Near the horizon and infinity, the perturbative solutions contain logarithmic terms that encode changes in the ingoing and outgoing wave phases induced by the background deformation. We isolate these terms analytically and test their phase interpretation. At the horizon, Frobenius and Laurent expansions allow the singular Green integrals to be evaluated by analytic continuation, retaining the finite contributions from both homogeneous branches. At infinity, we analytically isolate the logarithmic contribution and fit the remaining large-radius dependence of the Green-function integral to its non-logarithmic asymptotic form. This separates the finite outgoing amplitude from the residual particular-solution contributions. The remaining source terms are integrated numerically, and the full regularization procedure is detailed in the Supplemental Material.

\subsection{Calculation of the energy fluxes}

To calculate the additional gravitational energy fluxes generated by the ultralight scalar cloud, we need to relate them to the Weyl scalar perturbations $\Psi_{0,4}^{(2,1)}$. As explained in the main text, the relation between the perturbations of $\Psi_4$ and the gravitational-wave energy flux at spatial infinity stays the same, i.e., $d^2E^{\infty}/(dtd\Omega)=r^2|\Psi_4^{(1)}|^2/(4\pi\omega^2)$, where the right-hand side is evaluated at infinity, and $\Psi_4^{(1)}$ represents a generic linear perturbation of $\Psi_4$.
Then, after linearization and harmonic decomposition, one obtains 
\begin{align}  \label{eq:infinity_flux}
    \dot{E}^{\infty}
    &=\lim_{r\rightarrow\infty}
    \sum_{\ell,m}\frac{1}{4\pi\omega^2 r^{6}}
    \Big(\left|{}_{-2}\mathcal{R}_{\ell m}\right|^2 \nonumber\\
    &+2\zeta^2\operatorname{Re}
    \left[{}_{-2}\bar{\mathcal{R}}_{\ell m}\;
    {}_{-2}\mathscr{R}_{\ell m}\right]+\cdots\Big)\,,
\end{align}
where we only include terms up to $\mathcal{O}(\zeta^2)$. 

The calculation for the gravitational-wave energy flux into the horizon is more involved, because the background spacetime is no longer Kerr and the Einstein equations in Eq.~\eqref{eq:EOM} are sourced. One standard way to obtain the relation between this energy flux and the gravitational-wave metric perturbation into the horizon encoded in $\Psi_0$ is to use the first law of black hole thermodynamics, which allows us to relate the energy change to the horizon area change \cite{Chandrasekhar_1983}. The latter can be determined from several spin coefficients or optical scalars that control the expansion (or contraction), rotation, and shear of null congruences on the horizon. Via a partial metric reconstruction near the horizon, these spin coefficients can be solved from a perturbed $\Psi_0$. In the Supplemental Material, we follow Refs.~\cite{Hawking:1972hy, Chandrasekhar_1983, Poisson:2004cw, Yang:2026erj, Yang:2026csy} to show how this procedure is modified due to the ultralight scalar cloud. In the end, we obtain the horizon flux formula in the form:
\begin{align}  \label{eq:horizon_flux}
    \dot{E}^{H}
    &=\lim_{r\rightarrow r_+}\sum_{\ell,m}
    \frac{\omega\Delta^4}{64\pi k(2Mr_+)^3(\varkappa_0^2+k^2)}
    \Big[\left|{}_{2}\mathcal{R}_{\ell m}\right|^2 \\
    &+\zeta^2\left(2\operatorname{Re}
    \left[{}_{2}\bar{\mathcal{R}}_{\ell m}\;
    {}_{2}\mathscr{R}_{\ell m}\right]
    +\mathcal{F}[{}_{2}\mathcal{R}_{\ell m},{}_{s_b}R^{ab}_{\ell_b}]\right)+\cdots\Big]\,, \nonumber
\end{align}
where $\varkappa_0=\sqrt{M^2-a^2}/(2Mr_+)$ is the surface gravity of the Kerr spacetime. The term $\mathcal{F}$ is a functional in ${}_{2}\mathcal{R}_{\ell m}$ and the radial part ${}_{s_b}R^{ab}_{\ell_b}$ of the background metric correction that corresponds to the correction to the horizon flux formula derived in the Supplemental Material. For circular equatorial orbits around a stationary and axisymmetric black hole, $\dot{L}_{\ell m}^{\infty}=m\dot{E}_{\ell m}^{\infty}/\omega$ and $\dot{L}_{\ell m}^{H}=m\dot{E}_{\ell m}^{H}/\omega$, where $\dot{E}_{\ell m}^{\infty}$ and $\dot{E}_{\ell m}^{H}$ are the $(\ell,m)$ contributions to $\dot{E}^{\infty}$ and $\dot{E}^{H}$, respectively. Thus, we focus on the energy fluxes in the calculations above. 

In evaluating Eqs.~\eqref{eq:infinity_flux} and \eqref{eq:horizon_flux}, we compare the gravitational-wave fluxes at the same coordinate radius $r_0$. The scalar cloud deforms the background geometry, changing the four-velocity and orbital frequency of a circular orbit at that radius. Consequently, the particle’s stress-energy tensor and the frequency of its emitted radiation change, with $\omega=m\omega_z=m(\Omega+\zeta^2\delta\omega_z)$. We consistently include these changes to the particle source, together with the frequency-dependent changes to the angular modes, radial wave response, and conversion from wave amplitudes to energy fluxes. The resulting flux correction therefore accounts for the altered orbital motion as well as the cloud’s effect on wave propagation, with each contribution included once.

\subsection{Environmental conservative effects and orbital evolution}

Here, we show how the environmental conservative effects due to a deformed background are calculated and how the orbital evolution is computed. 

For a circular, equatorial orbit with four-velocity $u^\mu=u^t(1,0,0,\omega_z)$, the radial geodesic equation and four-velocity normalization give
\begin{subequations}
\begin{align}
    & g_{tt,r}+2\omega_z g_{t\phi,r}
    +\omega_z^2g_{\phi\phi,r}=0\,,
    \label{eq:supp_circular_condition} \\
    & u^t=\left[-g_{tt}-2\omega_z g_{t\phi}
    -\omega_z^2g_{\phi\phi}\right]^{-1/2}\,,
    \label{eq:supp_ut} \\
    & \mathcal{E}\equiv\frac{E_{\rm orb}}{m_p}=-u_t
    =-u^t\left(g_{tt}+\omega_z g_{t\phi}\right)\,.
    \label{eq:supp_specific_energy}
\end{align}
\end{subequations}
All metric components and their derivatives are evaluated at $(r,\theta)=(r_0,\pi/2)$, and a comma followed by $r$ denotes a radial partial derivative.
At fixed $r_0$, we expand $\omega_z=\Omega+\zeta^2\delta\omega_z$, $u^t=U+\zeta^2\delta u^t$, and $\mathcal{E}=\mathcal{E}_0+\zeta^2\delta\mathcal{E}$.
Here, $\Omega$ is the Kerr orbital frequency defined in
Eq.~\eqref{eq:four_vector_GR}, $U$ is the corresponding
$t$-component of the four-velocity, and
\begin{equation}
    \mathcal{E}_0
    =\frac{1-2M/r_0+a\sqrt{M/r_0^3}}
    {\sqrt{1-3M/r_0+2a\sqrt{M/r_0^3}}}
\end{equation}
is the Kerr specific orbital energy. Expanding Eqs.~\eqref{eq:supp_circular_condition}--\eqref{eq:supp_specific_energy} then yields
\begin{align}
    & \delta\omega_z
    =-\frac{h_{tt,r}+2\Omega h_{t\phi,r}
    +\Omega^2h_{\phi\phi,r}}
    {2\left(g_{t\phi,r}+\Omega g_{\phi\phi,r}\right)}\,, \nonumber\\
    & \delta\mathcal{E}
    =-U\left(h_{tt}+\Omega h_{t\phi}
    +g_{t\phi}\delta\omega_z\right)
    -\left(g_{tt}+\Omega g_{t\phi}\right)\delta u^t\,, \nonumber\\
    & \delta u^t
    =\frac{U^3}{2}
    \Big[h_{tt}+2\Omega h_{t\phi}+\Omega^2h_{\phi\phi}
    +2\left(g_{t\phi}+\Omega g_{\phi\phi}\right)
    \delta\omega_z\Big]\,,
\end{align}
where we have dropped the superscripts of $g_{\mu\nu}^{(0,0)}$ and $h_{\mu\nu}^{(2,0)}$ for simplicity. 

For investigating the orbital evolution, we first decompose the total orbital-energy loss $\mathcal{F}$, where $\mathcal{F}=-\dot{E}_{\rm orb}/\epsilon^2$ with $\dot{E}_{\rm orb}$ defined in Eq.~\eqref{eq:energy_balance_channels}, as
\begin{align} \label{eq:supp_total_flux}
    & \mathcal{F}
    =\mathcal{F}_0
    +\zeta^2\left(
    \delta\mathcal{F}_h+\delta\mathcal{F}_s
    \right)\,,
    \nonumber\\
    & \mathcal{F}_0
    =\dot{E}_{h,H}^{\mathrm{GR}}
    +\dot{E}_{h,\infty}^{\mathrm{GR}}\,,
    \nonumber\\
    & \delta\mathcal{F}_{h/s}
    =\dot{E}_{h/s,H}^{\mathrm{Env}}
    +\dot{E}_{h/s,\infty}^{\mathrm{Env}}\,.
\end{align}
Using the dimensionless variables $x=r_0/M$ and $\hat{t}=t/M$ introduced in the main text, we expand $x_{\rm Env}=x_{\rm GR}+\zeta^2\delta x$ and $\phi_{\rm Env}=\phi_{\rm GR}+\zeta^2\delta\phi$. We then expand Eq.~\eqref{eq:supp_adiabatic_evolution} and evolve these coefficients at equal $\hat{t}$:
\begin{subequations} \label{eq:radial_phase_variation}
\begin{align}
    \frac{d\delta x}{d\hat{t}}
    &=\epsilon\left(
    v_0'\delta x
    +\frac{\mathcal{F}_0\delta\mathcal{E}'}
    {\mathcal{E}_0'^2}
    -\frac{\delta\mathcal{F}_h+\delta\mathcal{F}_s}
    {\mathcal{E}_0'}
    \right)\,,
    \label{eq:supp_radial_variation}\\
    \frac{d\delta\phi}{d\hat{t}}
    &=\hat{\Omega}'\delta x+\delta\hat{\omega}_z\,,
    \label{eq:supp_phase_variation}
\end{align}
\end{subequations}
where $v_0=-\mathcal{F}_0/\mathcal{E}_0'$, $\hat{\Omega}=M\Omega$, $\delta\hat{\omega}_z=M\delta\omega_z$, and a prime denotes $d/dx$. All coefficient functions are evaluated on the reference trajectory $x_{\rm GR}(\hat{t})$. The terms proportional to $\delta x$ account for evaluating the Kerr radial and orbital velocities at the shifted radius. The terms involving $\delta\mathcal{E}'$ and $\delta\hat{\omega}_z$ describe the background-induced changes to the energy--radius and frequency--radius relations, while $\delta\mathcal{F}_h$ and $\delta\mathcal{F}_s$ supply the environmental corrections to orbital-energy loss.

In our evolution, the reference vacuum-GR orbit starts at $r_0\approx 10.6M$, and the orbit under environmental effects is matched to the former's initial frequency and phase, rather than its
coordinate radius. Thus,
\begin{equation} \label{eq:supp_frequency_matching}
    \delta x(0)
    =-\frac{\delta\hat{\omega}_z(x_i)}{\hat{\Omega}'(x_i)}\,,
    \qquad \delta\phi(0)=0\,,
\end{equation}
where $x_i=x_{\rm GR}(0)$ is the initial dimensionless
radius of the reference vacuum-GR orbit. All the evolutions use the same elapsed time and end when the GR orbit
reaches $r_0=5.1M$. Linearity of Eq.~\eqref{eq:radial_phase_variation} allows the gravitational, scalar, and conservative phase contributions to be computed separately and added. The conservative contribution includes the initial radial displacement in Eq.~\eqref{eq:supp_frequency_matching}. The scalar-only curve in Fig.~\ref{fig:dephasing} retains $\delta\mathcal F_s$ but sets $\delta\mathcal F_h=\delta\mathcal{E}=\delta\omega_z=0$.

\newpage
\section*{Supplemental Material}
\setcounter{subsection}{0}

\subsection{Modified Teukolsky equations}

Here, we review the modified Teukolsky equations derived in \cite{Li:2022pcy} and their adaptation to the EMRI modeling in \cite{Li:2025ffh, LaHaye:2025ley}. The action of the class of systems we are interested in has the following form:
\begin{equation} \label{eq:action}
    S=\int d^4 x\sqrt{-g}\left(\frac{R}{16\pi G}
    +\mathcal{L}_{\bGR}
    +\mathcal{L}_{m}\right)-m_p\int d\tau\,,
\end{equation}
where $\mathcal{L}_{\bGR}$ is the Lagrangian for beyond-GR corrections that admit an effective-field-theory description, i.e., $\mathcal{L}_{\bGR}$ is proportional to a coupling constant, such that the theory reduces to GR as the coupling goes to zero. In this work, we do not consider any beyond-GR corrections, so $\mathcal{L}_{\bGR}=0$, although the same framework formulated in this work has been applied to model EMRIs in higher-derivative gravity \cite{Yang:2026erj, Yang:2026csy}. The term $\mathcal{L}_m$ denotes the Lagrangian for any set of fields or matter $\{\varphi_i\}$ that minimally couple to GR, and we assume the amplitudes of $\{\varphi_i\}$ can be treated perturbatively. For simplicity, let us consider only a single extra field/matter $\varphi$. The last term is the action for a point particle of mass $m_p$ representing the secondary in an EMRI system, and $\tau$ is its proper time. 

Varying the action in Eq.~\eqref{eq:action} with respect to the metric and each extra field $\varphi$, one finds the Einstein equations
\begin{equation}
    G_{\mu\nu}
    =8\pi G\left(T_{\mu\nu}^{m}+T^{p}_{\mu\nu}\right)
\end{equation}
with the stress-energy tensors
\begin{subequations} \label{eq:stress}
\begin{align} 
    & T_{\mu\nu}^{m}
    =-\frac{2}{\sqrt{-g}} \frac{\delta}{\delta g^{\mu\nu}}
    \left[\sqrt{-g}\mathcal{L}_{m}\right]\,, \\
    & T_{\mu\nu}^{p}
    =m_p\int u_\mu u_\nu\frac{\delta^{(4)}
    \left(x^\mu-x_p^\mu(\tau)\right)}{\sqrt{-g}}d\tau\,,
    \label{eq:stress_particle}
\end{align}    
\end{subequations}
where $x_p^{\mu}$ is the particle's worldline, and $u^{\mu}=dx^\mu/d\tau$ is the particle's four-velocity. For circular, equatorial, prograde orbits around rotating black holes in vacuum GR, Eq.~\eqref{eq:stress_particle} reduces to
\begin{equation} \label{eq:stress_particle_circular}
   T_{\mu\nu}^{p}
   =u_{\mu}u_{\nu}\frac{m_p}{r_0^2 u^t}\delta\left[r-r_0\right]
   \delta[\theta-\pi/2]\delta\left[\phi-\Omega t\right]\,,
\end{equation}
with
\begin{align} \label{eq:four_vector_GR}
    & u^a=u^t(1,0,0,\Omega)\,,\;
    u^t=\frac{1+a\sqrt{M/r_0^3}}{\sqrt{1-3M/r_0+2a\sqrt{M/r_0^3}}}\,, \nonumber\\
    & \Omega=\frac{\sqrt{M}}{r_0^{3/2}+a\sqrt{M}}\,.
\end{align} 
For the ultralight scalar cloud example considered in the main text, $T_{\mu\nu}^{m}$ is replaced by $T_{\mu\nu}^{\Phi}$ in Eq.~\eqref{eq:stress_cloud}.

Linearizing the Einstein equations in Eq.~\eqref{eq:EE} with the expansion in Eq.~\eqref{eq:expansion} and projecting them onto the Newman-Penrose (NP) basis, Refs.~\cite{Li:2022pcy, Hussain:2022ins, Li:2025ffh, LaHaye:2025ley} found that the Weyl scalar perturbation $\Psi_{0}^{(2,1)}$ satisfies the following modified Teukolsky equation:
\begin{equation} \label{eq:MTF_Psi0_Sup}
    H_{0}^{(0,0)}\Psi_0^{(2,1)}
    =\mathcal{S}_{\geo}^{(2,1)}+\mathcal{S}_{m}^{(2,1)}
    +\mathcal{S}_{p}^{(2,1)}\,,
\end{equation}
where the operator $H_{0}$ is defined as 
\begin{equation} \label{eq:def_operators}
    \begin{aligned}
        & H_0=\mathcal{E}_2F_2-\mathcal{E}_1F_1-3\Psi_2\,, \\
        & \mathcal{E}_1=E_1-\Psi_2^{-1}\delta\Psi_2\,,\quad
        \mathcal{E}_2=E_2-\Psi_2^{-1}D\Psi_2\,,
    \end{aligned}
\end{equation}
with the auxiliary operators $F_{1,2}$, $J_{1,2}$, and $E_{1,2}$
\begin{align} \label{eq:auxiliary_operators}
    & F_1\equiv\bar{\delta}_{[-4,0,1,0]}\,,\quad
    && F_2\equiv\boldsymbol{\Delta}_{[1,0,-4,0]}\,, \nonumber\\
    & J_1\equiv D_{[-2,0,-4,0]}\,,\quad
    && J_2\equiv\delta_{[0,-2,0,-4]}\,, \nonumber\\
    & E_1\equiv\delta_{[-1,-3,1,-1]}\,,\quad
    && E_2\equiv D_{[-3,1,-1,-1]}\,.
\end{align}
At $\mathcal{O}(\zeta^0,\epsilon^0)$, $H_{0}^{(0,0)}$ reduces to the Teukolsky operator acting on $\Psi_0$ in vacuum GR \cite{Teukolsky:1973ha}.

For convenience, we have defined the derivative operators
\begin{subequations} \label{eq:NP_operators}
\begin{align}
    D_{[p,q,u,v]}
    &=D+p\varepsilon+q\bar{\varepsilon}+u\rho+v\bar{\rho}\,, \\
    \boldsymbol{\Delta}_{[p,q,u,v]}
    &=\boldsymbol{\Delta}+p\mu+q\bar{\mu}+u\gamma+v\bar{\gamma}\,, \\
    \delta_{[p,q,u,v]}
    &=\delta+p\bar{\alpha}+q\beta+u\bar{\pi}+v\tau\,, \\
    \bar{\delta}_{[p,q,u,v]}
    &=\bar{\delta}+p\alpha+q\bar{\beta}+u\pi+v\bar{\tau}\,,
\end{align}
\end{subequations}
where $p,q,u$, and $v$ are integer coefficients, $\{D,\boldsymbol{\Delta},\delta,\bar{\delta}\}$ are the directional derivatives associated with the NP tetrad basis vectors $\{l^{\mu},n^{\mu},m^{\mu},\bar{m}^{\mu}\}$, and the remaining quantities are spin coefficients in the NP formalism \cite{Newman:1961qr, Chandrasekhar_1983, Li:2022pcy}. In the Kinnersley tetrad, one can rewrite these operators in terms of the Chandrasekhar operators $\{\mathcal{D}_{n},\mathcal{D}_{n}^{\dagger},\mathcal{L}_{n},\mathcal{L}_{n}^{\dagger}\}$ \cite{Li:2025ffh}, which will be helpful in our subsequent calculations,
\begin{align} 
    D_{[p,q,u,v]}
    &=\mathcal{D}_0-\left(\frac{v}{\Gamma}
    +\frac{u}{\bar{\Gamma}}\right)\,, \nonumber\\
    \boldsymbol{\Delta}_{[p,q,u,v]}
    &=-\frac{\Delta}{2\Gamma\bar{\Gamma}}
    \left(\mathcal{D}^{\dagger}_{-(u+v)/2}
    +\frac{q+v}{\Gamma}+\frac{p+u}{\bar{\Gamma}}\right)\,, 
    \nonumber\\
    \delta_{[p,q,u,v]}
    &=\frac{1}{\sqrt{2}\Gamma}
    \left[\mathcal{L}^{\dagger}_{(q-p)/2}-ia\sin{\theta}
    \left(\frac{p+u}{\Gamma}
    +\frac{v}{\bar{\Gamma}}\right)\right]\,, \nonumber\\
    \bar{\delta}_{[p,q,u,v]}
    &=\frac{1}{\sqrt{2}\bar{\Gamma}}
    \left[\mathcal{L}_{(q-p)/2}+ia\sin{\theta}
    \left(\frac{p+u}{\bar{\Gamma}}
    +\frac{v}{\Gamma}\right)\right]\,,
\end{align}
where $\Delta=r^2-2Mr+a^2$ and $\Gamma=r+ia\cos{\theta}$. In Boyer-Lindquist coordinates $(t,r,\theta,\phi)$, the Chandrasekhar operators are given by \cite{Chandrasekhar_1983, Dolan:2023enf}
\begin{align}
    & \mathcal{D}_n
    =\partial_r-\frac{iK}{\Delta}+n\frac{\Delta'}{\Delta}\,,\;
    \mathcal{D}^{\dagger}_n
    =\partial_r+\frac{iK}{\Delta}+n\frac{\Delta'}{\Delta}\,, \nonumber\\
    & \mathcal{L}_n
    =\partial_\theta+Q+n\cot{\theta}\,,\;
    \mathcal{L}^{\dagger}_n
    =\partial_\theta-Q+n\cot{\theta}\,,
\end{align}
where $K=\omega(r^2+a^2)-am$, $Q=m\csc{\theta}-a\omega\sin{\theta}$, and $\Delta'=2(r-M)$ is the radial derivative of $\Delta$.

The three source terms on the right-hand side of Eq.~\eqref{eq:MTF_Psi0_Sup} are defined as
\begin{subequations} \label{eq:source_Psi0}
\begin{align}
    \mathcal{S}_{\geo}^{(2,1)}
    =& \;-H_0^{(2,0)}\Psi_0^{(0,1)}-H_0^{(0,1)}\Psi_0^{(2,0)}\,, 
    \label{eq:S_geo} \\
    \mathcal{S}_{m}^{(2,1)}
    =& \;\mathcal{E}_2^{(0,0)}S_{2}^{m(2,1)}
    +\mathcal{E}_2^{(0,1)}S_{2}^{m(2,0)}
    -\mathcal{E}_1^{(0,0)}S_{1}^{m(2,1)} \nonumber\\
    & \;-\mathcal{E}_1^{(0,1)}S_{1}^{m(2,0)}\,, 
    \label{eq:S_m} \\
    \mathcal{S}_{p}^{(2,1)}
    =& \;\mathcal{E}_2^{(0,0)}S_{2}^{p(2,1)}
    +\mathcal{E}_2^{(2,0)}S_{2}^{p(0,1)}
    -\mathcal{E}_1^{(0,0)}S_{1}^{p(2,1)} \nonumber\\
    & \;-\mathcal{E}_1^{(2,0)}S_{1}^{p(0,1)}\,, 
    \label{eq:S_p}
\end{align}
\end{subequations}
where
\begin{subequations} \label{eq:source_bianchi_psi0}
\begin{align}
    S_1=& \;\delta_{[-2,-2,1,0]}\Phi_{00}
    -D_{[-2,0,0,-2]}\Phi_{01} \nonumber\\
    & \;+2\sigma\Phi_{10}-2\kappa\Phi_{11}-\bar{\kappa}\Phi_{02}\,, \\
    S_2=& \;\delta_{[0,-2,2,0]}\Phi_{01}
    -D_{[-2,2,0,-1]}\Phi_{02} \nonumber\\
    & \;-\bar{\lambda}\Phi_{00}+2\sigma\Phi_{11}-2\kappa\Phi_{12}\,.
\end{align}
\end{subequations}
The NP Ricci scalars $\Phi_{ij}$ are projections of the Ricci tensor $R_{\mu\nu}$ onto the NP tetrad $e^{\mu}_{a}$, i.e.,
\begin{equation}\label{eq:RicciScalars}
    \Phi_{ij}\sim R_{\mu\nu}e^{\mu}_{a}e^{\nu}_{b}\,,
\end{equation}
where $i,j \in \{0,1,2\}$ and $a,b$ are tetrad indices,
with complete expressions provided in \cite{Newman:1961qr, Chandrasekhar_1983, Li:2022pcy}. The quantities $\kappa$, $\sigma$, and $\lambda$ are spin coefficients in the NP formalism.

The first source term $\mathcal{S}_{\geo}^{(2,1)}$ is regarded as purely ``geometrical'' since it only depends on the corrections to the central black hole's geometry [i.e., the terms at $\mathcal{O}(\zeta^2,\epsilon^0)$] and gravitational radiation in vacuum GR [i.e., the terms at $\mathcal{O}(\zeta^0,\epsilon^1)$]. The second and third source terms,  $\mathcal{S}_{m}^{(2,1)}$ and $\mathcal{S}_{p}^{(2,1)}$, are directly driven by the matter and point-particle stress-energy tensors, $T_{\mu\nu}^{m}$ and $T_{\mu\nu}^{p}$, respectively. The terms $S_{i}^{m}$ and $S_{i}^{p}$ ($i=1,2$) are defined in the same way as $S_{i}$ in Eq.~\eqref{eq:source_bianchi_psi0}, with $\Phi_{ij}$ determined by $T_{\mu\nu}^{m}$ and $T_{\mu\nu}^{p}$, respectively. Since the stress-energy tensors $T_{\mu\nu}^{m}$ and $T_{\mu\nu}^{p}$ enter at $\mathcal{O}(\zeta^2)$ and $\mathcal{O}(\epsilon^1)$, the sources $S_{i}^{m}$ and $S_{i}^{p}$ start at $\mathcal{O}(\zeta^2)$ and $\mathcal{O}(\epsilon^1)$, respectively. Finally, as we have shown in \cite{Li:2022pcy, LaHaye:2025ley}, both $S_{i}^{m(2,1)}$ and $S_{i}^{p(2,1)}$ only depend on the matter or metric fields below $\mathcal{O}(\zeta^2,\epsilon^1)$, so $\Psi_0^{(2,1)}$ decouples from other geometrical quantities at the same order, similar to the standard Teukolsky equations in \cite{Teukolsky:1973ha}.

Analogously, the modified Teukolsky equation of $\Psi_4^{(2,1)}$ is given by
\begin{equation} \label{eq:MTF_Psi4_Sup}
    H_{4}^{(0,0)}\Psi_4^{(2,1)}
    =\mathcal{T}_{\geo}^{(2,1)}+\mathcal{T}_{m}^{(2,1)}
    +\mathcal{T}_{p}^{(2,1)}\,.
\end{equation}
The source terms are defined as
\begin{subequations} \label{eq:source_Psi4}
\begin{align}
    \mathcal{T}_{\geo}^{(2,1)}
    =& \;-H_4^{(2,0)}\Psi_4^{(0,1)}-H_4^{(0,1)}\Psi_4^{(2,0)}\,, 
    \label{eq:T_geo} \\
    \mathcal{T}_{m}^{(2,1)}
    =& \;\mathcal{E}_4^{(0,0)}S_{4}^{m(2,1)}
    +\mathcal{E}_4^{(0,1)}S_{4}^{m(2,0)}
    -\mathcal{E}_3^{(0,0)}S_{3}^{m(2,1)} \nonumber\\
    &\;-\mathcal{E}_3^{(0,1)}S_{3}^{m(2,0)}\,, 
    \label{eq:T_m} \\
    \mathcal{T}_{p}^{(2,1)}
    =& \;\mathcal{E}_4^{(0,0)}S_{4}^{p(2,1)}+
    \mathcal{E}_4^{(2,0)}S_{4}^{p(0,1)}
    -\mathcal{E}_3^{(0,0)}S_{3}^{p(2,1)} \nonumber\\
    &\;-\mathcal{E}_3^{(2,0)}S_{3}^{p(0,1)}\,,
    \label{eq:T_p}
\end{align}
\end{subequations}
where $S_{3,4}^{m}$ and $S_{3,4}^{p}$ are defined in the same way as $S_{3,4}$ below with $\Phi_{ij}$ defined by their corresponding stress-energy tensors $T^{m}_{\mu\nu}$ and $T^{p}_{\mu\nu}$, respectively:
\begin{subequations} \label{eq:source_bianchi_psi4}
\begin{align}
    S_3=& \;-\mathbf{\Delta}_{[0,2,2,0]}\Phi_{21}
    +\bar{\delta}_{[2,2,0,-1]}\Phi_{22} \nonumber\\
    & \;+2\nu\Phi_{11}+\bar{\nu}\Phi_{20}-2\lambda\Phi_{12}\,, \\
    S_4=& \;-\mathbf{\Delta}_{[0,1,2,-2]}\Phi_{20}
    +\bar{\delta}_{[2,0,0,-2]}\Phi_{21}  \nonumber\\
    & \;+2\nu\Phi_{10}-2\lambda\Phi_{11}+\bar{\sigma}\Phi_{22}\,,
\end{align}
\end{subequations}
with $\lambda$ and $\nu$ being spin coefficients. The operators $H_{4}$ and $\mathcal{E}_{3,4}$ are defined as 
\begin{equation} \label{eq:H_in_Teuk_ORG}
\begin{aligned}
    & H_4=\mathcal{E}_4F_4-\mathcal{E}_3F_3-3\Psi_2\,, \\
    & \mathcal{E}_3=E_3-\Psi_2^{-1}\bar{\delta}\Psi_2\,,\quad
    \mathcal{E}_4=E_4-\Psi_2^{-1}\mathbf{\Delta}\Psi_2\,,
\end{aligned}
\end{equation}
with
\begin{align}
    & F_3\equiv\delta_{[0,4,0,-1]}\,,\quad
    && F_4\equiv D_{[4,0,-1,0]}\,, \nonumber\\
    & J_3\equiv\mathbf{\Delta}_{[4,0,2,0]}\,,\quad
    && J_4\equiv\bar{\delta}_{[2,0,4,0]}\,, \nonumber\\
    & E_3\equiv\bar{\delta}_{[3,1,1,-1]}\,,\quad 
    && E_4\equiv\mathbf{\Delta}_{[1,1,3,-1]}\,.
\end{align}
As a final remark, in the derivation of the modified Teukolsky equations in Eqs.~\eqref{eq:MTF_Psi0_Sup} and \eqref{eq:MTF_Psi4_Sup}, we have adopted the tetrad gauge \cite{Li:2022pcy, Li:2025ffh}
\begin{equation} \label{eq:psi1_psi3}
    \Psi_{1,3}^{(0,1)}=\Psi_{1,3}^{(2,0)}=\Psi_{1,3}^{(2,1)}=0\,.
\end{equation}

\subsection{Reduction of the sources}

Here, we show how to reduce the sources in Eqs.~\eqref{eq:source_Psi0} and \eqref{eq:source_Psi4} into an almost separable form. We take the ultralight scalar cloud case as an example. First, we make a harmonic decomposition of all the fields as follows:
\begin{align} \label{eq:harmonic_scalar_metric}
    & \Phi^{(1,0)}
    =\sum_{\ell_c,m_c}{}_{0}R_{\ell_c m_c}^{\Phi}\,
    {}_{0}S_{\ell_c m_c}\,
    e^{-i\omega_c t+im_c\phi}\,, \nonumber\\
    & \Phi^{(1,1)}
    =\sum_{\ell_s,m_s}{}_{0}\mathcal{R}_{\ell_s m_s}^{\Phi}\,
    {}_{0}S_{\ell_s m_s}\,
    e^{-i\omega_s t+im_s\phi}\,, \nonumber\\
    & h_{ab}^{(2,0)}
    =\sum_{\ell_b}\mathcal{N}^{ab}
    {}_{s_b}R_{\ell_b}^{ab}\,
    {}_{s_b}Y^{ab}_{\ell_b 0}\,, \nonumber\\
    & h_{ab}^{(0,1)}
    =\sum_{\ell_g,m_g}\mathcal{N}^{ab}
    {}_{s_g}\mathcal{R}_{\ell_g m_g}^{ab}\,
    {}_{s_g}Y^{ab}_{\ell_g m_g}\,
    e^{-i\omega_g t+im_g\phi}\,,
\end{align} 
while a similar decomposition has been made for perturbations to the Weyl scalars $\Psi_{0,4}$:
\begin{align} \label{eq:harmonic_Weyl_scalar}
    {}_{s_b}\Psi^{(2,0)}
    =& \;\sum_{\ell_b}{}_{s_b}R_{\ell_b}\,
    {}_{s_b}Y_{\ell_b 0}\,, \nonumber\\
    {}_{s_g}\Psi^{(0,1)}
    =& \;\sum_{\ell_g,m_g}\bar{\Gamma}^{s_g-2}
    {}_{s_g}\mathcal{R}_{\ell_g m_g}\,
    {}_{s_g}S_{\ell_g m_g}\,
    e^{-i\omega_g t+im_g\phi}\,, \nonumber\\
    {}_{s}\Psi^{(2,1)}
    =& \;\sum_{\ell,m}\bar{\Gamma}^{s-2}
    {}_{s}\mathscr{R}_{\ell m}\,
    {}_{s}S_{\ell m}\,e^{-i\omega t+im\phi}\,,
\end{align}
where $\Psi_0={}_{2}\Psi$ and $\Psi_4={}_{-2}\Psi$. The angular functions ${}_{s}Y_{\ell m}={}_{s}Y_{\ell m}(\theta)$ and ${}_{s}S_{\ell m}={}_{s}S_{\ell m}(\theta)$ are the $(\ell,m)$ mode of spin-weighted spherical and spheroidal harmonics of spin weight $s$, respectively. Both $h_{ab}^{(2,0)}$ and $h_{ab}^{(0,1)}$ are in the unnormalized Kinnersley tetrad $\{l_+^{\mu},l_-^{\mu},m_+^{\mu},m_-^{\mu}\}$, defined as \cite{Dolan:2023enf}:
\begin{align} \label{eq:unnormalized_Kinnersley}
    l_{\pm}^\mu 
    & =\left[\pm\left(r^2+a^2\right)/\Delta,1,0,\pm a/\Delta\right]\,, \nonumber\\
    m_{\pm}^\mu 
    & =[\pm ia\sin\theta,0,1,\pm i\csc\theta]\,,
\end{align}
which is related to the Kinnersley tetrad $\{l,n,m,\bar{m}\}$ as:
\begin{align} \label{eq:normalized_Kinnersley}
    & l^\mu=l_{+}^\mu\,,
    && n^\mu=-\frac{\Delta}{2\Sigma}l_{-}^\mu\,, \nonumber\\
    & m^\mu=\frac{1}{\sqrt{2}\Gamma}m_{+}^\mu\,,
    && \bar{m}^\mu=\frac{1}{\sqrt{2}\bar{\Gamma}}m_{-}^\mu\,,
\end{align}
where $\Sigma=\Gamma\bar{\Gamma}$. The directional derivatives along the unnormalized Kinnersley tetrad in Eq.~\eqref{eq:unnormalized_Kinnersley} are the Chandrasekhar operators $\mathcal{D}_0$, $\mathcal{D}^{\dagger}_0$, $\mathcal{L}^{\dagger}_0$, and $\mathcal{L}_0$, respectively. The function $\mathcal{N}^{ab}$ is the normalization factor used by Ref.~\cite{Dolan:2021ijg, Dolan:2023enf} when decomposing the reconstructed metric onto spin-weighted spherical harmonics:
\begin{align}
    & \mathcal{N}^{l_+m_+}=\mathcal{N}^{l_-m_-}
    =\bar{\mathcal{N}}^{l_+m_-}=\bar{\mathcal{N}}^{l_-m_+}
    =\frac{1}{\Gamma}\,, \nonumber\\
    & \mathcal{N}^{l_+l_-}=\frac{1}{\Gamma\bar{\Gamma}\Delta}\,,
\end{align}
and we have used the same normalization factor for the decomposition of $h_{ab}^{(2,0)}$ for convenience, so the expressions of $\mathcal{O}(\zeta^0,\epsilon^1)$ Newman-Penrose quantities in terms of $h_{ab}^{(0,1)}$ can be reused for $\mathcal{O}(\zeta^2,\epsilon^0)$ quantities after switching the perturbation order. The factors of $\Gamma$ in the decompositions of $\Psi_4^{(0,1)}$ and $\Psi_4^{(2,1)}$ ensure the separability of the equations. Since we focus on the fundamental mode of the dipolar cloud in this work, we have fixed $\ell_c=m_c=1$ in the decomposition of $\Phi^{(1,0)}$.

Then, to reconstruct perturbations of NP quantities, we use the tetrad gauge in \cite{Campanelli:1998jv, Loutrel:2020wbw, Wagle:2023fwl} at $\mathcal{O}(\zeta^0,\epsilon^1)$:
\begin{align} \label{eq:perturbed_tetrad}
    & l^{\mu(0,1)}
    =\frac{1}{2}h_{ll}^{(0,1)}n^{\mu}\,,\nonumber \\
    & n^{\mu(0,1)}
    =\frac{1}{2}h_{nn}^{(0,1)}l^{\mu}
    +h_{ln}^{(0,1)}n^{\mu}\,, \\
    & m^{\mu(0,1)}=h_{nm}^{(0,1)}l^{\mu}
    +h_{lm}^{(0,1)}n^{\mu}
    -\frac{1}{2}h_{m\bar{m}}^{(0,1)}m^{\mu}
    -\frac{1}{2}h_{mm}^{(0,1)}\bar{m}^{\mu}\,,\nonumber 
\end{align}
where the same relation between the tetrad at $\mathcal{O}(\zeta^2,\epsilon^0)$ and the background metric correction $h_{ab}^{(2,0)}$ holds, with $h_{ab}^{(0,1)}$ replaced by $h_{ab}^{(2,0)}$. For simplicity, we have dropped the order-counting superscript of quantities at $\mathcal{O}(\zeta^0,\epsilon^0)$, and we will use this convention unless there is an ambiguity. In Eq.~\eqref{eq:perturbed_tetrad}, the metric perturbation is in the normalized Kinnersley tetrad, which can be converted to the unnormalized basis using Eq.~\eqref{eq:normalized_Kinnersley}. Given the tetrad conventions above, one can then directly use Eq.~(B4) and the linearization of Eq.~(B5) in \cite{Wagle:2023fwl} to compute the perturbed spin coefficients and Weyl scalars, respectively. These two sets of equations come from the commutation relations and Ricci identities in the NP formalism, respectively.

Finally, to restore the gauge in deriving the modified Teukolsky equations in Eq.~\eqref{eq:psi1_psi3}, we need to perform additional type I and II tetrad rotations in \cite{Chandrasekhar_1983}, i.e.,
\begin{align} \label{eq:tetrad_rot_fix_psi1}
    & l^{\mu(p,q)}
    \rightarrow l^{\mu(p,q)}+\bar{b}^{(p,q)}m^{\mu}
    +b^{(p,q)}\bar{m}^{\mu}\,, \nonumber\\
    & n^{\mu(p,q)}
    \rightarrow n^{\mu(p,q)}+\bar{a}^{(p,q)}m^{\mu}
    +a^{(p,q)}\bar{m}^{\mu}\,, \nonumber\\
    & m^{\mu(p,q)}
    \rightarrow m^{\mu(p,q)}+a^{(p,q)}l^{\mu}+b^{(p,q)}n^{\mu}\,,
\end{align}
where $(p,q)$ is either $(0,1)$ or $(2,0)$, and the rotation parameters $a^{(p,q)}$ and $b^{(p,q)}$ are \cite{Li:2022pcy, Wagle:2023fwl}
\begin{equation} \label{eq:tetrad_rot_fix_psi1_params}
    a^{(p,q)}=-\frac{\bar{\Psi}_{3}^{(p,q)}}{3\bar{\Psi}_2}\,,\quad
    b^{(p,q)}=-\frac{\Psi_{1}^{(p,q)}}{3\Psi_2}\,.
\end{equation}

After obtaining all the NP quantities, we can then examine the structure of the stress-energy tensors in Eq.~\eqref{eq:stress} and investigate how they enter the sources of the modified Teukolsky equations. Linearizing the scalar-field stress-energy tensor $T^{\Phi}_{\mu\nu}$ in Eq.~\eqref{eq:stress_cloud}, we observe that it generally has the structure:
\begin{align}
    T^{\Phi(2,0)}_{\mu\nu}
    & \sim\left[\Phi^{(1,0)}\right]
    \left[\bar{\Phi}^{(1,0)}\right]\,, \nonumber\\
    T^{\Phi(2,1)}_{\mu\nu}
    & \sim\left[\Phi^{(1,0)}\right]
    \left[\bar{\Phi}^{(1,1)}\right]
    +\left[\Phi^{(1,1)}\right]
    \left[\bar{\Phi}^{(1,0)}\right] \nonumber\\
    & \;+\left[h^{(0,1)}\right]\left[\Phi^{(1,0)}\right]
    \left[\bar\Phi^{(1,0)}\right]\,,
\end{align}
where we use $[f]$ to denote quantities linear in $f$. One can observe that since $\Phi$ enters at $\mathcal{O}(\zeta)$ and $T^{\Phi}_{\mu\nu}$ is quadratic in $\Phi$, $T^{\Phi}_{\mu\nu}$ enters at $\mathcal{O}(\zeta^2)$. Both the background scalar field $\Phi^{(1,0)}$ and the scalar radiation $\Phi^{(1,1)}$ have been computed in \cite{Dyson:2025dlj, Li:2025ffh}. The perturbed NP Ricci scalars are then related to $T^{\Phi}_{\mu\nu}$ as follows:
\begin{subequations}
\begin{align}
    & \Phi_{ij}^{\Phi(2,0)}
    \sim T^{\Phi(2,0)}_{\mu\nu}e_{a}^{\mu}e_{b}^{\nu}\,, \\
    & \Phi_{ij}^{\Phi(2,1)}
    \sim T^{\Phi(2,1)}_{\mu\nu}e_{a}^{\mu}e_{b}^{\nu}
    +T^{\Phi(2,0)}_{\mu\nu}e_{a}^{\mu(0,1)}e_{b}^{\nu}\,,
\end{align}    
\end{subequations}
where $i,j\in\{0,1,2\}$ and $e_{a}^{\mu}$ is the tetrad basis. Considering also the operators acting on $\Phi_{ij}$ in Eqs.~\eqref{eq:source_Psi0}, \eqref{eq:source_bianchi_psi0}, \eqref{eq:source_Psi4}, and \eqref{eq:source_bianchi_psi4}, one can then divide up $\mathcal{S}_{m}^{(2,1)}$ and $\mathcal{T}_{m}^{(2,1)}$ into the $A$ and $B$ sectors above after linearization, where they directly couple to the vacuum-GR gravitational radiation $h_{\mu\nu}^{(0,1)}$ and the scalar radiation $\Phi^{(1,1)}$, respectively. For the $A$ sector, the vacuum-GR metric perturbation $h_{\mu\nu}^{(0,1)}$ is reconstructed from the Teukolsky solutions $\Psi_{0,4}^{(0,1)}$ via the Lorenz reconstruction approach in \cite{Dolan:2021ijg, Dolan:2023enf}. For the $B$ sector, the scalar radiation $\Phi^{(1,1)}$ has been obtained in \cite{Li:2025ffh}.

In contrast, the point-particle stress-energy tensor $T_{\mu\nu}^{p}\propto m_p$, so it enters at $\mathcal{O}(\epsilon)$. On a deformed background, the particle's four-velocity and orbital frequency are modified, so one finds
\begin{equation} \label{eq:stress_particle_expand}
    T_{\mu\nu}^{p(2,1)}\sim T_{\mu\alpha}^{p(0,1)}h_{\nu}^{\alpha(2,0)}\,,
\end{equation}
where $T_{\mu\nu}^{p(0,1)}$ is given in Eq.~\eqref{eq:stress_particle} with $m_p$ factored out and evaluated on a Kerr background. The perturbed NP Ricci scalars driven by $T_{\mu\nu}^{p}$ then take the form
\begin{subequations} \label{eq:NP_Ricci_particle_expand}
\begin{align}
    & \Phi_{ij}^{p(0,1)}
    \sim T^{p(0,1)}_{\mu\nu}e_{a}^{\mu}e_{b}^{\nu}\,, \\
    & \Phi_{ij}^{p(2,1)}
    \sim T^{p(2,1)}_{\mu\nu}e_{a}^{\mu}e_{b}^{\nu}
    +T^{p(0,1)}_{\mu\nu}e_{a}^{\mu(2,0)}e_{b}^{\nu}\,.
\end{align}
\end{subequations}
After applying the operators in Eqs.~\eqref{eq:source_Psi0}, \eqref{eq:source_bianchi_psi0}, \eqref{eq:source_Psi4}, and \eqref{eq:source_bianchi_psi4}, the sources $\mathcal{S}_{p}^{(2,1)}$ and $\mathcal{T}_{p}^{(2,1)}$ then couple the background metric correction $h^{(2,0)}_{\mu\nu}$ to the Dirac delta function and its derivatives evaluated at the particle's location. In a later section, we explain in detail how $h^{(2,0)}_{\mu\nu}$ is obtained. Finally, the geometrical sources $\mathcal{S}_{\geo}^{(2,1)}$ and $\mathcal{T}_{\geo}^{(2,1)}$ couple the background metric correction $h_{\mu\nu}^{(2,0)}$ to the vacuum-GR gravitational radiation $h_{\mu\nu}^{(0,1)}$, which can now be computed, given the data obtained above for the other sources. 

To reduce these sources into an almost separable format, we follow Refs.~\cite{Dolan:2023enf, Li:2025ffh} to convert all the NP directional derivatives and spin coefficients to the Chandrasekhar operators $\left\{\mathcal{D}_{n},\mathcal{D}^{\dagger}_{n},\mathcal{L}_{n},\mathcal{L}^{\dagger}_{n}\right\}$ and the functions $\left\{\Delta,\sin{\theta},\Gamma,\bar{\Gamma}\right\}$ using Eq.~\eqref{eq:NP_operators}. In this case, one can leverage their commutation relations, as provided in Appendix A of \cite{Li:2025ffh}, to separate the radial and angular parts of each source term. In addition, we regroup all the operators and functions to ensure that the angular Chandrasekhar operators $\{\mathcal{L}_{s},\mathcal{L}^{\dagger}_{s}\}$ are in normal order, i.e., $\mathcal{L}_{s}$ and $\mathcal{L}^{\dagger}_{s}$ only act on spin-weighted spheroidal harmonics of spin weight $s$ and $-s$, respectively. Furthermore, any two consecutive angular Chandrasekhar operators in an expression are enforced to be in the order $\mathcal{L}_{s-1}\mathcal{L}_{s}$, $\mathcal{L}^{\dagger}_{-s+1}\mathcal{L}_{s}$, $\mathcal{L}_{s+1}\mathcal{L}^{\dagger}_{-s}$, and
$\mathcal{L}^{\dagger}_{-s-1}\mathcal{L}^{\dagger}_{-s}$. 

For nonrotating BHs, $\{\mathcal{L}_{s},\mathcal{L}^{\dagger}_{s}\}\rightarrow\{\hat{\mathcal{L}}_{s},\hat{\mathcal{L}}^{\dagger}_{s}\}$, with
\begin{subequations} \label{eq:ladder}
\begin{align}
    \hat{\mathcal{L}}_{s}\left[{}_{s}Y_{\ell m}\right]
    =& \sqrt{(\ell+s)(\ell-s+1)}\;{}_{s-1}Y_{\ell m}\,, \\
    \hat{\mathcal{L}}^{\dagger}_{-s}\left[{}_{s}Y_{\ell m}\right]
    =& -\sqrt{(\ell-s)(\ell+s+1)}\;{}_{s+1}Y_{\ell m}\,,
\end{align}
\end{subequations}
so $\{\hat{\mathcal{L}}_{s},\hat{\mathcal{L}}^{\dagger}_{s}\}$ are similar to the ladder operators for a hydrogen atom in quantum mechanics. For rotating black holes, such ladder operators are not known, but $\{\mathcal{L}_{s},\mathcal{L}^{\dagger}_{s}\}$ play a similar role as ladder operators. Since
\begin{equation} 
    \mathcal{L}_{s}
    =\hat{\mathcal{L}}_{s}-a\omega\sin{\theta}\,,\quad
    \mathcal{L}^{\dagger}_{s}
    =\hat{\mathcal{L}}^{\dagger}_{s}+a\omega\sin{\theta}\,,
\end{equation}
one can make a spheroidal-spherical expansion 
\begin{equation} \label{eq:spheroidal_spherical}
    {}_{s}S_{\ell m}
    =\sum_{\ell'}{}_{s}c_{\ell\ell'm}\,{}_{s}Y_{\ell' m}\,,
\end{equation}
where ${}_{s}c_{\ell\ell'm}$ are spherical-spheroidal mixing coefficients. The terms $\mathcal{L}_{s}\left[{}_{s}S_{\ell m}\right]$ and $\mathcal{L}^{\dagger}_{-s}\left[{}_{s}S_{\ell m}\right]$ can then be expressed as sums of ${}_{s-1}Y_{\ell' m}$ and ${}_{s+1}Y_{\ell' m}$ over $\ell'$, respectively. Thus, if there are no singular angular functions in the source after arranging all the angular Chandrasekhar operators into normal order, the source must be regular in $\theta$ and can be projected onto spin-weighted spheroidal harmonics. We observe that all the sources in Eqs.~\eqref{eq:source_Psi0} and \eqref{eq:source_Psi4} satisfy this property. 

After the above procedures, all the sources in Eqs.~\eqref{eq:source_Psi0} and \eqref{eq:source_Psi4} are reduced into an almost separable form. Taking $\mathcal{S}_{\geo}^{(2,1)}=\sum_m \mathcal{S}_{\geo}^{m}e^{-i\omega t+im\phi}$ as an example, we get
\begin{align} \label{eq:s_geo_regroup}
    \mathcal{S}_{\geo}^{m}
    =& \;\sum_{I_{\rm g}}
    \Gamma^{\alpha}\bar{\Gamma}^{\beta}
    \left\{\mathcal{F}_{I_{\rm g}}
    \hat{\mathcal{O}}_{e}^{\rm g}
    \left[{}_{s_b}{R}^{ab}_{\ell_b}\right]
    \hat{\mathcal{O}}_{f}^{\rm g}
    \left[{}_{s_g}\mathcal{R}^{cd}_{\ell_g m}\right]
    \right\} \nonumber\\
    & \;\times\left\{\sin^{\kappa}{\theta}\,
    {}_{s_b}Y_{\ell_b 0}^{ab}\,
    {}_{s_g}Y_{\ell_g m}^{cd}\right\}\,,
\end{align}
where $I_{\rm g}=\{\alpha,\beta,\kappa,a,b,c,d,e,f,s_b,s_g,\ell_b,\ell_g\}$ is a set of integer indices to sum over, with $\kappa$ being nonnegative. The function $\mathcal{F}_{I_{\rm g}}=
\mathcal{F}_{I_{\rm g}}(r)$ represents radial functions independent of the radial parts ${}_{s_b}R^{ab}_{\ell_b}$ and ${}_{s_g}\mathcal{R}^{ab}_{\ell_g m}$. The radial operators $\hat{\mathcal{O}}_{e}^{\rm g}$ and $\hat{\mathcal{O}}_{f}^{\rm g}$ are sequences of $\{\mathcal{D}_{0},\mathcal{D}^{\dagger}_{0}\}$. For ${}_{s_g}\mathcal{R}^{ab}_{\ell_g m_g}$, one can, in principle, find reduction rules for their derivatives using the equations of Hertz-like functions generating the reconstructed metric in \cite{Dolan:2021ijg, Dolan:2023enf}, but these rules are not provided explicitly and are involved to derive, so we choose to keep the derivatives of ${}_{s_g}\mathcal{R}^{ab}_{\ell_g m_g}$, as the order of them is at most three for the theories considered above. 

One further issue in differentiating ${}_{s_g}\mathcal{R}^{ab}_{\ell_g m_g}$ is the treatment of distributions supported at the orbit. Let $h$ denote a spherical-harmonic radial mode of the reconstructed metric obtained from \cite{Dolan:2021ijg, Dolan:2023enf}, and let $h_-$ and $h_+$ be its interior and exterior radial branches. Define $[f]_0=f(r_0^+)-f(r_0^-)$, where $r_0^+$ and $r_0^-$ denote approach to the orbit from larger and smaller radii, respectively.

The metric matching conditions in \cite{Dolan:2023enf} imply $[h]_0=0$. Its first-derivative jump is fixed by the distributional Einstein equation. For example, in the tetrad and harmonic normalization of Ref.~\cite{Dolan:2023enf},
Eq.~(127) of that reference gives
\begin{equation} \label{eq:metric_first_jump_reference}
    \left[\partial_r h^{\ell_g m_g}_{l_+l_+}\right]_0
    =-\frac{16\pi m_p}{u^t\Delta(r_0)}
    (u_{\mu}l_+^{\mu})^2
    {}_{0}Y_{\ell_g m_g}(\pi/2)\,,
\end{equation}
where $u^{\mu}$ is the four-velocity of the point particle, and the tetrad contraction above is evaluated on the orbit. The other components have correspondingly projected Einstein-equation matching conditions. In our numerical implementation, the jumps are extracted from the existing spectral representations of the two radial branches inside and outside the orbit. As a representative check, for $m_g=2$ and $r_0=10.1M$, the fitted first-derivative jump of $\Delta^2 h^{\ell_g m_g}_{l_+l_+}$ differs from Eq.~\eqref{eq:metric_first_jump_reference} by relative residuals of $6.27\times10^{-6}$, $4.60\times10^{-5}$, $2.56\times10^{-4}$, and $5.36\times10^{-4}$ for $\ell_g=2,4,6,8$, respectively. The resulting distributional derivatives of the reconstructed metric are then
\begin{align}
    \partial_r^2h
    &=(\partial_r^2h)_{\rm pw}+[h']_0\delta(r-r_0),\nonumber\\
    \partial_r^3h
    &=(\partial_r^3h)_{\rm pw}
    +[h'']_0\delta(r-r_0)+[h']_0\delta'(r-r_0),
    \label{eq:metric_interface_derivatives}
\end{align}
where ${\rm pw}$ denotes ordinary differentiation on $r<r_0$ and
$r>r_0$ piecewise, and $f'(r)=\partial_rf(r)$.

As we obtain the Weyl scalar perturbations $\Psi_{0,4}^{(0,1)}$ from the \texttt{Teukolsky} package of the BHPToolkit \cite{BHPToolkit} directly, the distributional derivatives of $\Psi_{0,4}^{(0,1)}$  require a separate treatment. The radial part ${}_{s_g}\mathcal{R}_{\ell_g m_g}$ of $\Psi_{0,4}^{(0,1)}$ satisfies the following ordinary-GR point-particle Teukolsky equation:
\begin{align}
    & \left[\Delta^{-s}\frac{d}{dr}
    \left(\Delta^{s+1}\frac{d}{dr}\right)+{}_{s}V_{\ell m}\right]
    {}_{s}\mathcal{R}_{\ell m}(r)
    ={}_{s}\mathfrak{S}^{\ell m}(r)\,, \nonumber\\
    & {}_{s}\mathfrak{S}^{\ell m}(r)
    =\sum_{n=0}^{2}{}_{s}\mathfrak{C}_n^{\ell m}
    \delta^{(n)}(r-r_0)\,,
    \label{eq:GR_source_contact_definition}
\end{align}
where $s=\pm2$ and ${}_{s}V_{\ell m}(r)$ is defined in Eq.~\eqref{eq:MTF_radial}. We drop the label $\mathrm{g}$ of the mode numbers here for simplicity. The constants ${}_{s}\mathfrak{C}_n^{\ell m}$, $n\in\{0,1,2\}$, are the known coefficients of the delta distribution and its first two derivatives in the vacuum-GR point-particle source, and we obtain them using the circular-orbit source expressions of the \texttt{Teukolsky} package in the BHPToolkit \cite{BHPToolkit}. Writing ${}_{s}\mathcal{R}_{\ell m}={}_{s}\mathcal{R}^{\rm pw}_{\ell m}+{}_{s}Q_{\ell m}\delta(r-r_0)$ and matching the three distributions gives
\begin{align} \label{eq:radial_interface_matching}
    {}_{s}Q^{\ell m}
    &=\frac{{}_{s}\mathfrak{C}_2^{\ell m}}
    {\Delta}\,, \nonumber\\
    [{}_{s}\mathcal{R}^{\rm pw}_{\ell m}]_0
    &=\frac{{}_{s}\mathfrak{C}_1^{\ell m}
    -(s-1)\Delta'{}_{s}Q^{\ell m}}{\Delta}\,, \\
    [\partial_r{}_{s}\mathcal{R}^{\rm pw}_{\ell m}]_0
    &=\frac{{}_{s}\mathfrak{C}_0^{\ell m}
    -s\Delta'[{}_{s}\mathcal{R}^{\rm pw}_{\ell m}]_0
    -({}_{s}V_{\ell m}-2s){}_{s}Q^{\ell m}}{\Delta}\,, \nonumber
\end{align}
where the radial functions $\Delta$, $\Delta'$, and ${}_{s}V_{\ell m}$ are evaluated at $r_0$. The relations above reproduce the distributional matching structure of Eqs.~(D21)--(D25) in Ref.~\cite{Dolan:2023enf}, after identifying the source and radial conventions.

For the derivatives of ${}_{s_b}R^{ab}_{\ell_b}$, we use the stationary Einstein equations in Eq.~\eqref{eq:constraint}, defined later, to reduce them. The derivatives on ${}_{s}Y_{\ell m}$ are reduced with the ladder operator relations in Eq.~\eqref{eq:ladder}. The source $S^{m}_{\geo}$ is almost separable as all the terms in the sum are multiplications of purely radial and angular terms [i.e., the first and second bracket of Eq.~\eqref{eq:s_geo_regroup}, respectively], except that the factor $\Gamma^{\alpha}\bar{\Gamma}^{\beta}$ is non-separable if $\alpha,\beta<0$. The other source terms can be simplified to a similar form but with different contributions from the scalar cloud profile, scalar radiation, background metric correction, and vacuum-GR gravitational radiation.

\subsection{Extraction of radial sources}

Here, we show how to reduce the sources in a form similar to $\mathcal{S}_{\geo}^{m}$ in Eq.~\eqref{eq:s_geo_regroup} into a completely separable form. As discussed above, the most important term that needs further reductions is the factor $\Gamma^{\alpha}\bar{\Gamma}^{\beta}$ when $\alpha,\beta<0$.
Following \cite{Spiers:2024src, Li:2025ffh}, we resolve this issue by making a Fourier series decomposition of this factor. After this decomposition, the sources can then be projected onto spin-weighted spheroidal harmonics without evaluating the projection integral at every radial grid point. 

Specifically, we first rewrite the factor $\Gamma^{\alpha}\bar{\Gamma}^{\beta}$ into the following form:
\begin{equation} \label{eq:gamma_regroup}
     \Gamma^{\alpha}\bar{\Gamma}^{\beta}
     =\begin{cases}
     \Sigma^{\beta}\Gamma^{\alpha-\beta}
     & \text{if } \alpha\geq\beta \\
     \Sigma^{\alpha}\bar{\Gamma}^{\beta-\alpha}
     & \text{if } \alpha < \beta
     \end{cases}\,,
\end{equation}
where $\Sigma=\Gamma\bar{\Gamma}=r^2+a^2\cos^2{\theta}$. For each case, the exponent of $\Gamma$ or $\bar{\Gamma}$ is positive, so we can make a binomial expansion of them, i.e.,
\begin{align}
    (r\pm ia\cos{\theta})^n=
    \sum_{k=0}^{n}\binom{n}{k}r^{n-k}(\pm ia\cos{\theta})^k\,.
\end{align}
For the factor of $\Sigma$, if its exponent is nonnegative, one can also make a binomial expansion. When the exponent is negative, we follow \cite{Spiers:2024src, Li:2025ffh} and make a Fourier series expansion to facilitate angular projections. Given that $\Sigma$ is a periodic and even function of $\theta$, we obtain
\begin{equation} \label{eq:sigma_expansion}
    \Sigma^{-n}
    =(r^2+a^2\cos^2{\theta})^{-n}
    =\sum_{p=0}^{\infty}f_p(r)\cos{p\theta}\,,
\end{equation}
where $f_p(r)$ can be calculated analytically at each $p$ \cite{Spiers:2024src, Li:2025ffh}. For all the sources considered in this work, the most negative power of $\Sigma$ we need to consider is $\Sigma^{-7}$. To ensure that this series representation is accurate after truncating at some value $p=p_{\max}$, we have set $p_{\max}\geq2k$ and use $64$ digits of precision for $r$ when evaluating $f_p(r)$. As tested in \cite{Spiers:2024src, Li:2025ffh}, this should be sufficient for the purpose of LISA data analysis.

After decomposing the factors of $\Gamma^{\alpha}\bar{\Gamma}^{\beta}$, the source in Eq.~\eqref{eq:s_geo_regroup} is then a sum of terms that are multiplications of purely angular and radial terms, i.e.,
\begin{align}
    \mathcal{S}_{\geo}^{m}
    &= \sum_{\mathcal{I}_{\rm g}}
    \left\{\mathcal{F}_{\mathcal{I}_{\rm g}}
    \hat{\mathcal{O}}_{e}^{\rm g}
    \left[{}_{s_b}{R}^{ab}_{\ell_b}\right]
    \hat{\mathcal{O}}_{f}^{\rm g}
    \left[{}_{s_g}\mathcal{R}^{cd}_{\ell_g m}\right]
    \right\} \nonumber\\
    & \mathrel{\times}
    \left\{\sin^{\kappa}\theta
    \cos^{k}{\theta}\cos{p\theta}
    {}_{s_b}Y_{\ell_b 0}^{ab}\,
    {}_{s_g}Y_{\ell_g m}^{cd}
    \right\}
\end{align}
where $\mathcal{I}_{\rm g}=\{\kappa,a,b,c,d,e,f,k,p,s_b,s_g,\ell_b,\ell_g\}$, and the ranges of $k$ and $p$ depend on the exponents $\alpha$ and $\beta$ of $\Gamma^{\alpha}\bar{\Gamma}^{\beta}$. One can then use the orthogonality of spin-weighted spheroidal harmonics to extract each radial $(\ell,m)$ mode of $\mathcal{S}_{\geo}^{m}$, and similarly for other sources, with the results given in Eq.~\eqref{eq:radial_source} and the angular projection coefficients defined as:
\begin{align} \label{eq:angular_coeff}
    {}_{s}\mathscr{C}_{\mathcal{J}_{\rm g}}^{\ell m}
    =& \;2\pi\int_{0}^{\pi}\left\{\sin^{\kappa}{\theta}
    \cos^k{\theta}\cos{p\theta}\,
    {}_{s_b}Y_{\ell_b 0}^{ab}\,
    {}_{s_g}Y_{\ell_g m}^{cd}\right\} \nonumber\\
    & \;{}_{s}\bar{S}_{\ell m}
    \sin{\theta}d\theta\,, \nonumber\\
    {}_{s}\mathscr{C}_{\mathcal{J}_A}^{\ell m}
    =& \;2\pi\int_{0}^{\pi}\Big\{\sin^{\kappa}{\theta}
    \cos^k{\theta}\cos{p\theta}\,
    \hat{\mathcal{Q}}^{A}_{i}
    \left[{}_{0}S_{\ell_c m_c}\right]
    \nonumber\\
    & \;\hat{\mathcal{Q}}^{A}_{j}
    \left[{}_{0}\bar{S}_{\ell_c m_c}\right]
    {}_{s_g}Y_{\ell_g m}^{ab}\Big\}\,
    {}_{s}\bar{S}_{\ell m}
    \sin{\theta}d\theta\,, \nonumber\\
    {}_{s}\mathscr{C}_{\mathcal{J}_B}^{\ell m}
    =& \;2\pi\int_{0}^{\pi}\Big\{\sin^{\kappa}{\theta}
    \cos^k{\theta}\cos{p\theta}\,
    \hat{\mathcal{Q}}_{i}^{B}
    \left[{}_{0}S_{\ell_c m_c}\right]
    \nonumber\\
    & \;\hat{\mathcal{Q}}_{j}^{B}
    \left[{}_{0}\bar{S}_{\ell_s m_s}\right]\Big\}\,
    {}_{s}\bar{S}_{\ell m}
    \sin{\theta}d\theta\,, \nonumber\\
    {}_{s}\mathscr{C}_{\mathcal{J}_B}^{\dagger \ell m}
    =& \;2\pi\int_{0}^{\pi}\Big\{\sin^{\kappa}{\theta}
    \cos^k{\theta}\cos{p\theta}\,
    \hat{\mathcal{Q}}_{i}^{\dagger B}
    \left[{}_{0}\bar{S}_{\ell_c m_c}\right]
    \nonumber\\
    & \;\hat{\mathcal{Q}}_{j}^{\dagger B}
    \left[{}_{0}S_{\ell_s m_s}\right]\Big\}\,
    {}_{s}\bar{S}_{\ell m}
    \sin{\theta}d\theta\,, \nonumber\\
    {}_{s}\mathscr{C}_{\mathcal{J}_p}^{\ell m} 
    =& \;\int_{0}^{\pi}{}_{s_b}Y_{\ell_b 0}^{ab}\,
    \delta^{(i)}(\theta-\pi/2)\,
    {}_{s}\bar{S}_{\ell m}\sin{\theta}
    d\theta\,.
\end{align}
The index sets are defined to be 
\begin{align} \label{eq:index_set}
    & \mathcal{J}_{\rm g}
    =\{\kappa,a,b,c,d,k,p,s_b,s_g,\ell_b,\ell_g\}\,,\;
    \mathcal{I}_{\rm g}
    =\mathcal{J}_{\rm g}\cup\{e,f\}\,, \nonumber\\
    & \mathcal{J}_A
    =\{\kappa,a,b,i,j,k,p,s_g,\ell_g\}\,,\;
    \mathcal{I}_A
    =\mathcal{J}_A\cup\{d,e,f\}\,, \nonumber\\
    & \mathcal{J}_B
    =\{\kappa,i,j,k,p,\ell_c,\ell_s\}\,,\;
    \mathcal{I}_{B}
    =\mathcal{J}_{B}\cup\{d,e\}\,, \nonumber\\
    & \mathcal{J}_p
    =\{a,b,i,s_b,\ell_b\}\,,\;
    \mathcal{I}_{\rm p}
    =\mathcal{J}_{\rm p}\cup\{d,n\}\,.
\end{align}
In Eq.~\eqref{eq:angular_coeff}, the angular operators $\hat{\mathcal{Q}}$ and $\hat{\mathcal{Q}}^{\dagger}$, with their indices and labels hidden, are sequences of $\{\mathcal{L}_{n},\mathcal{L}^{\dagger}_{n}\}$ in the normal order specified above. The factor of $2\pi$ comes from the integral in $\phi$.

One caveat in evaluating Eq.~\eqref{eq:angular_coeff} is that there are generally at most two $\theta$-derivatives acting on $T_{\mu\nu}^{p}\propto\delta(\theta-\pi/2)$. We can simplify them by using the distributional properties of the Dirac delta function and its derivative, i.e.,
\begin{align} \label{eq:deltaidentity}
    & f(x)\,\delta^{(n)}(x-x_0) \nonumber\\
    & =\sum_{k=0}^{n}(-1)^k \binom{n}{k}\delta^{(n-k)}(x-x_0)\,\left.\frac{d^k f}{dx^k}\right|_{x=x_0}\,,
\end{align}
so the coefficients in front of $\delta(\theta-\pi/2)$ and its derivatives can be evaluated at $\theta=\pi/2$. In obtaining $\mathscr{C}_{\mathcal{J}_p}^{\ell m}$, we have also reduced the $\theta$-derivatives acting on ${}_{s_b}Y_{\ell_b 0}^{ab}$, due to using Eq.~\eqref{eq:deltaidentity}, by recursively applying the rule,
\begin{align}
    \partial_\theta{}_{s}Y_{\ell m}
    =& \;\frac{1}{2}(\mathcal{L}_{s}+\mathcal{L}_{-s}^{\dagger})
    {}_{s}Y_{\ell m}\,,
\end{align}
and Eq.~\eqref{eq:ladder}. In this case, all the derivatives acting on ${}_{s_b}Y_{\ell_b 0}^{ab}$ are removed in the ${}_{s}\mathscr{C}_{\mathcal{J}_p}^{\ell m}$ of Eq.~\eqref{eq:angular_coeff}.

For the factor $\sin^\kappa{\theta}\cos^{k}{\theta}\cos{p\theta}$, we use trigonometric identities to reduce it into a sum of $\cos{q\theta}$ and $\sin{q\theta}$, where $q$ is a positive integer. For the target angular mode ${}_{\pm 2}\bar{S}_{\ell m}$, we expand it in terms of spin-weighted spherical harmonics using Eq.~\eqref{eq:spheroidal_spherical}. Since the spheroidicity $\gamma=a\omega$ is tiny for a particle in a circular, equatorial orbit (i.e., the largest value $\gamma\approx 0.35$ occurs when $r_0=5.1M$ and $|m|=5$ in our case), this is a very effective treatment. Given $\ell_{\max}=5$, we have set $\ell'_{\max}=18$ in Eq.~\eqref{eq:spheroidal_spherical}. In this case, one does not need to evaluate the integrals in Eq.~\eqref{eq:angular_coeff} at every particle radius but only the spherical-spheroidal mixing coefficients, the latter of which can be efficiently calculated and tabulated. Furthermore, for the terms like $\hat{\mathcal{Q}}_{i}^{B}
\left[{}_{0}S^{\Phi}_{\ell_c m_c}\right]$ and $\hat{\mathcal{Q}}_{j}^{B}\left[{}_{0}\bar{S}^{\varphi}_{\ell_s m_s}\right]$, we use their angular equations to reduce the order of derivatives such that there is at most one $\mathcal{L}_{n}$ or $\mathcal{L}_{n}^{\dagger}$ in the angular operators $\hat{\mathcal{Q}}_{i}^{B}$ and $\hat{\mathcal{Q}}_{j}^{B}$. This treatment introduces an artificial factor of $\csc{\theta}$ due to the use of the angular equation, but the source term itself is angular finite as argued above. This factor also cancels with $\sin\theta$ in the projection integrals, so we adopt this treatment to reduce the number of evaluations. 

After those simplifications, we then collect all the independent angular integrals in Eq.~\eqref{eq:angular_coeff} and evaluate them. These simplifications are crucial for efficient computations involving a large set of modes. For example, before the simplifications above, the source term $\mathcal{S}_{m}^{(2,1)}$ contained $\sim 10^6$ terms, and there were $\sim 10^4$ modes to evaluate, given the range of $r_0,\ell_g,\ell_s,\ell,m$ considered in this work. After the simplifications, there are now only $\sim 500$ independent terms at each mode, though this number varies with the cutoff in the Fourier series decomposition of negative powers of $\Gamma$ and $\bar{\Gamma}$. The procedures above also help reduce the computational cost, if one wants to increase the resolution of the sources, i.e., more points in the $r_0$ grid or more harmonic modes.

\subsection{Stationary scalar-cloud background}

\begin{figure*}[t]
    \centering
    \includegraphics[width=\linewidth]{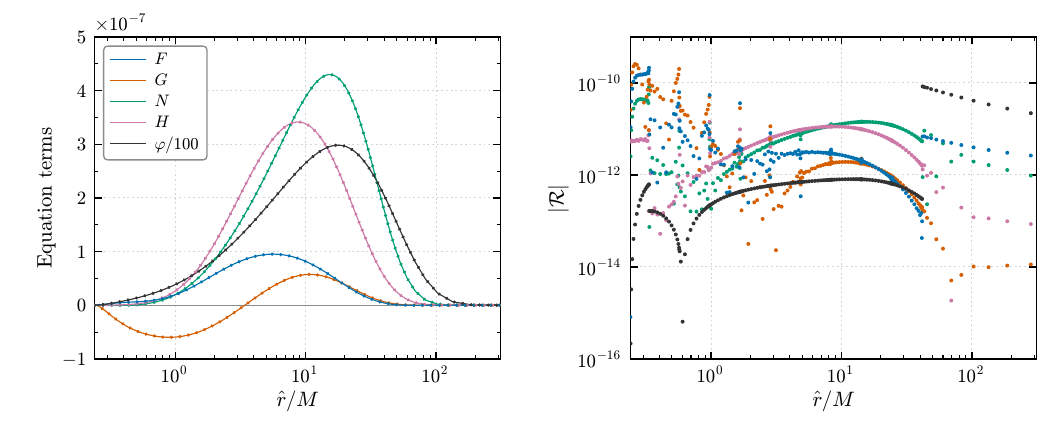}
    \caption{Numerical check of the five coupled Einstein--scalar equations in Eq.~\eqref{eq:constraint} at the equator $\theta=\pi/2$ for the background configuration considered in the main text, except for $M_c\approx0.35\%M$. \textbf{Left}: Left-hand sides (thin solid curves) and right-hand sides (dotted curves) of the five equations in Eq.~\eqref{eq:constraint}, distinguished by color and labeled $N$, $H$, $G$, $F$, and $\varphi$ in their order of appearance in that equation. The scalar-equation terms are divided by $100$ for display purposes in this panel only. \textbf{Right}: Absolute value of the residuals $|\mathcal{R}|$ of Eq.~\eqref{eq:constraint}. Both panels use the quasi-isotropic radial coordinate ${\hat r}$ normalized by $M$.} 
    \label{fig:background_EE}
\end{figure*}

Here, we show how to calculate the cloud-induced metric correction $h_{\mu\nu}^{(2,0)}$ entering the modified Teukolsky source. We first construct a stationary and axisymmetric solution of the Einstein--Klein--Gordon system. In quasi-isotropic coordinates, we write the metric as
\begin{align}
    ds^2 
    =& -N^2 dt^2+F^2\left(d{\hat r}^2+{\hat r}^2 d\theta^2\right) \nn
    &+ G^2{\hat r}^2\sin^2\theta
    \left(d\phi+H dt\right)^2\,,
\end{align}
where the radial coordinate ${\hat r}$ in quasi-isotropic coordinates should not be confused with the Boyer-Lindquist one. The scalar field is synchronized with the horizon, i.e., $\omega_c=m_c\omega_+$, where $\omega_+$ is the horizon angular velocity, and has the form
\begin{align}
    \Phi=\varphi(r,\theta)e^{-i\omega_c t+im_c\phi}\,.
\end{align}
Although $\Phi$ has a harmonic dependence on $t$ and $\phi$, its stress-energy tensor is stationary and axisymmetric. Projecting Eq.~\eqref{eq:stress_cloud} onto the spatial hypersurfaces gives
\begin{widetext}
\begin{subequations}
\begin{align}
    & \mathcal{E} 
    \equiv T_{\mu\nu}^{\Phi}n^\mu n^\nu
    =\left[\frac{(\omega_c+m_c H)^2}{N^2}
    +\frac{m_c^2}{G^2 {\hat r}^2\sin^2\theta}\right]\varphi^2
    +F^{-2}(\pa\varphi)^2+\mu_{\Phi}^2\varphi^2 \,,\\
    & J_\varphi
    \equiv-T_{\mu\nu}^{\Phi}n^\nu\gamma^\mu_\phi 
    =\frac{2m_c(\omega_c+m_c H)}{N}\varphi^2  \,,\\
    & S_\varphi^{~\varphi}
    \equiv T_{\mu\nu}^{\Phi}\gamma^{\mu\phi}\gamma^\nu_\phi
    =\left[\frac{(\omega_c+m_c H)^2}{N^2}
    +\frac{m_c^2}{G^2 {\hat r}^2\sin^2\theta}\right]\varphi^2
    -F^{-2}(\pa\varphi)^2-\mu_{\Phi}^2\varphi^2 \,,\\
    & S_k^{~k}
    \equiv T_{\mu\nu}^{\Phi}\gamma^{\mu k}\gamma^\nu_k
    =\left[\frac{3(\omega_c+m_c H)^2}{N^2}
    -\frac{m_c^2}{G^2 {\hat r}^2\sin^2\theta}\right]\varphi^2
    -F^{-2}(\pa\varphi)^2-3\mu_{\Phi}^2\varphi^2\,,
\end{align}
\end{subequations}
where $\gamma_{ij}$ is the spatial part of the metric, $n^\mu$ is the unit vector normal to the spatial hypersurface, and $(\pa u)(\pa v)\equiv\pa_{\hat r} u\,\pa_{\hat r} v+{\hat r} ^{-2}\pa_\theta u\,\pa_\theta v$. The Einstein and Klein-Gordon equations then reduce to the coupled elliptic system \cite{Grandclement:2014msa, Garcia:2023ntf}:
\begin{figure*}[t!]
    \centering
    \includegraphics[width=\linewidth]{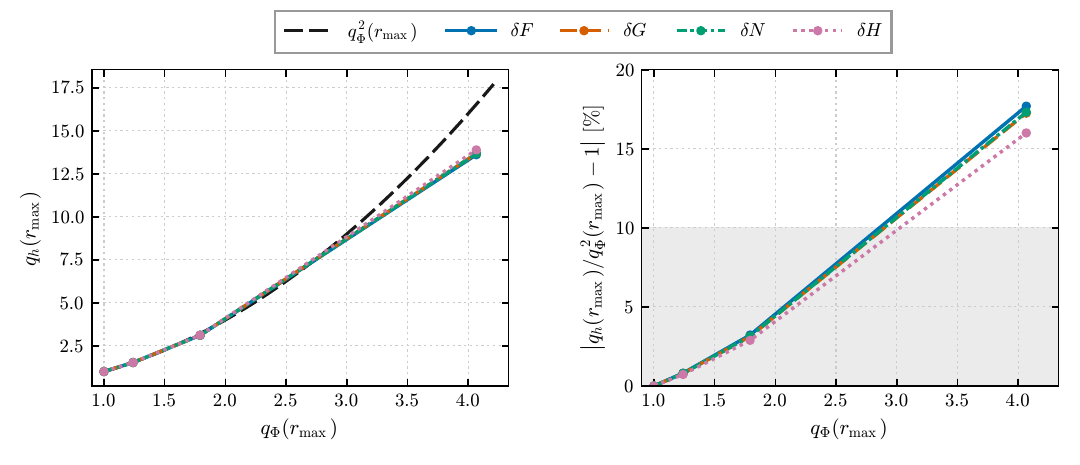}
    \caption{The relative change of the metric function corrections with respect to the scalar field amplitude changes. \textbf{Left:} The variable $q_{h}(r_{\max})$ as a function of $q_{\Phi}(r_{\max})$, where both variables are defined in Eq.~\eqref{eq:def_q} with $h$ being $\delta F$ (blue), $\delta G$ (orange), $\delta N$ (green), or $\delta H$ (purple). The reference configuration has the cloud mass $M_c\approx0.35\%M$, while the configurations for comparison, marked by solid circles, have cloud masses $M_c/M\approx0.54\%,1.14\%,5.86\%$ from left to right, respectively. The black curve $q_{\Phi}^2(r_{\max})$ is shown as a reference. \textbf{Right:} The characteristic deviation from quadratic scaling $\left|q_{h}(r_{\max})/q_{\Phi}^2(r_{\max})-1\right|$ in percentage as a function of $q_{\Phi}(r_{\max})$. The shaded region shows when $\left|q_{h}(r_{\max})/q_{\Phi}^2(r_{\max})-1\right|\leq 10\%$.} 
    \label{fig:background_ratio}
\end{figure*}
\begin{subequations} \label{eq:constraint}
\begin{align}
    & N\triangle_2N
    +N\frac{(\pa N)(\pa G)}{G} 
    -\frac{G^2 {\hat r}^2\sin^2\theta}{2}(\pa H)^2 
    =4\pi F^2N^2(\mathcal{E}+S_k^{~k})\,, 
    \label{eq:constraint_alpha} \\
    & N\left(\triangle_2
    -\frac{1}{{\hat r}^2\sin^2\theta}\right)\tilde{H}
    -{\hat r}\sin\theta(\pa H)(\pa N)
    +3N {\hat r}\sin\theta(\pa H)(\pa\ln G) 
    =\frac{16\pi F^2N^2 J_\varphi}{G^2 {\hat r}\sin\theta}\,, \\
    & N\left(2\triangle_2-\tilde{\triangle}_2\right)
    \left(N G\right)
    =8\pi F^2 GN^2(S_k^{~k}-S_\phi^{~\phi})\,,\\ 
    & N\tilde\triangle_2F
    +F\tilde\triangle_2N
    -\frac{N}{F}(\pa F)^2
    -\frac{3}{4}FG^2 {\hat r}^2\sin^2\theta\frac{(\pa H)^2}{N}
    = 8\pi F^3N S_\phi^{~\phi}\,, \\
    & N^2\left(\triangle_2
    -\frac{m_c^2}{{\hat r}^2\sin^2\theta}\right)\varphi
    + F^2\left(\omega_c+m_c H\right)^2\varphi
    +\left(1-\frac{F^2}{G^2}\right)
    \frac{m_c^2N^2\varphi}{{\hat r}^2\sin^2\theta}
    +N(\pa\varphi)(\pa N)
    +N^2(\pa\varphi)(\pa\ln G)
    = F^2N^2\mu_{\Phi}^2\varphi\,.
    \label{eq:constraint_phi}
\end{align}
\end{subequations}
Here, $\tilde{H}\equiv H\hat{r}\sin\theta$, while $\triangle_2$ and $\tilde{\triangle}_2$ denote the axisymmetric 3D and 2D flat-space Laplacians, respectively:
\begin{align}
    \triangle_2
    \equiv\pa_{\hat r}^2
    +2{\hat r}^{-1}\pa_{\hat r}
    +{\hat r}^{-2}\pa_\theta^2
    +\frac{1}{{\hat r}^2\tan\theta}\pa_\theta\,,\quad
    \tilde{\triangle}_2
    \equiv\pa_{\hat r}^2
    +{\hat r}^{-1}\pa_{\hat r}
    +{\hat r}^{-2}\pa_\theta^2\,.
\end{align}
\end{widetext}

We regularize Eq.~\eqref{eq:constraint} near the horizon by rewriting the system to remove inverse powers of the lapse $N$. We solve the resulting equations with the spectral code \texttt{Celepha{\"i}s}~\cite{Kuan:2026pss}. At the horizon, we impose $N=0$. Asymptotic flatness requires $N=F=G=1$ and $H=\varphi=0$ at spatial infinity. The radial grid comprises several Chebyshev subdomains, including a compactified outer domain, while Fourier series represent the angular dependence. In particular, outside the horizon, we have 10 domains with the radius of each domain being twice that of the preceding one and the first one spanning over $[{\hat r}_+,\,2{\hat r}_+]$ for ${\hat r}_+$ the horizon radius in the quasi-isotropic coordinate. We then append a compactified layer as the outermost domain. In each domain, we use 17 spectral coefficients in both the radial and polar directions and 16 in the azimuthal direction. We terminate the Newton iteration when the equation residual, an absolute tolerance on the equations as implemented and scaled, between successive iterations falls below $10^{-8}$.

Figure~\ref{fig:background_EE} compares the left- and right-hand sides of Eq.~\eqref{eq:constraint} at the equator $\theta=\pi/2$ and shows their absolute residuals for the same black hole and scalar cloud configuration considered in the main text, except for $M_c\approx0.35\%M$. The maximum absolute residual across the five equations over the displayed radial grid is approximately $2.6\times10^{-10}$. After matching the mass and spin, we transform the obtained background metric to Boyer-Lindquist coordinates and subtract the corresponding Kerr metric from the numerical solution to obtain the correction $h_{\mu\nu}^{(2,0)}$ used in the modified Teukolsky source.

To ensure that the correction $h_{\mu\nu}^{(2,0)}$ we obtained is in the linear regime, we iterate the above solver at several scalar field amplitudes, or, equivalently, cloud masses. We then check whether the relative changes of the metric function amplitudes scale quadratically in the relative change of the scalar field amplitude. Specifically, we define
\begin{align} \label{eq:def_q}
    q_{\Phi}(r)
    =\frac{\varphi(r)}{\varphi_{\mathrm{ref}}(r)}\,,\;
    q_{h}(r)
    =\frac{h(r)}{h_{\mathrm{ref}}(r)}\,,
\end{align}
where $h$ represents the metric function corrections $\delta N$, $\delta H$, $\delta F$, and $\delta G$. The function $\delta N$ is the residual of $N$ after subtracting its vacuum-GR value, and similarly for other functions. The fields $\varphi_{\mathrm{ref}}$ and $h_{\mathrm{ref}}$ are the scalar field and metric function corrections of the reference configuration in Fig.~\ref{fig:background_EE} with the cloud mass $M_c\approx0.35\%M$. We further define the radial position $r_{\max}$ of each field to be the location of its peak in the reference configuration. 

In the left panel of Fig.~\ref{fig:background_ratio}, we plot $q_{h}(r_{\max})$ against $q_{\Phi}(r_{\max})$ for configurations with the same parameters, except for the cloud mass (or scalar field amplitude): $M_c\approx0.54\%M$, $M_c\approx1.14\%M$, and $M_c\approx5.86\%M$, where the curve $q_{\Phi}^2(r_{\max})$ is shown as a reference. In the right panel, we plot the characteristic deviation from quadratic scaling, $\left|q_{h}(r_{\max})/q_{\Phi}^2(r_{\max})-1\right|$, in percentage as a function of $q_{\Phi}(r_{\max})$. From both panels, we observe that $q_{h}(r_{\max})$ is quadratic in $q_{\Phi}(r_{\max})$ for most of the displayed regions until $q_{\Phi}(r_{\max})\approx 3$, where $\left|q_{h}(r_{\max})/q_{\Phi}^2(r_{\max})-1\right|\approx 10\%$ and $M_c\approx 3.19\% M$. 

\begin{figure*}[t!]
    \centering
    \includegraphics[width=\linewidth]{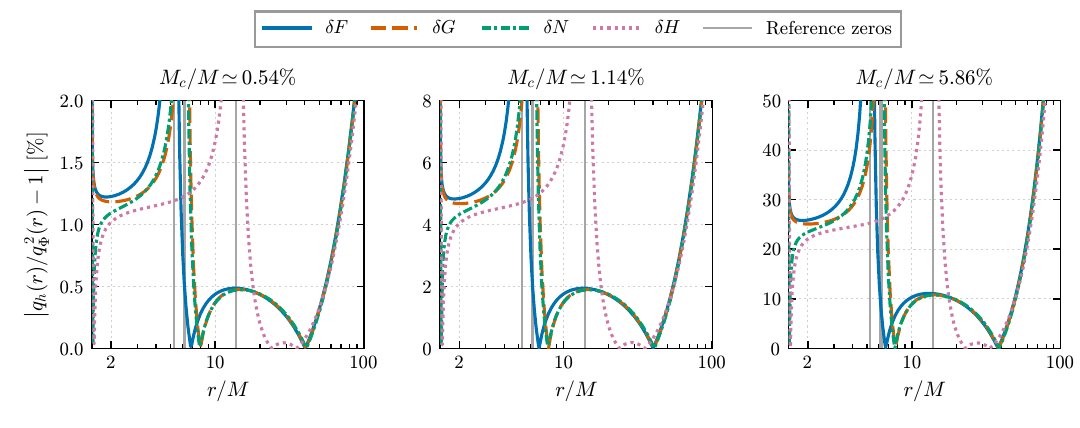}
    \caption{The local deviation from quadratic scaling, $\left|q_h(r)/q_{\Phi}^2(r)-1\right|$, expressed in percent, as a function of the Boyer–Lindquist radius $r/M$ on the equatorial plane. From left to right, the configurations have cloud masses $M_c/M\approx 0.54\%,\,1.14\%,\,5.86\%$, compared with the reference configuration with $M_{c}/M\approx 0.35\%$. Blue, orange, green, and purple curves correspond to $h=\delta F,\delta G,\delta N,\delta H$, respectively. Solid gray vertical lines mark zeros of the reference metric corrections, where the relative deviations become singular.} 
    \label{fig:background_ratio_full_range}
\end{figure*}

One may wonder whether this quadratic scaling is only valid at $r_{\max}$, or also at other radial locations. To answer this question, we show the local deviation from quadratic scaling, $\left|q_{h}(r)/q_{\Phi}^2(r)-1\right|$, with $r\in [r_+,100M]$, in Fig.~\ref{fig:background_ratio_full_range}. For the configuration with cloud mass $M_c\approx0.54\%M$, $\left|q_{h}(r)/q_{\Phi}^2(r)-1\right|<2\%$ throughout most of the displayed domain, except near the zero-crossing locations of the reference metric corrections, marked by gray solid lines, where this measure is not well defined. These checks together indicate that our Einstein-scalar configurations do indeed solve the field equations and are in the perturbative regime.

\subsection{Regularization of the solutions}

Here, we detail the distributional and boundary prescriptions used to evaluate the Green's function in Eq.~\eqref{eq:Green_function}. We treat particle-localized sources, near-horizon terms, and near-infinity terms separately.

\subsubsection{Distributions localized at the particle's location}

First, the radial source contains distributions supported on the worldline of the small compact object. Besides the explicit point-particle contribution
${}_{s}\mathscr{S}^{\ell m}_{p}$, these singular distributions arise from radial differentiation of the metric and Weyl scalar perturbations at $\mathcal{O}(\zeta^0,\epsilon^1)$ in the other source terms. After moving every smooth radial multiplier onto the distribution coefficients using the distributional properties of the Dirac delta function, we write the localized part as
\begin{align} \label{eq:localized_source}
    {}_{s}\mathscr{S}^{\ell m}_{\rm loc}(r)
    &=\sum_{n=0}^{2}{}_{s}C_{n,\rm loc}^{\ell m}(r_0)
     \delta^{(n)}(r-r_0)\,, \\
    {}_{s}C_{n,\rm loc}^{\ell m}
    &={}_{s}C_{n,p}^{\ell m}+{}_{s}C_{n,\rm int}^{\ell m}\,.
    \nonumber
\end{align}
Here ${}_{s}C_{n,p}^{\ell m}$ denotes the direct contribution from the point-particle source in ${}_{s}\mathscr{S}^{\ell m}_{p}$, while ${}_{s}C_{n,\rm int}^{\ell m}$ comes from differentiation of the metric and Weyl scalar fields at $\mathcal{O}(\zeta^0,\epsilon^1)$. For a smooth weight $f$ and an integration interval containing $r_0$ in its interior, the localized contribution is evaluated analytically:
\begin{equation}
    \int f(r)\,{}_{s}\mathscr{S}^{\ell m}_{\rm loc}(r)\,dr
    =\sum_{n=0}^{2}(-1)^n{}_{s}C_{n,\rm loc}^{\ell m}
    f^{(n)}(r_0)\,.
\end{equation}
For Eq.~\eqref{eq:Green_function}, the weights are $f=\Delta^s{}_{s}\mathcal{R}^{\rm in}_{\ell m}$ or $f=\Delta^s{}_{s}\mathcal{R}^{\rm up}_{\ell m}$. 

\subsubsection{Near-horizon regularization for $\Psi_0^{(2,1)}$}

Second, the near-horizon Green's function integrals for $\Psi_0^{(2,1)}$ require a separate treatment. With $y=r-r_+$, let us define the normalized source ${}_{2}\hat{\mathscr S}^{\ell m}=\Delta^2e^{ikr_*}{}_{2}\mathscr S^{\ell m}$.
Near the horizon, we find the source ${}_{2}\hat{\mathscr S}^{\ell m}$ to have the following decomposition:
\begin{equation} \label{eq:horizon_source_laurent}
    {}_{2}\hat{\mathscr S}^{\ell m}(y)
    =\frac{\mathcal P_{\ell m}}{y}
    +{}_{2}\hat{\mathscr S}_{\mathrm{reg}}^{\ell m}(y)\,,
\end{equation}
where $\mathcal P_{\ell m}$ is a constant coefficient, and the second term is regular as $y\to0$. To examine the behavior of the Green's function integrals in Eq.~\eqref{eq:Green_function} near the horizon, we follow Refs.~\cite{Yang:2026erj, Yang:2026csy} to make a Frobenius series expansion of the homogeneous solutions:
\begin{align} \label{eq:flux-horizon-frobenius}
    {}_{2}\mathcal{R}^{\rm in}_{\ell m}
    &={}_{2}B_{H}\Delta^{-2}e^{-ikr_*}P_{H}(y)\,,\\
    {}_{2}\mathcal{R}^{\rm up}_{\ell m}
    &={}_{2}D_{\rm in}\Delta^{-2}e^{-ikr_*}P_{\rm in}(y)
    +{}_{2}D_{\rm out}e^{ikr_*}P_{\rm out}(y)\,,\nonumber
\end{align}
where $\{P_{H},P_{\rm in},P_{\rm out}\}$ are regular power series in $y$, and we have kept the terms up to $y^{30}$. Their coefficients follow recursively from the homogeneous part of Eq.~\eqref{eq:MTF_radial}, with normalization matched to the BHPToolkit solutions \cite{BHPToolkit}. 

The ingoing-branch contribution to the integrand of the second integral in Eq.~\eqref{eq:Green_function} (i.e., $\propto {}_{2}D_{\mathrm{in}}$) takes the form
\begin{align} \label{eq:horizon_singular_part}
    & {}_{2}D_{\rm in}e^{-ikr_*}P_{\rm in}(y)
    \,{}_{2}\mathscr{S}^{\ell m}(y)
    ={}_{2}D_{\rm in}\mathcal{C}
    y^{\eta}e^{-2iky}B(y)\,, \nonumber\\
    & \mathcal{C}=(2M)^{4ikM}e^{-2ikr_+}\,,\;
    \eta=-2-2ikc_{+}\,, \nonumber\\
    & B(y)=P_{\rm in}(y)\,{}_{2}\hat{\mathscr{S}}^{\ell m}(y+r_+)
    (y+d)^{-2+2ikc_-}\,, 
\end{align}
where $d=r_+-r_-$ and $c_{\pm}=2Mr_{\pm}/d$. Given the pole in Eq.~\eqref{eq:horizon_source_laurent}, we perform a Laurent expansion of $B(y)$ :
\begin{equation}
    B(y)=\frac{B_{-1}}{y}
    +\sum_{n=0}^{N_B}B_ny^n\,,\quad
    B_{-1}=\mathcal P_{\ell m}d^{-2+2ikc_-}\,,
\end{equation}
where $P_{\rm in}(0)=1$.

At the horizon, we evaluate the integrals of these moments by analytic continuation:
\begin{align} \label{eq:near_horizon_moments}
    & \mathrm{AC}
    \left[\int_0^{y_c}y^{\nu-1}e^{-2iky}\,dy\right]_{\nu=\nu_n}
    =(2ik)^{-\nu_n}\gamma_L(\nu_n,2iky_c)\,, \nonumber\\
    & \nu_n=\eta+n+1\,,
\end{align}
where $\gamma_L$ is the lower incomplete gamma function, and $\mathrm{AC}$ denotes analytic continuation from $\mathrm{Re}\,\nu>0$ at fixed $k$ and $y_c>0$. The cutoff $y_c$ marks the outer boundary of the near-horizon integration interval. We retain 26 Laurent coefficients, corresponding to $n=-1,0,\ldots,24$. The pole in Eq.~\eqref{eq:horizon_source_laurent} contributes the additional moment with $\nu_{-1}=\eta$. For real, nonzero $k$, $\nu_n=n-1-2ikc_+$ has a nonzero imaginary part and therefore avoids the poles of $\gamma_L(\nu_n,z)$ at nonpositive integer $\nu_n$. We use consistent principal branches for the complex power and incomplete gamma function, with $\arg(2ik)=\mathrm{sgn}(k)\pi/2$. The regularized moments are consequently finite at fixed $y_c>0$, although the original improper integrals for $n=-1,0,1$ do not converge. The first Green's function integral in Eq.~\eqref{eq:Green_function} has the same behavior near the horizon and is evaluated with the same strategy. 

For the outgoing-branch contribution to the second integral in Eq.~\eqref{eq:Green_function} (i.e., $\propto {}_{2}D_{\mathrm{out}}$), the pole $\mathcal P_{\ell m}/y$ in Eq.~\eqref{eq:horizon_source_laurent} generates a logarithmic term in the near-horizon solution. Combining both terms in Eq.~\eqref{eq:Green_function} and factoring out the leading ingoing behavior gives
\begin{equation} \label{eq:horizon_finite_part}
    \Delta^2e^{ikr_*}{}_{2}\mathscr R_{\ell m}
    =\mathscr Z_{\ell m}^{H}
    +b_0^{H}\log(y/L_H)+\cdots\,,
\end{equation}
where $y=r-r_+$, $L_H$ is a reference length, and $\mathscr Z_{\ell m}^{H}$ denotes the finite horizon amplitude of the environmental correction $\Psi_0^{(2,1)}$. The coefficient $b_0^{H}$ is determined by the source pole in Eq.~\eqref{eq:horizon_source_laurent}. The finite horizon amplitude
$\mathscr Z_{\ell m}^{H}(L_H)$ is obtained by combining the finite contributions from both regularized Green's function integrals in Eq.~\eqref{eq:Green_function}, with the source matching and boundary conditions imposed.

The near-horizon logarithm is consistent with the phase response induced by the background correction to the surface gravity. For the Kerr metric, $r_*\sim\log(y/L_H)/(2\varkappa_0)$ near the horizon, where $\varkappa_0$ is the Kerr surface gravity. At fixed $k=\omega-m\omega_+$, the shift $\varkappa_0\rightarrow \varkappa_0+\zeta^2\varkappa^{(2,0)}$ therefore produces a relative correction $\zeta^2C_{\delta\kappa}\log(y/L_H)$
to the ingoing phase factor, where
\begin{equation}
    \label{eq:horizon_log_coeff}
    C_{\delta\kappa}
    =\frac{ik\,\varkappa^{(2,0)}}{2\varkappa_0^2}\,.
\end{equation}

To compare this prediction with the source-derived logarithmic amplitude coefficient $b_{0}^{H}$, we multiply the angularly dependent background coefficient by the full GR horizon angular field and project onto the same normalized angular basis:
\begin{equation}
    b_{0}^{\delta\kappa}
    =\int d\Omega\,{}_2\overline{S}_{\ell m}\,
    C_{\delta\kappa}
    \sum_{\ell'}\mathcal Z_{\ell'm}^{H}\,{}_2S_{\ell'm}\,,
\end{equation}
with mode labels on $b_0^H$, $b_0^{\delta\kappa}$
and $C_{\delta\kappa}$ suppressed. We then quantify the agreement via the relative difference $\left|b_{0}^{H}-b_{0}^{\delta\kappa}\right|/\left|b_{0}^{H}\right|$. Across the tested orbital radii, the maximum relative difference $\lesssim1.4\times10^{-7}$ for the diagonal modes $\ell=|m|$,  strongly supporting the surface-gravity interpretation of this logarithmic contribution. For all modes with $|m|\geq2$, including the non-diagonal modes $\ell>|m|$, the maximum relative difference remains $\lesssim 3.38\%$. Agreement is less accurate for $|m|=1$, where the extracted source poles also show greater sensitivity to the fitting interval, and our construction of the sources becomes less accurate. For real $k$ and $\varkappa^{(2,0)}$, $C_{\delta\kappa,m}$ is purely imaginary, so the associated logarithmic term does not contribute to the interference term $\mathrm{Re}\left[{}_{2}\bar{\mathcal{R}}_{\ell m}\;{}_{2}\mathscr{R}_{\ell m}\right]$ for the $\mathcal{O}(\zeta^2,\epsilon^2)$ contribution to the horizon flux in Eq.~\eqref{eq:horizon_flux}. Outside the near-horizon interval, the remaining contributions are finite and evaluated numerically.

\subsubsection{Near-infinity regularization for $\Psi_4^{(2,1)}$}

Third, extracting the outgoing amplitude of $\Psi_4^{(2,1)}$ from Eq.~\eqref{eq:Green_function} requires regularization at large radius. Although the scalar cloud profile decays exponentially, the corresponding background metric deformation has a long-range power-law tail. Coupling this deformation to the outgoing vacuum-GR perturbation results in the following asymptotic expansion of the geometrical source ${}_{-2}\mathscr S_{\geo}^{\ell m}$:
\begin{equation} \label{eq:psi4_source_asymp}
    {}_{-2}\mathscr{S}_{\geo}^{\ell m}
    =e^{i\omega r_*}
    \left(t_4r^4+t_3r^3+t_2r^2+\cdots\right)\,.
\end{equation}
For this source, we seek an asymptotic particular solution
of Eq.~\eqref{eq:MTF_radial} in the form
\begin{align}
    {}_{-2}\mathscr R_{\ell m}^{p,\infty}
    &\sim {}_{-2}\mathscr R_{\ell m}^{p,\log}
    +e^{i\omega r_*}\sum_{j\geq1}
    a_j^\infty r^{3-j}\,, \nonumber\\
    {}_{-2}\mathscr R_{\ell m}^{p,\log}
    &=\frac{b_0^\infty}{{}_{-2}D_\infty}
    {}_{-2}\mathcal R_{\ell m}^{\rm up}(r)
    \log(r/L_\infty)\,,
    \label{eq:infinity_log_puncture}
\end{align}
where $L_\infty$ is a reference length and ${}_{-2}D_\infty$ is the outgoing amplitude defined in Eq.~\eqref{eq:hom_asymp}. Since ${}_{-2}\mathcal R_{\ell m}^{\rm up}$ solves the homogeneous part of Eq.~\eqref{eq:MTF_radial}, applying $\mathcal H_4$ to ${}_{-2}\mathscr R_{\ell m}^{p,\log}$ produces no explicit logarithmic terms and yields the leading contribution $2i\omega b_0^\infty r^4e^{i\omega r_*}$. Setting $b_0^\infty=t_4/(2i\omega)$ therefore solves the source term $\propto r^4$ in Eq.~\eqref{eq:psi4_source_asymp}. The coefficients $a_j^\infty$ then account for the remaining lower powers of $r$ in the asymptotic source after this subtraction.

Now, we can subtract the source generated by ${}_{-2}\mathscr{R}_{\ell m}^{p,\infty}$ from ${}_{-2}\mathscr{S}_{\geo}^{\ell m}$ and solve the rest of the contribution via the Green's function. Using Eq.~\eqref{eq:MTF_radial}, one can show that
\begin{align}
    & {}_{-2}\mathcal{R}_{\ell m}^{\rm in}
    \Delta^{-2}\mathcal{H}_4
    \left[{}_{-2}\mathscr{R}_{\ell m}^{p,\infty}\right] \nonumber\\
    & =\partial_r\left({}_{-2}W_{\ell m}
    \left[{}_{-2}\mathcal{R}_{\ell m}^{\rm in},{}_{-2}\mathscr{R}_{\ell m}^{p,\infty}\right]\right)
    \equiv \partial_r\mathcal{W}_{\ell m}\,,
\end{align}
where ${}_{s}W_{\ell m}[u,v]$ means replacing ${}_{s}\mathcal{R}^{\rm in}_{\ell m}$ and ${}_{s}\mathcal{R}^{\rm up}_{\ell m}$ appearing in the Wronskian ${}_{s}W_{\ell m}$ of Eq.~\eqref{eq:Green_function} by $u$ and $v$. Then after subtracting the outer boundary contribution generated by ${}_{-2}\mathscr R_{\ell m}^{p,\infty}$, the regularized first Green's function integral in Eq.~\eqref{eq:Green_function} takes the form
\begin{align} \label{eq:infinity_reg}
    & \int_{r_+}^{r_m}\Delta^{-2}
    {}_{-2}\mathcal{R}_{\ell m}^{\rm in}\,
    {}_{-2}\mathscr{S}_{\ell m}\,dr'
    -\mathcal{W}_{\ell m}(r_m)\nonumber\\
    & +\int_{r_m}^{\infty}\Delta^{-2}
    {}_{-2}\mathcal{R}_{\ell m}^{\rm in}
    \left({}_{-2}\mathscr{S}_{\ell m}
    -\mathcal{H}_4\left[{}_{-2}\mathscr{R}_{\ell m}^{p,\infty}\right]\right)dr'\,,
\end{align}
where $r_m$ is the matching radius beyond which we regularize the integral by subtracting the source generated by $\mathscr{R}_{\ell m}^{p,\infty}$, i.e., $\mathcal{H}_4[{}_{-2}\mathscr{R}_{\ell m}^{p,\infty}]$, from ${}_{-2}\mathscr{S}_{\ell m}$.  

We then fit the remaining radial dependence of the log-subtracted Green's function integral to the boundary contributions generated by the non-logarithmic terms of ${}_{-2}\mathscr R_{\ell m}^{p,\infty}$, together with a constant determining the finite outgoing amplitude. We retain five such terms, generated by $e^{i\omega r_*}\{r^2,r,1,r^{-1},r^{-2}\}$, in addition to the fitted constant, and use the interval $300\leq r/M\leq389$. We have tested similar expansions with three to eight non-logarithmic terms across all the harmonic modes. For the five-term fit, changing the fitting interval from $300\leq r/M\leq389$ to $250\leq r/M\leq389$ changes the mode-summed $\mathcal{O}(\zeta^2,\epsilon^2)$ correction to the flux at infinity due to the geometrical source ${}_{-2}\mathscr{S}_{\geo}^{\ell m}$ by at most $0.897\%$ over the full orbital-radius range.

In total, the leading contribution to ${}_{-2}\mathscr R_{\ell m}$ at infinity takes the form
\begin{equation} \label{eq:psi4_MTF_infty}
    r^{-3}e^{-i\omega r_*}{}_{-2}\mathscr{R}_{\ell m}
    =\mathscr Z_{\ell m}^{\infty}
    +b_0^{\infty}\log(r/L_{\infty})+\cdots\,,
\end{equation}
where $\mathscr Z_{\ell m}^{\infty}$ is the finite outgoing amplitude obtained from the regularized Green's function solution, and the mode dependence of $b_0^{\infty}$ is suppressed. Note that the reference length scale $L_{\infty}$ is fixed as a combination with $\mathscr{Z}^{\infty}_{\ell m}$. Similar to the logarithmic term at the horizon, we find that the logarithmic contribution in Eq.~\eqref{eq:psi4_MTF_infty} also corresponds to a correction to the asymptotic tortoise coordinate caused by the deformed background. To examine this interpretation, we write
\begin{equation}
    b_0^{\infty}=i\omega\alpha_{\rm eff}\mathcal{Z}^{\infty}_{\ell m}\,,
\end{equation}
where $\alpha_{\rm eff}$ is a generally complex mode-dependent coefficient and $\mathcal{Z}^{\infty}_{\ell m}$ is the amplitude of $r^{3}e^{i\omega r_{*}}$ in the expansion of ${}_{-2}\mathcal{R}_{\ell m}$ at infinity. At the perturbative order considered, a common, real value of $\alpha_{\rm eff}$ would correspond to a shift proportional to $\alpha_{\rm eff}\log(r/L_{\infty})$ in the tortoise coordinate. The resulting logarithm would then change the wave's phase without changing the $\mathcal{O}(\zeta^2,\epsilon^2)$ flux at null infinity.

Our numerical results provide support for this interpretation for the diagonal modes $|m|=\ell$, with $\ell=2,\ldots,5$ and $5.1\leq r_0/M\leq15.1$. Using three outgoing fitting terms over $300\leq r/M\leq389$, we find $\mathrm{Re}[\alpha_{\rm eff}]\in[0.02539,0.02624]$ in our normalization, while $|\mathrm{Im}[\alpha_{\rm eff}]|/|\mathrm{Re}[\alpha_{\rm eff}]|\leq5.14\%$. The clustering of the real parts and the comparatively small imaginary parts are consistent with a predominantly phase-like logarithmic contribution. This agreement becomes less uniform at larger orbital radii, where the extracted coefficients also show greater sensitivity to the fitting prescription. This suggests that extending the radial domain of the computed source may be necessary to resolve its asymptotic behavior more reliably. Nevertheless, the flux data at these larger orbital radii are not used in the dephasing calculation presented above, which follows an inward inspiral starting at $r_0=10.6M$.

To assess the logarithmic contamination of the gravitational-wave flux corrections at the horizon and infinity, we define
\begin{align}
    \epsilon_{\log}^{A}
    &=\frac{|Q_{\log}^{A}|}
    {|\dot E_{h,A}^{\mathrm{Env}}|}\,,
    \nonumber\\
    Q_{\log}^{\infty}
    &=\sum_{\ell,m}
    \frac{\operatorname{Re}
    [\bar{\mathcal Z}_{\ell m}^{\infty}b_0^\infty]}
    {2\pi\omega^2}
    \log\left(\frac{r_{\max}}{L_\infty}\right)\,,
    \nonumber\\
    Q_{\log}^{H}
    &=\sum_{\ell,m}
    \frac{\omega\operatorname{Re}
    [\bar{\mathcal Z}_{\ell m}^{H}b_0^H]}
    {32\pi k(r_+^2+a^2)^3(k^2+\varkappa_0^2)}
    \log\left(\frac{r_{\min}-r_+}{L_H}\right)\,,
\end{align}
where $A\in\{H,\infty\}$, and $r_{\max}=389M$ and $r_{\min}=r_++10^{-3}M$ are the outer and inner radii at which the infinity and horizon flux corrections are extracted, respectively. If the logarithmic terms correspond solely to the background-induced phase shifts discussed above, they leave the fluxes unchanged at the order considered, giving $Q_{\log}^{A}=0$. Residual departures from this condition, due to possible numerical contamination, give nonzero $Q_{\log}^{A}$, and $\epsilon_{\log}^{A}$ measures their relative contamination to the $\mathcal{O}(\zeta^2,\epsilon^2)$ flux corrections. Summing over all the modes considered in this work and setting $L_\infty=r_0$, we find that $\epsilon_{\log}^{\infty}$ ranges from $0.26$ to $0.74\%$ at the sampled radii $5.1\leq r_0/M\leq12.1$ and from $1.07$ to $2.07\%$ for $r_0\geq15.1M$. Near the sign-changing cancellation at $r_0=14.1M$, as shown in Fig.~\ref{fig:fluxes_total}, $\epsilon_{\log}^{\infty}$ increases to $16.49\%$, reflecting the small finite-flux denominator rather than an increase in $|Q_{\log}^{\infty}|$. At the horizon, adopting $L_H=M$ gives $\epsilon_{\log}^{H}$ between $4.23\times10^{-7}$ and $3.06\times10^{-6}$ across the sampled orbital radii, indicating an extremely small logarithmic contribution relative to the $\mathcal{O}(\zeta^2,\epsilon^2)$ horizon-flux correction at the stated evaluation radius.

In summary, our results support interpreting the leading logarithmic contributions at the horizon and infinity as phase shifts induced by the deformed background geometry. Such phase shifts leave the fluxes unchanged at $\mathcal{O}(\zeta^2,\epsilon^2)$. The residual logarithmic interference in our numerical results is generally small at the stated evaluation radii, although its relative contribution increases near the zero of the infinity-flux correction. A more systematic treatment remains for future work, including investigating whether an asymptotically adapted gauge can remove the associated logarithmic behavior, following the approach of Ref.~\cite{Spiers:2023cip}. Apart from these distributional and boundary contributions, the remaining regular source terms are evaluated by
numerical Green-function quadrature.

\subsection{Derivation of the horizon flux formula}

Here, we derive the precise form of Eq.~\eqref{eq:horizon_flux} by relating $\Psi_0$ to the shear of the horizon generators, and then relating the shear to the energy into the horizon. We closely follow the prescription in \cite{Hawking:1972hy, Chandrasekhar_1983, Poisson:2004cw}. 

Let us first transform our tetrad to a horizon-regular one. At $\mathcal{O}(\zeta^0,\epsilon^1)$, we use the Hawking-Hartle tetrad in ingoing Kerr coordinates 
\begin{align} \label{eq:ingoing_coord}
    & v=t+r_*\,,\quad
    \tilde\phi=\phi+\varphi_*\,,\nonumber\\
    & \frac{dr_*}{dr}=\frac{r^2+a^2}{\Delta}\,,\quad
    \frac{d\varphi_*}{dr}=\frac{a}{\Delta}\,,
\end{align}
with the integration constants fixed to
\begin{align}
    & r_*=r+\frac{2Mr_+}{r_+-r_-}\log\frac{r-r_+}{2M}
    -\frac{2Mr_-}{r_+-r_-}\log\frac{r-r_-}{2M}\,,\nonumber\\
    & \varphi_*=\frac{a}{r_+-r_-}\log\frac{r-r_+}{r-r_-}\,,
\end{align}
such that the future horizon is regular in $(v,r,\theta,\tilde\phi)$ and
\begin{align}
    \tilde{l}^\mu=\Lambda l^{\mu}\,,\quad
    \tilde{n}^\mu=\Lambda^{-1}n^{\mu}\,,\quad
    \Lambda=\frac{\Delta}{2(r^2+a^2)}\,,
\end{align}
or more explicitly,
\begin{align} \label{eq:tetrad_HH}
    & \tilde{l}^\mu
    =\left(1,\Lambda,0,\frac{a}{r^2+a^2}\right)\,,\;
    \tilde{n}^\mu
    =\left(0,-\frac{r^2+a^2}{\Sigma},0,0\right)\,,\nonumber\\
    & \tilde{m}^\mu
    =\frac{1}{\sqrt2\Gamma}
    \left(ia\sin\theta,0,1,i\csc{\theta}\right)\,.
\end{align}
Under such a rotation, $\Psi_0\rightarrow \Lambda^2\Psi_0$ \cite{Chandrasekhar_1983}, so on the horizon, we can decompose the rotated $\Psi_0^{(0,1)}$ and $\Psi_0^{(2,1)}$, denoted by $\psi$ and $\delta\psi$, respectively, as
\begin{align}
    \psi
    =\mathcal{P}_m\,\mathcal{Z}^H_{\ell m}\,
    {}_2S_{\ell m}\,,\quad
    \delta\psi
    =\mathcal{P}_m\,\mathscr{Z}^H_{\ell m}\,
    {}_2S_{\ell m}\,,
\end{align}    
with the finite radial amplitudes $\mathcal{Z}^H_{\ell m}$ and $\mathscr{Z}^H_{\ell m}$, where
\begin{align} \label{eq:horizon_amplitudes}
    \mathcal{Z}^H_{\ell m}
    &=\lim_{r\to r_+}\Delta^2e^{ikr_*}
    {}_2\mathcal R_{\ell m}\,,
\end{align}
and $\mathscr{Z}^H_{\ell m}$ is defined in Eq.~\eqref{eq:horizon_finite_part}. The function $\mathcal{P}_m=(4Mr_+)^{-2}e^{-im\varsigma_H}$, and the factor $e^{-im\varsigma_H}$ comes from rewriting the following phase in ingoing Kerr coordinates:
\begin{equation}
    e^{-ikr_*}e^{-i\omega t+im\phi}
    \rightarrow e^{-i\omega v+im\tilde{\phi}}
    e^{-im(\varphi_*-\omega_+r_*)}\,.
\end{equation}
The last factor has the finite horizon limit $e^{-im\varsigma_H}$, with
\begin{equation}
    \varsigma_H=-\frac{a}{2M}-\frac{a}{r_+}\log\frac{r_+-r_-}{2M}\,.
\end{equation}
For convenience, we use the superscript $f^H$ or subscript $f_H$ to denote the horizon value of any function $f$.

Due to the perturbations to the metric at $\mathcal{O}(\zeta^0,\epsilon^1)$ and $\mathcal{O}(\zeta^2,\epsilon^0)$, the horizon's location is shifted. To fix the horizon location at $r=r_+$, we first solve the following condition at $r=r_+$ to determine the perturbed horizon location in the current coordinates:
\begin{equation} \label{eq:horizon_generator}
    g^{\mu\nu}F_{,\mu}F_{,\nu}=0\,,
\end{equation}
where $F$ defines the horizon location:
\begin{equation}
    F=r-r_+-\epsilon\xi^{(0,1)}
    -\zeta^2\xi^{(2,0)}+\cdots=0\,,
\end{equation}
where $\xi^{(0,1)}$ and $\xi^{(2,0)}$ are $\mathcal{O}(\zeta^0,\epsilon^1)$ and $\mathcal{O}(\zeta^2,\epsilon^0)$ shifts to the horizon's location of the Kerr metric at $r=r_{+}$, respectively. We then perform a coordinate transformation in $r$: $r\rightarrow r+\epsilon\xi^{(0,1)}+\zeta^2\xi^{(2,0)}$. Expanding Eq.~\eqref{eq:horizon_generator} to $\mathcal{O}(\zeta^0,\epsilon^1)$ and evaluating it at $r=r_+$, we get
\begin{equation}
    \frac{r_+-r_-}{\Sigma_H}\xi^{(0,1)}
    -\frac{4Mr_+}{\Sigma_H}D_H\xi^{(0,1)}
    -\frac{4M^2r_+^2}{\Sigma_H^2}h_{ll}^{H(0,1)}=0\,,
\end{equation}
where $D_H=\partial_v+\omega_+\partial_{\tilde\phi}$, 
and $h_{ll}^{(0,1)}$ is in the Hawking-Hartle tetrad. After a harmonic decomposition, $D_H\rightarrow -ik$, so we get
\begin{align} \label{eq:horizon_lock_coord}
    \xi^{(0,1)}
    =\frac{2Mr_+ h_{ll}^{H(0,1)}}
    {2\Sigma_H(\varkappa_0+ik)}\,,
\end{align}
where we select the periodic solution without the growing
homogeneous displacement. The same expression holds for $\xi^{(2,0)}$ after replacing $h_{ll}^{H(0,1)}$ with $h_{ll}^{H(2,0)}$. On the other hand, our background metric solver has fixed the horizon location at its vacuum-GR value, and we verify that the maximal value of $h_{ll}^{H(2,0)}$ over various angular locations is $\lesssim 10^{-13}$, so we set $\xi^{(2,0)}=0$ in what follows.

Furthermore, to keep the horizon area formula below in Eq.~\eqref{eq:area_rho} consistent with the vacuum-GR one, we fix the normalization of the full $l^{\mu}$, including $\mathcal{O}(\zeta^2,\epsilon^0)$ and $\mathcal{O}(\zeta^0,\epsilon^1)$ corrections, on the horizon to be $l_H^{v}=1$, i.e.,
\begin{equation} \label{eq:tetrad_fix_norm}
    l_H^\mu
    =\left.\frac{g^{\mu\nu}F_{,\nu}}{g^{v\nu}F_{,\nu}}
    \right|_{F=0}\,.
\end{equation}
Although the Hawking-Hartle tetrad in Eq.~\eqref{eq:tetrad_HH} satisfies this condition, the perturbed tetrad we used above for evaluating the modified Teukolsky equations does not automatically satisfy it, so we perform the following tetrad rotations on the horizon
\begin{align} \label{eq:tetrad_rot_fix_norm}
    & l_H^{\mu} 
    \rightarrow \mathscr{A}_H\left(l_H^{\mu}
    +\bar{\mathfrak{b}}_H m_H^{\mu}
    +\mathfrak{b}_H\bar{m}_H^{\mu}
    +\mathfrak{b}_H\bar{\mathfrak{b}}_H n_H^{\mu}\right)\,, \nonumber\\
    & n_H^{\mu}\rightarrow \mathscr{A}_H^{-1}n_H^{\mu}\,, \quad
    m_H^{\mu}\rightarrow m_H^{\mu}+\mathfrak{b}_H\,n_H^{\mu}\,,
\end{align}
where $\mathfrak{b}_H$ and $\mathscr{A}_H$ are the rotation parameter of a type II tetrad rotation and the real boost factor of a type III tetrad rotation, respectively, in \cite{Chandrasekhar_1983}. Since we only need to fix the normalization at $\mathcal{O}(\zeta^2,\epsilon^0)$ and $\mathcal{O}(\zeta^0,\epsilon^1)$, we can take these parameters to be perturbative, i.e., $\mathscr{A}_H=1+\epsilon\mathscr{A}_{H}^{(0,1)}
+\zeta^2\mathscr{A}_{H}^{(2,0)}$ and $\mathfrak{b}=\epsilon\mathfrak{b}_{H}^{(0,1)}+\zeta^2\mathfrak{b}_{H}^{(2,0)}$ such that
\begin{equation} \label{eq:tetrad_rot_fix_norm_param}
    \mathscr{A}_H^{(p,q)}
    =-\tilde{n}^{\mu}_{H}d_{\mu}^{(p,q)}\,,\;
    \mathfrak{b}_H^{(p,q)}
    =\tilde{m}^{\mu}_{H}d_{\mu}^{(p,q)}\,,
\end{equation}
where $d^{\mu(p,q)}$ is the difference between the target tetrad $l_{\mathrm{T}}^{\mu(p,q)}$ in Eq.~\eqref{eq:tetrad_fix_norm} and the tetrad $l_{\mathrm{MTF}}^{\mu(p,q)}$ used in deriving the modified Teukolsky equation on the horizon:
\begin{equation}
    d^{\mu(p,q)}
    \equiv\left(l_{\mathrm{T}}^{\mu(p,q)}
    -l_{\mathrm{MTF}}^{\mu(p,q)}\right)_H\,.
\end{equation}

After the above coordinate transformations and tetrad rotations, the Weyl scalar perturbation $\Psi_0^{(2,1)}$ transforms as
\begin{align} \label{eq:psi0_horizon_adapted}
    & \Psi_0^{(2,1)}
    \rightarrow\Psi_0^{(2,1)}+\mathcal{G}^{(2,1)}\,,\nonumber\\
    & \mathcal G^{(2,1)}
    =\xi^{(0,1)}\partial_r\Psi_0^{(2,0)}
    +\delta\Psi_{0,\mathrm{frame}}^{(2,1)}\,,
\end{align}
where $\delta\Psi_{0,\mathrm{frame}}^{(2,1)}$ is due to the tetrad rotations in Eqs.~\eqref{eq:tetrad_rot_fix_norm} and \eqref{eq:tetrad_rot_fix_norm_param}:
\begin{align}
    \delta\Psi_{0,\mathrm{frame}}^{(2,1)}
    =12\mathfrak{b}_H^{(2,0)}\mathfrak{b}_H^{(0,1)}\Psi_2\,,
\end{align}
where we have used the property of the perturbed tetrad before the above rotations in Eq.~\eqref{eq:psi1_psi3}. Furthermore, $\mathscr{A}_H^{(2,0)}=\Psi_0^{(2,0)}=0$. The precise transformation rule of $\Psi_0$ under tetrad rotations can be found in \cite{Chandrasekhar_1983}.  In the following, we will work in the Hawking-Hartle tetrad in the ingoing Kerr coordinates of Eqs.~\eqref{eq:ingoing_coord} and \eqref{eq:tetrad_HH} at $\mathcal{O}(\zeta^0,\epsilon^0)$. At $\mathcal{O}(\zeta^2,\epsilon^0)$ and $\mathcal{O}(\zeta^0,\epsilon^1)$, we use the horizon-locking coordinates in Eq.~\eqref{eq:horizon_lock_coord} with the same tetrad for evaluating the modified Teukolsky equations above after the additional rotations in Eqs.~\eqref{eq:tetrad_rot_fix_norm} and \eqref{eq:tetrad_rot_fix_norm_param} to fix the normalization of $l_H^{\mu}$. 

Now, to derive the horizon flux formula, we start from the first law of black hole thermodynamics, which still applies to a black hole with minimally coupled scalar hair \cite{Herdeiro:2014goa}:
\begin{equation}
    \dot{E}_{H}-\omega_{+}\dot{L}_{H}
    =\frac{\varkappa}{8\pi}\dot{\mathcal{A}}\,.
\end{equation}
Here, $\dot{E}_{H}$ and $\dot{L}_{H}$ are the fluxes of energy and angular momentum through the event horizon, with positive values corresponding to transfer into the black hole. The quantities $\omega_{+}$, $\varkappa$ and $\mathcal{A}$ denote the horizon angular velocity, surface gravity and area, respectively. An overdot denotes differentiation with respect to the advanced-time coordinate $v$, so that $\dot{\mathcal{A}}=d\mathcal{A}/dv$. For harmonic perturbations of a stationary, axisymmetric background, each harmonic mode of the energy and angular momentum fluxes satisfies $\dot{L}^{\ell m}=m\dot{E}^{\ell m}/\omega$, so
\begin{equation} \label{eq:energy_area_flux}
    \dot{E}^{\ell m}_{H}
    =\frac{\omega\varkappa}{8\pi k}
    \dot{\mathcal{A}}^{\ell m}\,,
\end{equation}
where $k=\omega-m\omega_{+}$ and $\dot{\mathcal{A}}^{\ell m}$ denotes each harmonic contribution to the averaged horizon-area growth rate.

On the other hand, the horizon area change $\dot{\mathcal{A}}$ can be defined in terms of the expansion $\vartheta$ of the horizon generators \cite{Chandrasekhar_1983}. In the NP formalism, the horizon area element is 
\begin{equation}
    dA=\sqrt{q_H} \, d^2x\,,
\end{equation}
where $q_H$ is the determinant of the $2D$ metric $q_{\mu\nu}^{H}$ on the horizon, i.e.,
\begin{equation}
    q^{\mu\nu}_{H}=m^\mu_{H}\bar{m}^\nu_{H}+\bar{m}^\mu_{H}m^\nu_{H}\,,
\end{equation}
and the tetrad vectors $\{m^{\mu}_{H},\bar{m}^{\mu}_{H}\}$ are evaluated on the horizon. Given the expansion of the null congruences $\vartheta_{H}$ on the horizon,
\begin{equation}
    \vartheta_{H}=q^{\mu\nu}_{H}\nabla_\nu l_{\mu}^{H}
    =-(\rho_{H}+\bar{\rho}_{H})
    =-2\operatorname{Re}[\rho_{H}]\,,
\end{equation}
where $\rho$ is one of the spin coefficients, we find that $\dot{\mathcal{A}}$ is
\begin{equation} \label{eq:area_rho}
    \dot{\mathcal{A}}
    =\int_{S_v}\vartheta_{H}\,dA
    =-2\int_{S_v}\operatorname{Re}[\rho_H]\,dA\,,
\end{equation}
where $S_v$ is the horizon cross-section. On the other hand, $\varpi=\mathrm{Im}[\rho]$ is the twist of the null congruences. The twist enters, in general, into Raychaudhuri's equation [i.e.~Eq.~\eqref{eq:rho_reconstruct}], but due to the horizon properties in Eq.~\eqref{eq:horizon_properties}, the twist does not enter our equations, and the horizon area change is only determined by the expansion, which is proportional to $\mathrm{Re}[\rho]$.

To relate $\vartheta$ or $\rho$ to $\Psi_0$ on the horizon, we need to do a partial metric reconstruction. Following the prescription in \cite{Chandrasekhar_1983}, we use the following two Ricci identities:
\begin{subequations}
\begin{align}
    & D_{[-1,-1,-1,0]}\rho-\bar{\delta}_{[-3,-1,1,0]}\kappa
    -\sigma\bar{\sigma}+\bar{\kappa}\tau=\Phi_{00}\,, \\
    & D_{[-3,1,-1,-1]}\sigma-\delta_{[-1,-3,1,-1]}\kappa
    =\Psi_0\,.
\end{align}
\end{subequations}
In a horizon-adapted tetrad, $\kappa_{H}=0$ \cite{Poisson:2004cw}, so at the horizon, these two equations become
\begin{subequations}
\begin{align}
    & D_{[-1,-1,-1,0]}^{H}\rho_H-\sigma_H\bar{\sigma}_H=\Phi_{00}^{H}\,, 
    \label{eq:rho_reconstruct} \\
    & D_{[-3,1,-1,-1]}^{H}\sigma_{H}
    =\Psi_0^{H}\,, \label{eq:sigma_reconstruct}
\end{align}
\end{subequations}
where the spin coefficient $\sigma_H$ controls the shear of the horizon generators, and Eq.~\eqref{eq:rho_reconstruct} is Raychaudhuri's equation \cite{Poisson:2004cw}. 

For a stationary black hole, its null horizon generators have zero expansion, twist, and shear, so 
\begin{equation} \label{eq:horizon_properties}
    \sigma_H^{(0,0)}=\sigma_H^{(2,0)}
    =\rho_H^{(0,0)}=\rho_H^{(2,0)}=0\,.
\end{equation}
The above relation at $\mathcal{O}(\zeta^0,\epsilon^0)$ can be analytically verified from the spin coefficients of the Kerr spacetime in the Hawking-Hartle tetrad. At $\mathcal{O}(\zeta^2,\epsilon^0)$, we have also verified the above relation numerically, with the largest absolute residual $\lesssim 10^{-14}$. Furthermore, the term $\Phi_{00}^{H}$ in Eq.~\eqref{eq:rho_reconstruct} enters at $\mathcal{O}(\zeta^2,\epsilon^2)$, as $\Phi_{00}\propto T_{ll}\propto|l^\mu\nabla_\mu\Phi|^2$ and $l^{\mu}\nabla_{\mu}\Phi^{(1,0)}=0$ on the horizon. The latter condition comes from that our scalar frequency is synchronized to the horizon frequency for a stationary background. Thus, $\rho_H^{(0,1)}$ and $\rho_H^{(2,1)}$ satisfy
\begin{subequations}
\begin{align}
    & (D_H-\varkappa_0)\rho_H^{(0,1)}=0\,, \\
    & (D_H-\varkappa_0)\rho_H^{(2,1)}
    =-(D_H-\varkappa)^{(2,0)}\rho_H^{(0,1)}\,,
\end{align}
\end{subequations}
where we again drop the superscripts of $\mathcal{O}(\zeta^0,\epsilon^0)$ quantities for simplicity. Along each horizon generator, the homogeneous solution $\rho_H^{(0,1)}\propto e^{\varkappa_0v}$ grows exponentially as $v\to+\infty$. Requiring bounded late-time expansion eliminates this solution, giving $\rho_H^{(0,1)}=0$. The equation for $\rho_H^{(2,1)}$ is then homogeneous, and the same boundary condition gives $\rho_H^{(2,1)}=0$. Consequently, the horizon-area growth and associated energy flux begin at $\mathcal O(\epsilon^2)$. At this order, we obtain
\begin{subequations} \label{eq:rho_H_equations}
\begin{align}
    (D_H-\varkappa_0)\rho_H^{(0,2)}
    =&\; \left|\sigma_H^{(0,1)}\right|^2\,, \\
    (D_H-\varkappa_0)\rho_H^{(2,2)}
    =& \;\varkappa^{(2,0)}\rho_H^{(0,2)}
    +2\mathrm{Re}\left[\sigma_H^{(0,1)}\bar{\sigma}_H^{(2,1)}\right]\,,
\end{align}
\end{subequations}
where $\varkappa=2\,\mathrm{Re}[\varepsilon]$, and $\varepsilon$ is one of the spin coefficients. In particular, $\varkappa_0=\varkappa^{(0,0)}$ is the surface gravity of Kerr spacetime. In above, we have dropped the term $\Phi_{00}^{H(2,2)}$, as this stress-energy tensor contribution has already been incorporated in the scalar energy change at the horizon computed in \cite{Dyson:2025dlj, Li:2025ffh}. Furthermore, since we have locked the horizon angular frequency in our background metric solver, $D_H^{(2,0)}=0$. The equations of $\rho_H^{(0,2)}$ and $\rho_H^{(2,2)}$ can then be solved hierarchically once $\sigma_H^{(0,1)}$ and $\sigma_H^{(2,1)}$ are solved from Eq.~\eqref{eq:sigma_reconstruct}. 

To get $\sigma_H^{(0,1)}$ and $\sigma_H^{(2,1)}$, we expand Eq.~\eqref{eq:sigma_reconstruct} to $\mathcal{O}(\zeta^2,\epsilon^1)$, so
\begin{subequations}
\begin{align}
    & (D_H-\varkappa_0)\sigma_H^{(0,1)}
    =\Psi_{0,H}^{(0,1)}\,, \\
    & (D_H-\varkappa_0)\sigma_H^{(2,1)}
    =\Psi_{0,H}^{(2,1)}+\varkappa^{(2,0)}\sigma_H^{(0,1)}\,.
\end{align}
\end{subequations}
In above, we have set $\mathrm{Im}[\varepsilon^{(2,0)}]=0$, the maximal numerical residual of which is also $\lesssim 10^{-12}$. For the same reason as above, the homogeneous solutions of $\sigma_H^{(0,1)}$ and $\sigma_H^{(2,1)}$ are dropped. Thus, after harmonic decomposition, the solutions to $\sigma_H^{(0,1)}$ and $\sigma_H^{(2,1)}$ are
\begin{subequations} \label{eq:sigma_H_results}
\begin{align}
    & \sigma_H^{(0,1)}
    =-\frac{\psi}{\varkappa_0+ik}\,, \\
    & \sigma_H^{(2,1)}
    =-\frac{\delta\tilde{\psi}+\varkappa^{(2,0)}\sigma_{H}^{(0,1)}}
    {\varkappa_0+ik}\,,
\end{align}
\end{subequations}
where $\delta\tilde{\psi}$ is $\Psi_0^{(2,1)}$ in the Hawking-Hartle tetrad after the additional coordinate transformation in Eq.~\eqref{eq:psi0_horizon_adapted} to fix the horizon location.

\begin{figure*}[t!]
  \centering
  \includegraphics[width=\linewidth]{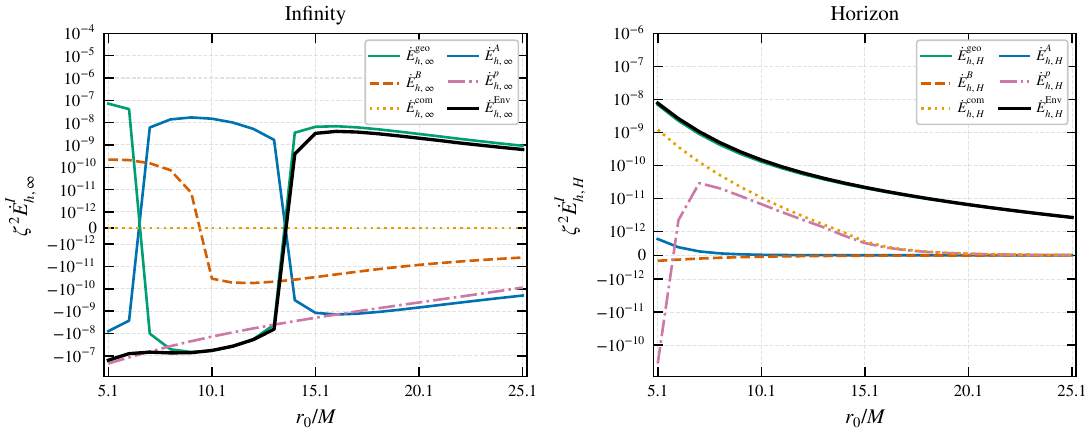}
  \caption{Source decomposition of the environmental correction to the gravitational-wave energy flux at infinity (left) and the horizon (right) for $M_c/M=10^{-4}$ ($\zeta=2.7\times10^{-4}$). The green, blue, orange, pink, and yellow curves show $\dot{E}_{h,H/\infty}^{I}$, $I\in\{\geo,A,B,p,\mathrm{com}\}$, respectively. The black curve gives the signed total. All curves include the factor $\zeta^2$, with $\epsilon^2$ scaled out.} 
  \label{fig:source_contributions}
\end{figure*}

Now, we can solve Eq.~\eqref{eq:rho_H_equations}. Following \cite{Chandrasekhar_1983}, we notice that the terms $D_H\rho_H^{(0,2)}$ and $D_H\rho_H^{(2,2)}$ can be removed after orbit averaging, 
\begin{equation}
    \left\langle D_H\rho_{H}^{(p,q)}\right\rangle_T
    =\frac{\rho_H^{(p,q)}(v+T)-\rho_H^{(p,q)}(v)}{T}\,,
\end{equation}
where $\langle \cdot \rangle$ denotes orbit averaging, and $T$ is the orbital period. Then, Eq.~\eqref{eq:rho_H_equations} can be easily solved, so using Eqs.~\eqref{eq:energy_area_flux}, \eqref{eq:area_rho}, and \eqref{eq:sigma_H_results}, we find
\begin{align} \label{eq:horizon_flux_GR}
    \dot E^{\ell m(0,2)}_{h,H}
    =& \;\frac{\omega}{4\pi k(\varkappa_0^2+k^2)}
    \int_H\left|\psi\right|^2\,dA\,, \nonumber\\
    =& \;\frac{\omega|\mathcal Z^H_{\ell m}|^2}
    {64\pi k(2Mr_+)^3(\varkappa_0^2+k^2)}\,,
\end{align}
where we have used that $dA=2Mr_+d\Omega$ for the Kerr spacetime. Given Eq.~\eqref{eq:horizon_amplitudes}, we find that Eq.~\eqref{eq:horizon_flux_GR} agrees with the $\mathcal{O}(\zeta^0)$ term in Eq.~\eqref{eq:horizon_flux}. Similarly, at $\mathcal{O}(\zeta^2,\epsilon^2)$, we obtain 
\begin{align} \label{eq:horizon_flux_env}
    \dot E^{\ell m(2,2)}_{h,H}
    =& \;\frac{\omega}{4\pi k(\varkappa_0^2+k^2)}\int_H
    \left[2\mathrm{Re}[\bar\psi\delta\tilde{\psi}]
    +\mathfrak w^{(2,0)}|\psi|^2\right]dA\,, \nonumber\\
    \mathfrak{w}^{(2,0)}
    =& \;\mathfrak{a}^{(2,0)}
    -\frac{2\varkappa_0\varkappa^{(2,0)}}{\varkappa_0^2+k^2} \nonumber\\
    & \;+\left(\frac{\omega^{(2,0)}}{\omega}
    -\frac{k^{(2,0)}}{k}
    -\frac{2kk^{(2,0)}}{\varkappa_0^2+k^2}\right)\,,
\end{align}
where $\mathfrak a^{(2,0)}=\tfrac12q_0^{\mu\nu}\delta q_{\mu\nu}^{(2,0)}$ is the fractional area-element correction coefficient, and $q_0^{\mu\nu}$ is the horizon screen projector evaluated on the Kerr background. The last term of $\mathfrak{w}^{(2,0)}$ comes from the corrections to the orbital energy and horizon frequency on a background deviating from Kerr spacetime. 

Finally, to get the correction term $\mathcal{F}$ to the horizon flux formula, we can define
\begin{align}
    & U_m=\sum_\ell\mathcal{Z}^H_{\ell m}\,{}_2S_{\ell m}\,,\;
    V_m=\sum_\ell\mathscr{Z}^H_{\ell m}\,{}_2S_{\ell m}\,,\nonumber\\
    & G_m=\mathcal{P}_m^{-1}\mathcal{G}_m^{(2,1)}\,,
\end{align}
where $\mathcal{G}_m^{(2,1)}$ is the $m$-mode of the transformation $\mathcal{G}^{(2,1)}$ in Eq.~\eqref{eq:psi0_horizon_adapted}. Then $\psi_m=\mathcal P_m U_m$ and
$\delta\tilde{\psi}_m=\mathcal P_m(V_m+G_m)$. With
$d\Omega=\sin\theta\,d\theta\,d\tilde\phi$, we find from Eq.~\eqref{eq:horizon_flux_env} that
\begin{align} \label{eq:horizon_flux_env_final}
    \dot{E}_{h,H}^{\ell m(2,2)}
    =& \frac{\omega}{64\pi k(2Mr_+)^3(\varkappa_0^2+k^2)}\times \nonumber\\
    & \;\Big(2\mathrm{Re}
    \left[\bar{\mathcal Z}^H_{\ell m}\mathscr Z^H_{\ell m}\right]
    +\mathfrak{F}_{\ell m}^H\Big)\,,
\end{align}
where the correction term $\mathfrak{F}_{\ell m}^{H}$ is
\begin{equation}
    \mathfrak{F}_{\ell m}^{H}
    =\operatorname{Re}\int d\Omega\,
    \bar{\mathcal{Z}}_{\ell m}^{H}\,{}_2\bar{S}_{\ell m}
    \left[2G_m+\mathfrak{w}^{(2,0)}U_m\right]\,.
\end{equation}
Comparing Eq.~\eqref{eq:horizon_flux_env_final} to Eq.~\eqref{eq:horizon_flux}, we then find that $\mathcal{F}=\Delta^{-4}\mathfrak{F}_{\ell m}^{H}$.

\subsection{Mode-by-mode fluxes and convergence tests}

Here, we analyze the contributions of different source sectors and harmonic modes to the gravitational-wave flux corrections induced by the ultralight scalar cloud. We then assess their sensitivity to the harmonic-mode cutoffs of both the input fields $\Phi^{(1,1)}$, $h_{\mu\nu}^{(0,1)}$, and $h_{\mu\nu}^{(2,0)}$, and the output radiative perturbations $\Psi_{0,4}^{(2,1)}$.

Using the source decomposition in Eq.~\eqref{eq:radial_source}, we separate the environmental gravitational-wave flux corrections into five contributions:
\begin{equation} \label{eq:supp_flux_sector_sum}
    \dot{E}_{h,H/\infty}^{\mathrm{Env}}
    =\sum_{I\in\{\geo,A,B,p,\mathrm{com}\}}
    \dot{E}_{h,H/\infty}^{I}\,,
\end{equation}
where $I=\geo,A,B,p$ denote contributions associated with the background metric deformation, the parts of $T_{\mu\nu}^{\Phi}$ independent of and dependent on $\Phi^{(1,1)}$, and the point-particle stress-energy tensor $T_{\mu\nu}^{p}$, respectively. Both $\dot{E}_{h,H/\infty}^{\geo}$ and $\dot{E}_{h,H/\infty}^{A}$ include their respective particle-localized contributions. The remaining term, $\dot{E}_{h,H/\infty}^{\mathrm{com}}$, accounts for the correction to the flux formula itself. Since the infinity flux formula is unchanged, $\dot{E}_{h,\infty}^{\mathrm{com}}=0$.

Figure~\ref{fig:source_contributions} shows these five contributions. At infinity, the negative $\dot{E}_{h,\infty}^{p}$ is the largest in magnitude until $r_0\approx 7M$, whereas $\dot{E}_{h,\infty}^{\geo}$ dominates at the remaining sampled radii. The contribution $\dot{E}_{h,\infty}^{A}$ has the opposite sign to $\dot{E}_{h,\infty}^{\geo}$ and partially offsets it, while $\dot{E}_{h,\infty}^{B}$ remains subdominant throughout the sampled range. Near $r_0=14M$, cancellation between the positive $\dot{E}_{h,\infty}^{\geo}$ and the negative $\dot{E}_{h,\infty}^{A}$ and $\dot{E}_{h,\infty}^{p}$ produces the sign change in the total correction. At the horizon, $\dot{E}_{h,H}^{\geo}$ dominates throughout the sampled range, with $\dot{E}_{h,H}^{\mathrm{com}}$ becoming significant only at small radii.

\begin{figure*}[t!]
  \centering
  \includegraphics[width=\linewidth]{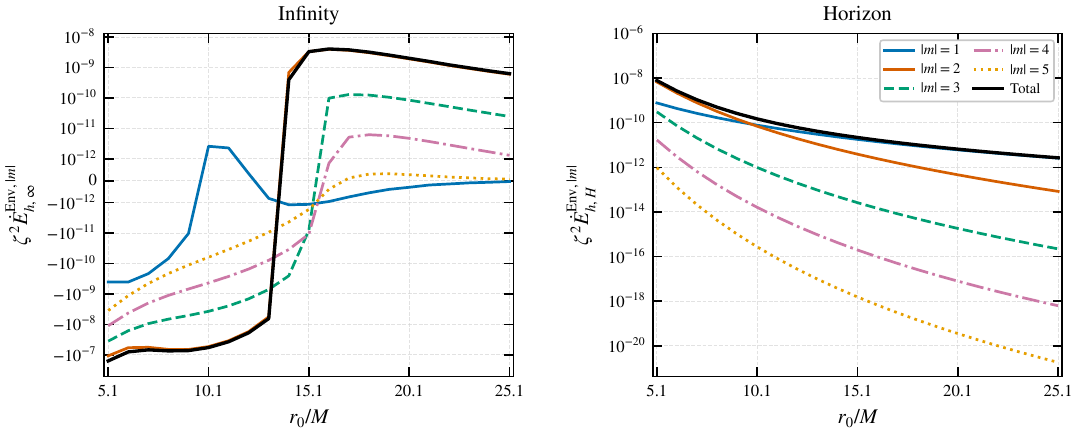}
  \caption{$|m|$-mode decomposition of the environmental correction to the gravitational-wave energy flux at infinity (left) and the horizon (right) for $M_c/M=10^{-4}$ ($\zeta=2.7\times10^{-4}$). The panels show $\dot{E}_{h,H/\infty}^{\mathrm{Env},|m|}$, with each $|m|$ contribution obtained by summing the signed $m$ and $-m$ modes over the allowed $\ell$. The blue, orange, green, pink, and yellow curves denote $|m|=1,\ldots,5$, respectively, and the black curve denotes the total $\dot{E}_{h,H/\infty}^{\mathrm{Env}}$. All curves include the factor $\zeta^2$, with $\epsilon^2$ scaled out.} 
  \label{fig:m_contributions}
\end{figure*}

Next, we group the gravitational-wave flux corrections by $|m|=1,\ldots,5$, summing both signs of $m$ over $\max(2,|m|)\leq\ell\leq5$. Figure~\ref{fig:m_contributions} shows the resulting contributions $\dot{E}_{h,H/\infty}^{\mathrm{Env},|m|}$. At infinity, $|m|=2$ dominates throughout the sampled orbital-radius range, with $|m|=3$ generally providing the next subdominant contribution. At the horizon, $|m|=2$ dominates at small orbital radii, but $|m|=1$ overtakes it near $r_0\approx10M$ and becomes increasingly dominant at larger radii. At both boundaries, the contributions from $|m|\geq3$ generally decrease in magnitude with increasing $|m|$.

With the output sum fixed at $2\leq\ell\leq5$, we first assess the sensitivity of the gravitational-wave flux corrections to the input-field cutoffs relative to the main-text baseline $(\ell_b^{\max},\ell_g^{\max},\ell_s^{\max})=(14,8,6)$ at five orbital radii, $r_0/M\in\{5.1,10.1,15.1,20.1,25.1\}$. These cutoffs correspond to $h_{\mu\nu}^{(2,0)}$, $h_{\mu\nu}^{(0,1)}$, and $\Phi^{(1,1)}$, respectively. Increasing $\ell_b^{\max}=14$ to $\ell_b^{\max}=16$ while holding the other cutoffs fixed changes $\dot{E}_{h,\infty}^{\mathrm{Env}}$ and $\dot{E}_{h,H}^{\mathrm{Env}}$ by at most $5.33\%$ and $1.08\%$, respectively. Similarly, increasing $\ell_g^{\max}=8$ to $\ell_g^{\max}=10$ gives maximum relative changes of $3.73\%$ and $0.61\%$. Changing $\ell_s^{\max}=6$ to $\ell_s^{\max}=8$, with the other source-sector contributions held at their baseline values, gives maximum fractional changes of $2.14\times10^{-9}$ and $7.89\times10^{-11}$ at infinity and the horizon, respectively. Finally, we extend the output-mode sum for $\Psi_{0,4}^{(2,1)}$ from $\ell_{\max}=5$ to $\ell_{\max}=6$ at the same five radii, keeping the input cutoffs fixed at $(14,8,6)$. The maximum fractional changes in $\dot E_{h,\infty}^{\mathrm{Env}}$ and $\dot E_{h,H}^{\mathrm{Env}}$ are $7.20\times10^{-3}$ and $7.44\times10^{-6}$, respectively, both at $r_0=5.1M$.


\end{document}